\documentclass[aps,prb,twocolumn,superscriptaddress,showpacs,english]{revtex4-1}
\usepackage[T1]{fontenc}
\usepackage[latin9]{inputenc}
\usepackage{babel}
\usepackage{amsmath}
\usepackage{amssymb}
\usepackage{graphicx}
\usepackage{xcolor}
\usepackage{esint}
\definecolor{mydarkgreen}{rgb}{0.0,0.5,0.0}
\usepackage[linktocpage=true,
  colorlinks=true, 
  pdfborder={0 0 0},
  linkcolor=blue,
  citecolor=red,
  filecolor=yellow,
  urlcolor=mydarkgreen,
  bookmarks,
  pdftitle={paper},
  pdfauthor={},
]{hyperref} 

\makeatletter

\usepackage{babel}
\usepackage{upgreek}

\makeatother

\begin{document}

\title{Ab Initio Theory of Superconductivity in a Magnetic Field I. : Spin
Density Functional Theory For Superconductors and Eliashberg Equations.}

\author{A. Linscheid}
\affiliation{Max Planck Institute of Microstructure Physics, Weinberg 2, D-06120 Halle, Germany.}
\author{A. Sanna}
\affiliation{Max Planck Institute of Microstructure Physics, Weinberg 2, D-06120 Halle, Germany.}
\author{F. Essenberger}
\affiliation{Max Planck Institute of Microstructure Physics, Weinberg 2, D-06120 Halle, Germany.}
\author{E.K.U. Gross}
\affiliation{Max Planck Institute of Microstructure Physics, Weinberg 2, D-06120 Halle, Germany.}

\begin{abstract}
We present a first-principles approach to describe
magnetic and superconducting systems and the phenomena of competition
between these electronic effects. We develop a density functional
theory: SpinSCDFT, by extending the Hohenberg-Kohn theorem and constructing
the non-interacting Kohn-Sham system. An exchange-correlation functional
for SpinSCDFT is derived from the Sham Schl\"uter connection
between the SpinSCDFT Kohn-Sham and a self-energy in Eliashberg approximation.
The reference Eliashberg equations for superconductors in the presence of magnetism are also derived
and discussed.
\end{abstract}
\maketitle

\section{Introduction}

In this work, we present how magnetic (M) and superconducting (SC)
properties can be computed on the same footing and from first principles 
by extending the Density Functional Theory (DFT)  framework. 
In developing this spin DFT for SC (SpinSCDFT) we will  
restrict ourselves to situations where currents are negligible
and only consider the effect of the Zeeman term of the Hamiltonian. 
Under this assumption we can exclude the occurrence of the Abrikosov vortex state
\cite{AbrikosovTheMagneticPropertiesOfSCAlloys1957}, that having a mesoscopic characteristic
length-scale would be beyond the present computational power for a fully ab-initio method.

The expulsion of static M fields from the bulk \cite{MeissnerOchsenfeld1933}
is one of the most spectacular properties of SC materials and illustrates
the profound competition between M and SC behavior. The SC-M interaction
generates in fact a large number of interesting phenomena on which
the scientific community has focused attention. 
Some of the most investigated are the 
Abrikosov vortices\cite{AbrikosovTheMagneticPropertiesOfSCAlloys1957}
and the variety of fascinating effects occurring in heterostructures\cite{SatapathyMagneticProximityEffectInSuperlattices2012},
such as stacked layers of M with SC material (see Ref.~\onlinecite{BuzdinProximityEffectsInSCFerromagnetHeterostructures2005}
for a review). Among these effects is the FFLO state, named after
Fulde, Ferrel \cite{FuldeFerrellSuperconductivityInAStrongSpinExchangeField1964},
Larkin and Ovchinnikov \cite{LarkinOvchinnikovInhomogeneousStateOfSuperconductors1965},
where strong exchange fields induce a SC state with a finite momentum
pairing. This state was recently observed experimentally \cite{BianchiPossibleFFLOSuperconductingStateCeCoIn5_2003,LortzCalorimetricEvidenceForAFFLOSuperconductingState2007}
in heavy Fermion SC, many years after its prediction. In addition,
triplet SC has been observed in several systems\cite{SigristUedaUnconventionalSC_RMP1991,JeromeOrganic1980,RiceSr2RuO41995,Luke_TimeRev_SrRuO3_1998,IshidaTripletSrRuO3_1998,SaxenaSCOnTheBorderOfItinerantElectronFerromagnetismInUGe2_2000,Bauer_CePtSi_2004}, and is usually associated to ferromagnetism.

Among the many effects generated by the interplay of magnetism and superconductivity,
some have an intrinsic microscopic nature and could be accessible to first-principle calculations,
in particular we refer to the sharp suppression of the critical temperature
due to paramagnetic impurities\cite{AbrikosovGorkovSCalloysParamagneticImpurities1960},
and the surprising evidence of coexisting phases between singlet
SC and local magnetism, in particular close to a magnetic phase boundary
\cite{FelnerCoexistenceOfMagnetismAndSCInRCeRuSeCuO1997,DikinEtAlCoexistenceOfSuperconductivityAndFerromagnetismInTwoDimensions2011,SaxenaSCOnTheBorderOfItinerantElectronFerromagnetismInUGe2_2000,Aoki2001}
where high$-T_{c}$ SC occurs \cite{manskeTheoryOfUnconventionalSC2004,LeeDopingMottInsulatorPhysicsOfHTcSC2006}.
We devote this work to set the ground for an ab-initio theory to describe these physical effects.

We will start our formulation from the Pauli Hamiltoninan (Sec.~\ref{sec:Hamiltonian-and-definitions}).
In Sec.~(\ref{sec:SpinSCDFT}), we formulate a density functional
theory (DFT), proving that the electronic density $n(\boldsymbol{r})$
, the spin magnetization $\boldsymbol{m}(\boldsymbol{r})$, the diagonal
of the nuclear $N$-body density matrix and the singlet and triplet
SC order parameters $\boldsymbol{\chi}(\boldsymbol{r},\boldsymbol{r}^{\prime})$
are uniquely connected with their respective external potentials.
With this extension of the Hohenberg-Kohn theorem \cite{HohenbergKohn1964}
we lay the foundation of the DFT for M and SC systems: SpinSCDFT.
In Sec.~\ref{sub:The-Kohn-Sham-System} we introduce the formally
non-interacting Kohn Sham (KS) system that reproduces the exact densities
of the interacting system. Similar to every DFT, SpinSCDFT relies
on the construction of an exchange correlation ($xc$) functional
that connects the KS with the interacting system. In this work, this
is achieved by establishing, in Sec.~\ref{sub:The-Sham-Schluter-Equation},
a Sham-Schl\"uter connection \cite{ShamSchlueterDFTOfTheEnergyGap1983}
via the Dyson equation of the interacting system. 

The interacting system is also being investigated directly by means of a magnetic extension 
of the Eliashberg method \cite{VonsovskySuperconductivityTransitionMetals,AllenTheoryOfSCTc1983,SannaPittalis2012,CarbottePropertiesOfBosonExchangeSC1990,EliashbergInteractionBetweenElAndLatticeVibrInASC1960}.
A derivation\footnote{Schossmann and Schachinger have derived Eliashberg equations including the vector potential
\cite{SchossmannStrongCouplingTheoryOfTheUpperCriticalMagneticField1986}. However,
they set out from a self-energy that is taken to be local in real space with an empirical electron phonon coupling.
It is not straightforward to generalize their approach to the case of ab-initio calculations,
where the pairing interactions are usually taken to be local in the space of normal state quasi particles.
Vonsovsky \textit{et al.} \cite{VonsovskySuperconductivityTransitionMetals} have derived Eliashberg equations,
treating the magnetic field perturbatively except for an on site splitting parameter.
They require the self-energy to be diagonal with respect to normal-state electronic
orbitals which is similar to the main results in this work.}
of this alternative approach in the present notation is given in Sec.~\ref{sec:The-Eliashberg-equations}.
Advantages and disadvantages of these two theoretical schemes, SpinSCDFT
and Eliashberg, will be discussed in the conclusions.

\section{Hamiltonian \label{sec:Hamiltonian-and-definitions}}

We assume that the interacting system is governed by the Pauli Hamiltonian
(we use Hartree atomic units throughout)
\begin{equation}
\hat{H}=\hat{T}_{{\rm {\scriptscriptstyle e}}}+\hat{T}_{{\rm {\scriptscriptstyle n}}}+\hat{V}_{{\scriptscriptstyle {\rm e}}}+\hat{U}_{{\rm {\scriptscriptstyle ee}}}+\hat{U}_{{\rm {\scriptscriptstyle en}}}+\hat{U}_{{\rm {\scriptscriptstyle nn}}}\,,\label{eq:FullInteractionHamiltonian}
\end{equation}
where $\hat{T}_{{\rm {\scriptscriptstyle e}}}$ ($\hat{T}_{{\rm {\scriptscriptstyle n}}}$)
is the kinetic energy operator of the electron (nuclei) and $\hat{U}_{{\rm {\scriptscriptstyle ee}}}$($\hat{U}_{{\rm {\scriptscriptstyle nn}}}$)
is the electron-electron (nuclei-nuclei) interaction, i.e.~usually
the Coulomb potential. $\hat{U}_{{\rm {\scriptscriptstyle en}}}$
is the Coulomb potential between electrons and nuclei. To break the
respective symmetries and allow the corresponding densities to adopt
non-zero values in a thermal average we include an external vector
potential $\boldsymbol{A}_{{\rm {\scriptscriptstyle ext}}}(\boldsymbol{r})$
and an external singlet/triplet pair potential $\boldsymbol{\varDelta}^{{\rm {\scriptscriptstyle ext}}}(\boldsymbol{r},\boldsymbol{r}^{\prime})$
in the Hamiltonian. These external fields will be set to zero at the
end of the derivation. Because we do not consider currents, the only
term in the Pauli Hamiltonian containing $\boldsymbol{A}_{{\rm {\scriptscriptstyle ext}}}(\boldsymbol{r})$
is: 
\begin{equation}
\hat{T}_{{\rm {\scriptscriptstyle e}}}=\int\hspace{-0.18cm}{\rm d}\boldsymbol{r}\,\hat{\psi}^{\dagger}(\boldsymbol{r})\cdot\Bigl(-\sigma_{0}\frac{\boldsymbol{\nabla}^{2}}{2}+\mathbf{S}\cdot\boldsymbol{B}_{{\scriptscriptstyle {\rm ext}}}(\boldsymbol{r})\Bigr)\cdot\hat{\psi}(\boldsymbol{r})
\end{equation}
with $\boldsymbol{B}_{{\scriptscriptstyle {\rm ext}}}(\boldsymbol{r})=\boldsymbol{\nabla}\times\boldsymbol{A}_{{\rm {\scriptscriptstyle ext}}}(\boldsymbol{r})$
and $\mathbf{S}=\frac{1}{2}(\begin{array}{ccc}
\sigma_{x} & \sigma_{y} & \sigma_{z}\end{array})^{{\rm T}}$, $\sigma_{x,y,z}$ being the Pauli matrices. We use the notation
$\hat{\psi}^{\dagger}(\boldsymbol{r})=\bigl(\begin{array}{cc}
\hat{\psi}^{\dagger}(\boldsymbol{r}\uparrow) & \hat{\psi}^{\dagger}(\boldsymbol{r}\downarrow)\end{array}\bigr)$ for the field operator where $\hat{\psi}^{\dagger}(\boldsymbol{r}\uparrow)$
creates an electron at location $\boldsymbol{r}$ with spin up. The
scalar potential part of $\hat{H}$ reads:

\begin{eqnarray}
\hat{V}_{{\scriptscriptstyle {\rm e}}} & = & \int\hspace{-0.18cm}{\rm d}\boldsymbol{r}\,\hat{\psi}^{\dagger}(\boldsymbol{r})\cdot\sigma_{0}\cdot\hat{\psi}(\boldsymbol{r})v_{{\scriptscriptstyle {\rm ext}}}(\boldsymbol{r})\nonumber \\
 &  & \!\!-\frac{1}{2}\int\hspace{-0.18cm}{\rm d}\boldsymbol{r}\int\hspace{-0.18cm}{\rm d}\boldsymbol{r}^{\prime}\Bigl(\hat{\boldsymbol{\chi}}(\boldsymbol{r},\boldsymbol{r}^{\prime})\cdot\boldsymbol{\varDelta}^{{\rm {\scriptscriptstyle ext}}\ast}(\boldsymbol{r},\boldsymbol{r}^{\prime})+{\rm h.c.}\Bigr)\,.
\end{eqnarray}
Here, the anomalous density operator is defined by 

\begin{eqnarray}
\hat{\boldsymbol{\chi}}(\boldsymbol{r},\boldsymbol{r}^{\prime}) & = & \hat{\psi}(\boldsymbol{r})\cdot\boldsymbol{\Phi}\cdot\hat{\psi}(\boldsymbol{r}^{\prime})\,.
\end{eqnarray}
$\hat{\boldsymbol{\chi}}(\boldsymbol{r},\boldsymbol{r}^{\prime})$
is a 4-vector of which the first component (proportional to $\Phi_{1}$)
is the singlet part of the order parameter, while the other components
(related to $\Phi_{2}$,$\Phi_{3}$ and $\Phi_{4}$) are the triplet
part. The 4 components of the singlet/triplet vector $\boldsymbol{\Phi}=(\mbox{i}\sigma_{y},-\sigma_{z},\sigma_{0},\sigma_{x})^{{\rm T}}$
are $2\times2$ spin matrices similar to the components of $\mathbf{S}$.
Similarly, the anomalous external potential
\begin{eqnarray}
\boldsymbol{\varDelta}^{{\rm {\scriptscriptstyle ext}}}(\boldsymbol{r},\boldsymbol{r}^{\prime}) & = & \left(\begin{array}{c}
\varDelta_{{\rm s}}^{{\rm {\scriptscriptstyle ext}}}(\boldsymbol{r},\boldsymbol{r}^{\prime})\\
\varDelta_{{\rm tx}}^{{\rm {\scriptscriptstyle ext}}}(\boldsymbol{r},\boldsymbol{r}^{\prime})\\
\varDelta_{{\rm ty}}^{{\rm {\scriptscriptstyle ext}}}(\boldsymbol{r},\boldsymbol{r}^{\prime})\\
\varDelta_{{\rm tz}}^{{\rm {\scriptscriptstyle ext}}}(\boldsymbol{r},\boldsymbol{r}^{\prime})
\end{array}\right)
\end{eqnarray}
is assumed to have singlet and triplet components.

\section{Spin SCDFT\label{sec:SpinSCDFT}}

The conventional density functional approach to the Many-Body problem
\cite{HohenbergKohn1964,KohnSham1965,RungeTDDFT1984,KreibichMultiCompDFT2001}
consists of two steps: \textit{first} establishing the Hohenberg Kohn
(HK) theorem, i.e.~ realize that a chosen set of densities is uniquely
connected with a set of external potentials; \textit{second}, construct
an auxiliary, non-interacting KS system to reproduce the densities
of the interacting system.

We follow Ref.~\onlinecite{LuedersSCDFTI2005} and consider a multi-component
DFT with the normal $n(\boldsymbol{r})$, the SC order parameter as
the anomalous density $\boldsymbol{\chi}(\boldsymbol{r},\boldsymbol{r}^{\prime})$,
that describes the electrons condensed into singlet and triplet states,
and $\varGamma(\boldsymbol{R}_{1}..\boldsymbol{R}_{N})$ the diagonal
of the nuclear $N$-body density matrix. In addition, we introduce
the magnetization $\boldsymbol{m}(\boldsymbol{r})$ as another electronic
density.

The HK proof $\bigl(n(\boldsymbol{r}),\boldsymbol{m}(\boldsymbol{r}),\boldsymbol{\chi}(\boldsymbol{r},\boldsymbol{r}^{\prime}),\varGamma(\boldsymbol{R}_{1}..\boldsymbol{R}_{N})\bigr)\leftrightarrow\bigl(v_{{\rm {\scriptscriptstyle ext}}}(\boldsymbol{r}),\boldsymbol{B}_{{\rm {\scriptscriptstyle ext}}}(\boldsymbol{r}),\boldsymbol{\varDelta}^{{\rm {\scriptscriptstyle ext}}}(\boldsymbol{r},\boldsymbol{r}^{\prime}),W_{{\rm {\scriptscriptstyle ext}}}(\boldsymbol{R}_{1}..\boldsymbol{R}_{N})\bigr)$
is a straightforward generalization of Mermin's HK proof in a finite
temperature ensemble\cite{MerminHoKoforFiniteT1965}. For this
reason we will not repeat it here. On the other hand the construction
of the KS system is done assuming that densities are always $v-$representable
i.e.~we assume the existence of the KS system. Being non-interacting
 it consists of independent equations for nuclei and electrons,
coupled only via the $xc$ potentials. Our focus will be on the electronic
system, discussed in detail in Sec.~\ref{sub:The-Electronic-Part}.
The nuclear part will be addressed in Sec.~\ref{sub:The-Nuclear-Part},
briefly, since it is usually enough to approximate the nuclear KS system
with its non SC counterpart\cite{LuedersSCDFTI2005,MarquesSCDFTIIMetals2005}.
The construction the $xc$ potentials will be discussed in Sec.~\ref{sub:The-Sham-Schluter-Equation}
and Sec.~\ref{sub:Derivation-xc-Potential}.

\subsection{The Kohn-Sham System\label{sub:The-Kohn-Sham-System}}

In this work we are mainly interested in the influence of a magnetic
field on the SC state. We briefly review the approximation steps
to arrive at the Fröhlich Hamiltonian starting from the formally exact
multi-component DFT. The reader may refer to the existing literature
for further details\cite{LuedersSCDFTI2005,KreibichMultiCompDFT2001}.
We introduce the KS Hamiltonian
\begin{equation}
\hat{H}_{{\scriptscriptstyle {\rm KS}}}=\hat{H}_{{\scriptscriptstyle {\rm KS}}}^{{\rm {\scriptscriptstyle e}}}+\hat{H}_{{\scriptscriptstyle {\rm KS}}}^{{\scriptscriptstyle {\rm n}}}\,,
\end{equation}
where we have separated the electronic $\hat{H}_{{\scriptscriptstyle {\rm KS}}}^{{\rm {\scriptscriptstyle e}}}$
\begin{eqnarray}
\hat{H}_{{\scriptscriptstyle {\rm KS}}}^{{\rm {\scriptscriptstyle e}}} & = & \int\hspace{-0.18cm}{\rm d}\boldsymbol{r}\,\hat{\psi}^{\dagger}(\boldsymbol{r})\cdot\sigma_{0}\Bigl(-\frac{\boldsymbol{\nabla}^{2}}{2}+v_{{\scriptscriptstyle {\rm s}}}(\boldsymbol{r})-\mu\Bigr)\cdot\hat{\psi}(\boldsymbol{r})\nonumber \\
 &  & -\frac{1}{2}\int\hspace{-0.18cm}{\rm d}\boldsymbol{r}\int\hspace{-0.18cm}{\rm d}\boldsymbol{r}^{\prime}\Bigl(\hat{\boldsymbol{\chi}}(\boldsymbol{r},\boldsymbol{r}^{\prime})\cdot\boldsymbol{\varDelta}^{{\scriptscriptstyle {\rm s}}\ast}(\boldsymbol{r},\boldsymbol{r}^{\prime})+{\rm h.c.}\Bigr)\nonumber \\
 &  & +\int\hspace{-0.18cm}{\rm d}\boldsymbol{r}\,\hat{\boldsymbol{m}}(\boldsymbol{r})\cdot\boldsymbol{B}_{{\scriptscriptstyle {\rm s}}}(\boldsymbol{r})\,,\label{eq:SPEQ-electronHamilRSpace}
\end{eqnarray}
from the nuclear $\hat{H}_{{\scriptscriptstyle {\rm KS}}}^{{\scriptscriptstyle {\rm n}}}$
\begin{eqnarray}
\hat{H}_{{\scriptscriptstyle {\rm KS}}}^{{\scriptscriptstyle {\rm n}}} & = & -\int\hspace{-0.18cm}{\rm d}\boldsymbol{R}\,\hat{\zeta}^{\dagger}(\boldsymbol{R})\frac{\boldsymbol{\nabla}_{\boldsymbol{R}}^{2}}{2M}\hat{\zeta}(\boldsymbol{R})+\nonumber \\
 &  & +\dotsint\hspace{-0.18cm}{\rm d}\boldsymbol{R}_{1}...{\rm d}\boldsymbol{R}_{N_{n}}\hat{\zeta}^{\dagger}(\boldsymbol{R}_{1})...\hat{\zeta}^{\dagger}(\boldsymbol{R}_{N_{n}})\times\nonumber \\
 &  & \times W_{{\scriptscriptstyle {\rm s}}}(\boldsymbol{R}_{1},...,\boldsymbol{R}_{N_{n}})\hat{\zeta}(\boldsymbol{R}_{1})...\hat{\zeta}(\boldsymbol{R}_{N_{n}}).\label{eq:SCDFT-KS nuclearHamil}
\end{eqnarray}
We write $v_{{\scriptscriptstyle {\rm s}}}(\boldsymbol{r})=v_{{\rm {\scriptscriptstyle ext}}}(\boldsymbol{r})+v_{{\scriptscriptstyle {\rm {\rm xc}}}}(\boldsymbol{r})$
with $v_{{\scriptscriptstyle {\rm {\rm xc}}}}(\boldsymbol{r})$ being
the scalar $xc$ potential (similar for $\boldsymbol{B}_{{\scriptscriptstyle {\rm s}}}(\boldsymbol{r})$
and $\boldsymbol{\varDelta}^{{\scriptscriptstyle {\rm s}}}(\boldsymbol{r},\boldsymbol{r}^{\prime})$).
$\hat{\boldsymbol{m}}(\boldsymbol{r})=\hat{\psi}^{\dagger}(\boldsymbol{r})\cdot\mathbf{S}\cdot\hat{\psi}(\boldsymbol{r})$
is the operator of the magnetic density. In the nuclear description,
$\hat{\zeta}^{\dagger}(\boldsymbol{R})$ creates the nuclear field
at location $\boldsymbol{R}$. Following L\"uders \textit{et al.}\cite{LuedersSCDFTI2005}
and Marques {\textit et al.}\cite{MarquesSCDFTIIMetals2005} we use the $N-$body potential
$W_{{\scriptscriptstyle {\rm s}}}(\boldsymbol{R}_{1},...,\boldsymbol{R}_{N_{n}})$
because in this way the nuclear KS system can be easily related to
the standard Born-Oppenheimer approximation. $M$ refers to the ionic
mass. Here, we neglect the spin of the nuclei and consider only one
atomic type (the generalization is straightforward).

\subsubsection{The Nuclear Part\label{sub:The-Nuclear-Part}}

Since SC occurs in the solid phase, we assume that ions can only perform
small oscillations about their equilibrium position. A discussion
that goes beyond this simple picture can be found in Ref.~\onlinecite{vanLeeuwenFirstPrinciplesApproachToTheElPhInteraction2004}
and \onlinecite{KreibichMultiCompDFT2001}. We expand $W_{{\scriptscriptstyle {\rm s}}}(\boldsymbol{R}_{1},...,\boldsymbol{R}_{N_{n}})$
in $\boldsymbol{u}_{i}=\boldsymbol{R}_{i}-{\boldsymbol{R}_{0}}_{i}$
around the equilibrium positions ${\boldsymbol{R}_{0}}_{i}$. The
nuclear degrees of freedom (up to harmonic order) are described by
the Hamiltonian $\hat{H}_{{\scriptscriptstyle {\rm KS}}}^{{\scriptscriptstyle {\rm ph}}}$
with $\hat{H}_{{\scriptscriptstyle {\rm KS}}}^{{\scriptscriptstyle {\rm n}}}=\hat{H}_{{\scriptscriptstyle {\rm KS}}}^{{\scriptscriptstyle {\rm ph}}}+\mathcal{O}(\boldsymbol{u}^{3})$
in second quantization 
\begin{equation}
\hat{H}_{{\scriptscriptstyle {\rm KS}}}^{{\scriptscriptstyle {\rm ph}}}=\sum_{q}\varOmega_{q}\bigl(\hat{b}_{q}^{\dagger}\hat{b}_{q}+\frac{1}{2}\bigr)\,.\label{eq:PhononZerothOrderSecondQuantizedForm}
\end{equation}
We use the notation $q=\boldsymbol{q},\lambda$ with Bloch vector
$\boldsymbol{q}$ and mode number $\lambda$. We further use the notation
$-q=-\boldsymbol{q},\lambda$ for all Bloch vector and band or mode
combinations. We point out that via the functional dependence of $W_{{\scriptscriptstyle {\rm s}}}[n,\boldsymbol{m},\boldsymbol{\chi},\varGamma]$
the KS phonon frequencies $\varOmega_{q}$ are in principle functionals
of the densities as well. $\hat{b}_{q}^{\dagger}$ creates a bosonic
KS phonon with quantum numbers $q$. Usually, approximating $W_{{\scriptscriptstyle {\rm s}}}$
with the Born-Oppenheimer energy surface, leads to phonon frequencies
 in excellent agreement with experiment \cite{BaroniPhononrev2001,GiannozziAbInitioCalculationOfPhDispersions1991}. 

The electron phonon scattering should be formally constructed from
the bare Coulomb interaction \cite{vanLeeuwenFirstPrinciplesApproachToTheElPhInteraction2004}.
However in order to have a proper description of the electronic screening
this is not feasible in practice. The solution is the substitution
of the many body electron phonon interaction with its Kohn Sham counterpart
$\hat{U}_{{\rm {\scriptscriptstyle en}}}\to\hat{H}_{{\scriptscriptstyle {\rm KS}}}^{{\scriptscriptstyle {\rm e-ph}}}$. 

\begin{equation}
\hat{H}_{{\scriptscriptstyle {\rm KS}}}^{{\scriptscriptstyle {\rm e-ph}}}=\sum_{q\, m}\int\hspace{-0.18cm}{\rm d}\boldsymbol{r}g_{q}^{a}(\boldsymbol{r})\hat{\psi}^{\dagger}(\boldsymbol{r})\cdot\sigma_{m}\cdot\hat{\psi}(\boldsymbol{r})\bigl(\hat{b}_{q}+\hat{b}_{-q}^{\dagger}\bigr)\,,\label{eq:KSElPhInteractionHamiltonian}
\end{equation}
where $m=0,z$ and $g_{q}^{0}(\boldsymbol{r})=\frac{\updelta v_{{\scriptscriptstyle {\rm s}}}(\boldsymbol{r})}{\updelta\boldsymbol{u}_{q}}$,
$g_{q}^{{\rm z}}(\boldsymbol{r})=\frac{\updelta{\boldsymbol{B}_{{\scriptscriptstyle {\rm s}}}}_{{\rm z}}(\boldsymbol{r})}{\updelta\boldsymbol{u}_{q}}$,
$\boldsymbol{u}$ being the phononic displacement vectors \cite{BaroniPhononrev2001,GiannozziAbInitioCalculationOfPhDispersions1991}.
This form incorporates most of the electronic influence on the bare
Coulomb interaction between electrons and nuclei. We consider this
as a good approximation for the dressed phonon vertex in the non-SC
state, see also Ref.~\onlinecite{vanLeeuwenFirstPrinciplesApproachToTheElPhInteraction2004}
for a further discussion. Note that $\langle\hat{H}_{{\scriptscriptstyle {\rm KS}}}^{{\scriptscriptstyle {\rm e-ph}}}\rangle$
is part of the $xc-$functional of the electronic KS system and will
be added later in our approximate functional using perturbation theory.
For later use in the derivation of the $xc$ potential, we define
the propagator of the non-interacting system of KS phonons
\begin{eqnarray}
D_{q,q^{\prime}}^{{\scriptscriptstyle {\rm 0}}}(\tau) & = & \langle\mathrm{T}\bigl(\hat{b}_{q}(\tau)+\hat{b}_{-q}^{\dagger}(\tau)\bigr)\bigl(\hat{b}_{q^{\prime}}(0)+\hat{b}_{-q^{\prime}}^{\dagger}(0)\bigr)\rangle_{{\scriptscriptstyle {\rm ph}}}\,,\nonumber \\
\\
D_{q,q^{\prime}}^{{\scriptscriptstyle {\rm 0}}}(\nu_{n}) & = & \updelta_{q,-q^{\prime}}\Bigl(\frac{1}{{\rm i}\nu_{n}+\varOmega_{q}}-\frac{1}{{\rm i}\nu_{n}-\varOmega_{q}}\Bigr)\,.
\end{eqnarray}
Here ${\rm T}$ is the usual time ($\tau$) ordering symbol of operators
$\hat{b}_{q}(\tau)+\hat{b}_{-q}^{\dagger}(\tau)$ in the Heisenberg
picture and $\langle\ldots\rangle_{{\scriptscriptstyle {\rm ph}}}$
means to evaluate the thermal average using the Hamiltonian $\hat{H}_{{\scriptscriptstyle {\rm KS}}}^{{\scriptscriptstyle {\rm ph}}}$
of Eq.~(\ref{eq:PhononZerothOrderSecondQuantizedForm}). The bosonic
Matsubara frequency is $\nu_{n}=\frac{2\pi n}{\beta}$.

\subsubsection{The Electronic Part\label{sub:The-Electronic-Part}}

The electronic KS Hamiltonian $\hat{H}_{{\scriptscriptstyle {\rm KS}}}^{{\rm {\scriptscriptstyle e}}}$
is not diagonal in the electronic field operator $\hat{\psi}(\boldsymbol{r})$
because Eq.~(\ref{eq:SPEQ-electronHamilRSpace}) involves terms proportional
to $\psi\psi$ and $\psi^{\dagger}\psi^{\dagger}$. Being a hermitian
operator, we can find an orthonormal set of eigenvectors of $\hat{H}_{{\scriptscriptstyle {\rm KS}}}^{{\rm {\scriptscriptstyle e}}}$
in which it is diagonal. Let $\hat{\gamma}_{k}^{\dagger}$ create
such a two component vector in spin space (the Hamiltonian is not
diagonal in spin so the spin degrees of freedom is in the set $\{k\}$),
then the SC KS system will take the form
\begin{equation}
\hat{H}_{{\scriptscriptstyle {\rm KS}}}^{{\rm {\scriptscriptstyle e}}}=E_{0}+\sum_{k}E_{k}\hat{\gamma}_{k}^{\dagger}\hat{\gamma}_{k}\quad E_{k}\geq0\;.\label{eq:SCDFT-KS digHamil}
\end{equation}
where $E_{0}$ is the ground state energy and the $E_{k}$ are all
positive. This form can be achieved \cite{noltingVielTeilchen2005}
by commuting the operators $\hat{H}_{{\scriptscriptstyle {\rm KS}}}^{{\rm {\scriptscriptstyle e}}}=\sum_{k}\tilde{E}_{k}\hat{a}_{k}^{\dagger}\hat{a}_{k}=\sum_{k|\tilde{E}_{k}<0}+\sum_{k|\tilde{E}_{k}\geq0}\tilde{E}_{k}\hat{a}_{k}^{\dagger}\hat{a}_{k}+\sum_{k|\tilde{E}_{k}<0}\vert\tilde{E}_{k}\vert\hat{a}_{k}\hat{a}_{k}^{\dagger}$
and then redefining the negative energy particle operators as holes
$\hat{a}_{k}=\hat{\gamma}_{k}^{\dagger}$. We use a notation that
is based on the one of Ref.~\onlinecite{NambuQParticlesGaugeInSuperconductivity1960},
\onlinecite{Anderson1958RPAInTheoryOfSuperconductivity} and \onlinecite{VonsovskySuperconductivityTransitionMetals}.
We introduce
\begin{equation}
\hat{\varPsi}(\boldsymbol{r})=\left(\begin{array}{c}
\hat{\psi}(\boldsymbol{r}\uparrow)\\
\hat{\psi}(\boldsymbol{r}\downarrow)\\
\hat{\psi}^{\dagger}(\boldsymbol{r}\uparrow)\\
\hat{\psi}^{\dagger}(\boldsymbol{r}\downarrow)
\end{array}\right)\,.
\end{equation}
Using this Nambu field operator $\hat{\varPsi}(\boldsymbol{r})$ the
KS Hamiltonian reads
\begin{eqnarray}
\hat{H}_{{\scriptscriptstyle {\rm KS}}}^{{\rm {\scriptscriptstyle e}}} & = & \int\hspace{-0.18cm}{\rm d}\boldsymbol{r}\int\hspace{-0.18cm}{\rm d}\boldsymbol{r}^{\prime}\hat{\varPsi}^{\dagger}(\boldsymbol{r})\cdot\frac{1}{2}\bar{H}_{\mbox{\ensuremath{{\scriptscriptstyle \text{KS}}}}}(\boldsymbol{r},\boldsymbol{r}^{\prime})\cdot\hat{\varPsi}(\boldsymbol{r}^{\prime})\label{eq:NambuAndersonKSHamiltonian}
\end{eqnarray}
where the KS Hamiltonian (first quantization Nambu form) is given
by 
\begin{eqnarray}
 &  & \hspace{-1cm}\bar{H}_{\mbox{\ensuremath{{\scriptscriptstyle \text{KS}}}}}(\boldsymbol{r},\boldsymbol{r}^{\prime})=\nonumber \\
 &  & \hspace{-0.75cm}\left(\begin{array}{cc}
\updelta(\boldsymbol{r}-\boldsymbol{r}^{\prime})H_{\mbox{\ensuremath{{\scriptscriptstyle \text{KS}}}}}^{{\rm {\scriptscriptstyle NS}}}(\boldsymbol{r}) & \boldsymbol{\Phi}\cdot\boldsymbol{\varDelta}^{{\scriptscriptstyle {\rm s}}}(\boldsymbol{r},\boldsymbol{r}^{\prime})\\
-\bigl(\boldsymbol{\Phi}\cdot\boldsymbol{\varDelta}^{{\scriptscriptstyle {\rm s}}}(\boldsymbol{r},\boldsymbol{r}^{\prime})\bigr)^{\ast} & -\updelta(\boldsymbol{r}-\boldsymbol{r}^{\prime})\bigl(H_{\mbox{\ensuremath{{\scriptscriptstyle \text{KS}}}}}^{{\rm {\scriptscriptstyle NS}}}(\boldsymbol{r})\bigr)^{{\rm T}_{{\rm s}}}
\end{array}\right)\,,\label{eq:KSHamiltonianNambuNotation}
\end{eqnarray}
with
\begin{eqnarray}
H_{\mbox{\ensuremath{{\scriptscriptstyle \text{KS}}}}}^{{\rm {\scriptscriptstyle NS}}}(\boldsymbol{r}) & = & \Bigl(-\frac{1}{2}\boldsymbol{\nabla}^{2}+v_{{\scriptscriptstyle \textrm{s}}}(\boldsymbol{r})-\mu\Bigr)\sigma_{0}-\mathbf{S}\cdot\mathbf{B}_{{\scriptscriptstyle \textrm{s}}}(\boldsymbol{r})\,.\label{eq:NormalPartNambuKSHamiltonian}
\end{eqnarray}
Note that the changed order of the electronic field operator implies
a transposition in spin space in the $(-1,-1)$ component that is
equivalent to using $\mathbf{S}^{\ast}$. In a similar transformation
the diagonal KS Hamiltonian Eq.~(\ref{eq:SCDFT-KS digHamil}) becomes
\begin{equation}
\hat{H}_{{\scriptscriptstyle {\rm KS}}}^{{\rm {\scriptscriptstyle e}}}=\sum_{k}\hat{\varPhi}_{k}^{\dagger}\cdot\frac{1}{2}\left(\begin{array}{cc}
E_{k} & 0\\
0 & -E_{k}
\end{array}\right)\cdot\hat{\varPhi}_{k}\label{eq:NambuAndersonDiagonalKSHamiltonian}
\end{equation}
with $\hat{\varPhi}_{k}=\left(\begin{array}{c}
\hat{\gamma}_{k}\\
\hat{\gamma}_{k}^{\dagger}
\end{array}\right)$. As a consequence of the rearrangement of the operators, in the Nambu-Anderson form should appear the trace of the Hamiltonian $\hat{H}_{{\scriptscriptstyle {\rm KS}}}^{{\rm {\scriptscriptstyle e}}}$. However, not being an operator, this cancels from thermal averages and has been disregarded.  
$\hat{\varPhi}_{k}$ is a two (not four) component vector because
the spin may not be a good quantum number in the SC KS system. We
can diagonalize the form in Eq.~(\ref{eq:NambuAndersonKSHamiltonian})
to the form Eq.~(\ref{eq:NambuAndersonDiagonalKSHamiltonian}) by
introducing a unitary transformation that we parameterise generically
with four complex spinor functions. This connection between $\hat{\varPsi}(\boldsymbol{r})$
and $\hat{\varPhi}_{k}$ is known as the Bogoliubov-Valatin transformation
\cite{ValatinTransform1958,BogolubovTransform1958}. We write it
in the form
\begin{eqnarray}
\hat{\varPsi}(\boldsymbol{r}) & = & \sum_{k}\left(\begin{array}{cc}
\vec{u}_{k}(\boldsymbol{r}) & \vec{v}_{k}^{\ast}(\boldsymbol{r})\\
\vec{v}_{k}(\boldsymbol{r}) & \vec{u}_{k}^{\ast}(\boldsymbol{r})
\end{array}\right)\cdot\hat{\varPhi}_{k}\nonumber \\
\hat{\varPhi}_{k} & = & \int\hspace{-0.18cm}\text{d}\boldsymbol{r}\left(\begin{array}{cc}
\vec{u}_{k}^{\ast}(\boldsymbol{r}) & \vec{v}_{k}^{\ast}(\boldsymbol{r})\\
\vec{v}_{k}(\boldsymbol{r}) & \vec{u}_{k}(\boldsymbol{r})
\end{array}\right)\cdot\hat{\varPsi}(\boldsymbol{r})\,.\label{eq:BogoliubovValatinTranformation}
\end{eqnarray}
Note that in the first case the matrix is $4\times2$ dimensional,
and in the second $2\times4$ because of the spinor property of the
$\vec{u}_{k}(\boldsymbol{r}),\vec{v}_{k}(\boldsymbol{r})$. In going
from Eq.~(\ref{eq:NambuAndersonKSHamiltonian}) to Eq.~(\ref{eq:NambuAndersonDiagonalKSHamiltonian}),
we identify
\begin{eqnarray}
 &  & \hspace{-0.5cm}\int\hspace{-0.18cm}{\rm d}\boldsymbol{r}\!\!\int\hspace{-0.18cm}{\rm d}\boldsymbol{r}^{\prime}\!\left(\!\begin{array}{cc}
\!\vec{u}_{k}^{\ast}(\boldsymbol{r}) & \!\vec{v}_{k}^{\ast}(\boldsymbol{r})\\
\!\vec{v}_{k}(\boldsymbol{r}) & \!\vec{u}_{k}(\boldsymbol{r})
\end{array}\!\right)\!\!\cdot\!\bar{H}_{\mbox{\ensuremath{{\scriptscriptstyle \text{KS}}}}}(\boldsymbol{r},\boldsymbol{r}^{\prime})\!\cdot\!\!\left(\!\begin{array}{cc}
\!\vec{u}_{k^{\prime}}(\boldsymbol{r}^{\prime}) & \!\vec{v}_{k^{\prime}}^{\ast}(\boldsymbol{r}^{\prime})\\
\!\vec{v}_{k^{\prime}}(\boldsymbol{r}^{\prime}) & \!\vec{u}_{k^{\prime}}^{\ast}(\boldsymbol{r}^{\prime})
\end{array}\!\right)\nonumber \\
 &  & \qquad=\left(\begin{array}{cc}
E_{k} & 0\\
0 & -E_{k}
\end{array}\right)\updelta_{kk^{\prime}}\,,\label{eq:DiagonalizeNambuKSHamiltonian}
\end{eqnarray}
which are the KS Bogoliubov de Gennes (KSBdG) equations for magnetic
system. Applying the inverse Bogoliubov-Valatin transformation from
the left we obtain two redundant vector equations of which we usually
consider the first for the positive eigenvalues $E_{k}$
\begin{eqnarray}
\int\hspace{-0.18cm}{\rm d}\boldsymbol{r}^{\prime}\bar{H}_{\mbox{\ensuremath{{\scriptscriptstyle \text{KS}}}}}(\boldsymbol{r},\boldsymbol{r}^{\prime})\cdot\left(\begin{array}{c}
\vec{u}_{k}(\boldsymbol{r}^{\prime})\\
\vec{v}_{k}(\boldsymbol{r}^{\prime})
\end{array}\right) & = & E_{k}\left(\begin{array}{c}
\vec{u}_{k}(\boldsymbol{r})\\
\vec{v}_{k}(\boldsymbol{r})
\end{array}\right)\,.\label{eq:KSBdGEquationsRealSpace}
\end{eqnarray}
This is the usual form of the KSBdG equations which generalize those
of Ref.~\onlinecite{OGK_SCDFT1988} and Ref.~\onlinecite{LuedersSCDFTI2005}. The equation in 
$\left(\begin{array}{cc}
\vec{v}_{k}^{\ast}(\boldsymbol{r}) & \vec{u}_{k}^{\ast}(\boldsymbol{r})
\end{array}\right)^{\rm{T}}$ leads to the equivalent negative eigenvalue $-E_{k}$ which reflects
the additional degrees of freedom that we have created in going to
the $2\times2$ Nambu formalism.

\paragraph{The Normal State KS Basis expansion}

The KSBdG equations~\ref{eq:KSBdGEquationsRealSpace} pose a challenging
integro differential problem. Sensible approximations can be obtained
by first performing an expansion into a basis set that is accessible
in practice and resembles closely to the true quasi particle structure
of the non-superconducting phase of the material under consideration.
With this in mind we consider the non-SC KS single particle equation:

\begin{equation}
\varepsilon_{i\sigma}\vec{\varphi}_{i\sigma}(\boldsymbol{r})=\Bigl(\!\bigl(\!-\frac{\boldsymbol{\nabla}^{2}}{2}\!+\! v_{{\scriptscriptstyle \textrm{s}}}^{{\scriptscriptstyle {\rm 0}}}(\boldsymbol{r})-\!\mu\bigr)\sigma_{0}-\!\boldsymbol{B}_{{\scriptscriptstyle \textrm{s}}z}^{{\scriptscriptstyle {\rm 0}}}(\boldsymbol{r})\frac{\sigma_{z}}{2}\!\Bigr)\!\!\cdot\!\vec{\varphi}_{i\sigma}(\boldsymbol{r})\label{eq:DefinitionKSOrbitalBasis}
\end{equation}
$v_{{\scriptscriptstyle \textrm{s}}}^{{\scriptscriptstyle {\rm 0}}}(\boldsymbol{r})$
and $\boldsymbol{B}_{{\scriptscriptstyle \textrm{s}}z}^{{\scriptscriptstyle {\rm 0}}}(\boldsymbol{r})$
are known functionals, like the local spin density approximation (LSDA)
\cite{vonBarthALocalEXPotentialForTheSpinPolarizedCase1972}. We
also assume that $\boldsymbol{B}_{{\scriptscriptstyle \textrm{s}}}^{{\scriptscriptstyle {\rm 0}}}$
is collinear and has components in $\sigma_{z}$ only. We use a pure
spinor notation for the orbitals, i.e.~$\vec{\varphi}_{i\sigma}(\boldsymbol{r})$
has only one non-vanishing component, e.g.~$\vec{\varphi}_{i\uparrow}(\boldsymbol{r})=\left(\begin{array}{c}
\varphi_{i}(\boldsymbol{r}\uparrow)\\
0
\end{array}\right)$. We use the indices $i,j$ for the quantum numbers of the basis and
thus distinguish from the quantum number $k$ of the SC KS system.
Later, in the Spin Decoupling Approximation \ref{sub:TheSDA} when
we assume the expansion coefficients to have only one non-vanishing
component each, this distinction will not be made. As a next step
we expand the Bogoliubov-Valatin transformations in these solutions
$\{\vec{\varphi}_{i\sigma}(\boldsymbol{r})\}$%
\footnote{Note that the $-1,-1$ component of the SC KS Hamiltonian Eq.~(\ref{eq:KSHamiltonianNambuNotation})
is the complex conjugated of the $1,1$. This comes from the property
$\bigl(H_{\mbox{\ensuremath{{\scriptscriptstyle \text{KS}}}}}^{{\rm {\scriptscriptstyle NS}}}(\boldsymbol{r})\bigr)^{{\rm T}_{{\rm s}}}=\bigl(H_{\mbox{\ensuremath{{\scriptscriptstyle \text{KS}}}}}^{{\rm {\scriptscriptstyle NS}}}(\boldsymbol{r})\bigr)^{\ast}$
of the Hamiltonian, $T_{\mathrm{s}}$ being a transposition in spin
space.%
}
\begin{equation}
\vec{u}_{k}(\boldsymbol{r})=\sum_{i\sigma}u_{k}^{i\sigma}\vec{\varphi}_{i\sigma}(\boldsymbol{r})\,,\quad\vec{v}_{k}(\boldsymbol{r})=\sum_{i\sigma}v_{k}^{i\sigma}\vec{\varphi}_{i\sigma}^{\ast}(\boldsymbol{r})\,.\label{eq:ExpansionUVk}
\end{equation}
Defining the matrix elements
\begin{eqnarray}
\hspace{-0.18cm}\hspace{-0.18cm}\mathcal{R}_{ij}^{\sigma\sigma^{\prime}} & = & \int\hspace{-0.18cm}\text{d}\boldsymbol{r}\vec{\varphi}_{i\sigma}^{\ast}\Bigl(\sigma_{0}\bigl(v_{{\scriptscriptstyle \textrm{s}}}(\boldsymbol{r})-v_{{\scriptscriptstyle \textrm{s}}}^{{\scriptscriptstyle {\rm 0}}}(\boldsymbol{r})\bigr)\nonumber \\
 &  & -\mathbf{S}\cdot\bigl(\boldsymbol{B}_{{\scriptscriptstyle \textrm{s}}}(\boldsymbol{r})-\boldsymbol{B}_{{\scriptscriptstyle \textrm{s}}}^{{\scriptscriptstyle {\rm 0}}}(\boldsymbol{r})\bigr)\Bigr)\vec{\varphi}_{j\sigma^{\prime}}\\
\hspace{-0.18cm}\hspace{-0.18cm}{\varDelta^{{\scriptscriptstyle {\rm s}}}}_{ij}^{\sigma\sigma^{\prime}} & = & \int\hspace{-0.18cm}\text{d}\boldsymbol{r}\int\hspace{-0.18cm}\text{d}\boldsymbol{r}^{\prime}\vec{\varphi}_{i\sigma}^{\ast}(\boldsymbol{r})\cdot\bigl(\boldsymbol{\Phi}\!\cdot\!\boldsymbol{\varDelta}^{{\scriptscriptstyle {\rm s}}}(\boldsymbol{r},\boldsymbol{r}^{\prime})\bigr)\!\cdot\vec{\varphi}_{j\sigma^{\prime}}^{\ast}(\boldsymbol{r}^{\prime})\\
\hspace{-0.18cm}\hspace{-0.18cm}\mathcal{E}_{ij}^{\sigma\sigma^{\prime}} & = & \varepsilon_{i\sigma}\delta_{ij}\delta_{\sigma\sigma^{\prime}}+\mathcal{R}_{ij}^{\sigma\sigma^{\prime}}\,,
\end{eqnarray}
and the singlet/triplet parts of the pair potential expansion coefficient
matrix
\begin{equation}
\begin{array}{ccc}
{\varDelta_{{\rm s}}^{{\scriptscriptstyle \text{s}}}}_{ij} & = & \frac{1}{2}\bigl({\varDelta^{{\scriptscriptstyle {\rm s}}}}_{ij}^{\uparrow\downarrow}-{\varDelta^{{\scriptscriptstyle {\rm s}}}}_{ij}^{\downarrow\uparrow}\bigr)\\
{\varDelta_{{\rm tx}}^{{\scriptscriptstyle \text{s}}}}_{ij} & = & \frac{1}{2}\bigl({\varDelta^{{\scriptscriptstyle {\rm s}}}}_{ij}^{\downarrow\downarrow}-{\varDelta^{{\scriptscriptstyle {\rm s}}}}_{ij}^{\uparrow\uparrow}\bigr)
\end{array}\:\begin{array}{ccc}
{\varDelta_{{\rm ty}}^{{\scriptscriptstyle \text{s}}}}_{ij} & = & \frac{1}{2}\bigl({\varDelta^{{\scriptscriptstyle {\rm s}}}}_{ij}^{\downarrow\downarrow}+{\varDelta^{{\scriptscriptstyle {\rm s}}}}_{ij}^{\uparrow\uparrow}\bigr)\\
{\varDelta_{{\rm tz}}^{{\scriptscriptstyle \text{s}}}}_{ij} & = & \frac{1}{2}\bigl({\varDelta^{{\scriptscriptstyle {\rm s}}}}_{ij}^{\uparrow\downarrow}+{\varDelta^{{\scriptscriptstyle {\rm s}}}}_{ij}^{\downarrow\uparrow}\bigr)
\end{array}
\end{equation}
we can finally cast Eq.~(\ref{eq:DiagonalizeNambuKSHamiltonian})
into a convenient form:
\begin{equation}
\left(\begin{array}{cc}
g_{k}^{{\scriptscriptstyle +}} & g_{k}^{{\scriptscriptstyle -}}\end{array}\right)^{\dagger}\!\!\cdot\!\left(\!\begin{array}{cc}
\mathcal{E} & \hspace{-0.18cm}\boldsymbol{\Phi}\cdot\boldsymbol{\varDelta}^{{\scriptscriptstyle {\rm s}}}\\
(\boldsymbol{\Phi}\cdot\boldsymbol{\varDelta}^{{\scriptscriptstyle {\rm s}}})^{\dagger} & \hspace{-0.18cm}-\mathcal{E}^{{\rm T}}
\end{array}\!\right)\!\!\cdot\!\left(\begin{array}{cc}
g_{k^{\prime}}^{{\scriptscriptstyle +}} & g_{k^{\prime}}^{{\scriptscriptstyle -}}\end{array}\right)=E_{k}\delta_{kk^{\prime}}\tau_{z}\,,\label{eq:KSBdGEquationsNSKSBasis}
\end{equation}
with 
\begin{eqnarray}
\hspace{-0.18cm}\hspace{-0.18cm}\hspace{-0.18cm}g_{k}^{{\scriptscriptstyle +}} & = & (\begin{array}{cccccccc}
u_{k}^{1\uparrow} & u_{k}^{1\downarrow} & u_{k}^{2\uparrow} & \ldots\vert & v_{k}^{1\uparrow} & v_{k}^{1\downarrow} & v_{k}^{2\uparrow} & \ldots\end{array})^{{\rm T}}\\
\hspace{-0.18cm}\hspace{-0.18cm}\hspace{-0.18cm}g_{k}^{{\scriptscriptstyle -}} & = & (\begin{array}{cccccccc}
{v_{k}^{1\uparrow}}^{\ast} & \!\!{v_{k}^{1\downarrow}}^{\ast} & \!\!{v_{k}^{2\uparrow}}^{\ast} & \!\!\!\ldots\vert & {u_{k}^{1\uparrow}}^{\ast} & \!\!{u_{k}^{1\downarrow}}^{\ast} & \!\!{u_{k}^{2\uparrow}}^{\ast} & \!\!\!\ldots\end{array})^{{\rm T}}\,.
\end{eqnarray}
The superscript $1,2,\ldots$ means we have ordered the Bloch vectors
and bands in some way. The precise way of ordering is unimportant.
Note that the set of $\{g_{k}^{{\scriptscriptstyle -}}\}$ solves
the eigenvalue equation similar to Eq.~(\ref{eq:KSBdGEquationsRealSpace})
with the negative eigenvalues $-E_{k}$ while the set $\{g_{k}^{{\scriptscriptstyle +}}\}$
corresponds to the eigenvectors with positive eigenvalues $E_{k}$.
The elements of the set $\{g_{k}^{{\scriptscriptstyle +}}\}$ are
the SC KS orbitals of SpinSCDFT in the normal KS orbital basis. We
may easily represent the densities using the normal state KS orbital
basis $\{\vec{\varphi}_{i\sigma}(\boldsymbol{r})\}$ for example
\begin{equation}
n(\boldsymbol{r})=\sum_{i\sigma j\sigma^{\prime}}\vec{\varphi}_{i}^{\ast}(\boldsymbol{r}\sigma)(n_{ij})_{\sigma\sigma^{\prime}}\vec{\varphi}_{j}(\boldsymbol{r}\sigma^{\prime})\,,
\end{equation}
and similar for $\boldsymbol{m}(\boldsymbol{r})$ and $\boldsymbol{\chi}(\boldsymbol{r},\boldsymbol{r}^{\prime})$
where $\boldsymbol{\chi}(\boldsymbol{r},\boldsymbol{r}^{\prime})$
is expanded in $\vec{\varphi}_{i}^{\ast}(\boldsymbol{r}\sigma)$ and
$\vec{\varphi}_{j}^{\ast}(\boldsymbol{r}^{\prime}\sigma^{\prime})$.
The coefficients read
\begin{eqnarray}
(n_{ij})_{\sigma\sigma^{\prime}} & = & (\sigma_{0})_{\sigma\sigma^{\prime}}\sum_{k}\bigl({(u_{k}^{i\sigma})}^{\ast}u_{k}^{j\sigma^{\prime}}f_{\beta}(E_{k})+\nonumber \\
 &  & +v_{k}^{i\sigma}{(v_{k}^{j\sigma^{\prime}})}^{\ast}f_{\beta}(-E_{k})\bigr)\,,\label{eq:SpinSCDFT_Ndens_ij}\\
(\boldsymbol{m}_{ij})_{\sigma\sigma^{\prime}} & = & (\mathbf{S})_{\sigma\sigma^{\prime}}\sum_{k}\bigl({(u_{k}^{i\sigma})}^{\ast}u_{k}^{j\sigma^{\prime}}f_{\beta}(E_{k})+\nonumber \\
 &  & +v_{k}^{i\sigma}{(v_{k}^{j\sigma^{\prime}})}^{\ast}f_{\beta}(-E_{k})\bigr)\,,\label{eq:MagneticDensityMatrixElements}\\
(\boldsymbol{\chi}_{ij})_{\sigma\sigma^{\prime}} & = & (\boldsymbol{\Phi})_{\sigma\sigma^{\prime}}\sum_{k}\bigl(u_{k}^{j\sigma^{\prime}}{(v_{k}^{i\sigma})}^{\ast}f_{\beta}(E_{k})+\nonumber \\
 &  & +u_{k}^{i\sigma}{(v_{k}^{j\sigma^{\prime}})}^{\ast}f_{\beta}(-E_{k})\bigr)\,.\label{eq:OrderParameterKSSpaceInTermsOfUV}
\end{eqnarray}
We want to stress that we have not performed any approximations so
far and the SC KS system reproduces the exact interacting densities
of the Hamiltonian of Eq.~(\ref{eq:FullInteractionHamiltonian}).

\paragraph{Singlet Superconductivity\label{par:Singlet-Superconductivity}}

Due to the antisymmetric structure of the fermionic wavefunction and
the effectively attractive interaction, in absence of magnetism, the
singlet solution always leads to a more stable SC state. Known SC
that feature a triplet pairing all share a very low critical temperature
less than a few Kelvin \cite{SigristUedaUnconventionalSC_RMP1991,JeromeOrganic1980,RiceSr2RuO41995,Luke_TimeRev_SrRuO3_1998,IshidaTripletSrRuO3_1998,SaxenaSCOnTheBorderOfItinerantElectronFerromagnetismInUGe2_2000,Bauer_CePtSi_2004}.
In presence of magnetism, as we have seen, the spin is not a good
quantum number and singlet/triplet components mix. Since the triplet
pairing channel seems to be rather unimportant for many systems, it
is of use to define a singlet approximation, in which it is completely
disregarded.

We therefore make the assumption that our pairing potential has only
the singlet component (marked as a subscript S in the KS potential).
In addition, we assume a collinear spin structure in the normal state
part of the Hamiltonian:

\begin{equation}
\boldsymbol{\Phi}\cdot\boldsymbol{\varDelta}^{{\scriptscriptstyle {\rm s}}}\approx\Phi_{1}\varDelta_{s}^{{\scriptscriptstyle {\rm s}}}\,,\quad\mathcal{E}_{ij}^{\sigma\sigma^{\prime}}\approx\mathcal{E}_{ij}^{\sigma}\delta_{\sigma\sigma^{\prime}}\,.
\end{equation}
Then, we observe that spin becomes a good quantum number in the SC
KS system. This follows because the KS Hamiltonian matrix elements
can be brought to a Block diagonal structure in Nambu and spin space
with two kind of eigenfunctions to each individual block. Consequently
we re-label the eigenvectors with $k\rightarrow k,\mu$ where the
size of the set of $k$ is reduced to half. Each block $\mu$ is diagonalized
as
\begin{eqnarray}
 &  & \hspace{-0.18cm}\hspace{-0.18cm}\left(\begin{array}{cc}
g_{k\mu}^{{\scriptscriptstyle +}} & g_{k,{\scriptscriptstyle -}\mu}^{{\scriptscriptstyle -}}\end{array}\right)^{\dagger}\!\!\!\cdot\!\left(\!\begin{array}{cc}
\mathcal{E}^{\mu} & \hspace{-0.18cm}{\rm sign}(\mu)\varDelta_{{\rm s}}^{{\scriptscriptstyle {\rm s}}}\\
{\rm sign}(\mu){\varDelta_{{\rm s}}^{{\scriptscriptstyle {\rm s}}}}^{\dagger} & \hspace{-0.18cm}-{\mathcal{E}^{-\mu}}^{{\rm T}}
\end{array}\!\right)\!\cdot\!\left(\begin{array}{cc}
g_{k^{\prime}\mu}^{{\scriptscriptstyle +}} & g_{k^{\prime},{\scriptscriptstyle -}\mu}^{{\scriptscriptstyle -}}\end{array}\right)\nonumber \\
 &  & \qquad=\delta_{kk^{\prime}}\left(\begin{array}{cc}
E_{k\mu}^{+} & 0\\
0 & E_{k\mu}^{-}
\end{array}\right)\label{eq:SingletKSBdGEquations}
\end{eqnarray}
with 
\begin{eqnarray}
g_{k\mu}^{{\scriptscriptstyle +}} & = & (\begin{array}{ccccc}
u_{k\mu}^{1\mu} & u_{k\mu}^{2\mu} & \ldots & \vert v_{k\mu}^{1-\mu} & \ldots\end{array})^{{\rm T}}\\
g_{k\mu}^{{\scriptscriptstyle -}} & = & (\begin{array}{ccccc}
{v_{k\mu}^{1,-\mu}}^{\ast} & {v_{k\mu}^{2,-\mu}}^{\ast} & \ldots & \vert{u_{k\mu}^{1\mu}}^{\ast} & \ldots\end{array})^{{\rm T}}\;.
\end{eqnarray}
$E_{k\mu}^{+}$ is an eigenvalue that may or may not be positive.
However, we have introduced the SC KS particles in Eq.~(\ref{eq:SCDFT-KS digHamil})
with a positive excitation energy $E_{k\mu}$ so this fact requires further commenting.
In the present situation where the matrix elements of the SC KS Hamiltonian
are block diagonal in Nambu and spin space we can show 
that if $g_{k\mu}^{{\scriptscriptstyle +}}$ has the eigenvalue
$E_{k\mu}^{+}$ the ``negative'' labeled eigenfunction $g_{k,-\mu}^{{\scriptscriptstyle -}}$
has the eigenvalue $-E_{k,-\mu}^{+}$
\footnote{The explicit calculation uses the fact
that $\mathcal{E}$ is hermitian and thus $\mathcal{E}^{\ast}=\mathcal{E}^{{\rm T}}$
and further that $\boldsymbol{\Phi}\cdot\boldsymbol{\varDelta}^{{\scriptscriptstyle {\rm s}}}$
is totally antisymmetric $(\boldsymbol{\Phi}\cdot\boldsymbol{\varDelta}^{{\scriptscriptstyle {\rm s}}})^{\dagger}=-(\boldsymbol{\Phi}\cdot\boldsymbol{\varDelta}^{{\scriptscriptstyle {\rm s}}})^{\ast}$.}.
Thus we still have the original redundancy in the eigenvalue spectrum but not in the same spin channel $\mu$.
Instead 
\begin{equation}
E_{k\mu}^{\pm}=-E_{k,-\mu}^{\mp}\,.
\end{equation}
We conclude that to every $k$ we have 4 eigenvalues of which 2 are
positive. These positive eigenvalues are identified with $E_{k\mu}$.
In the next Subsection \ref{sub:TheSDA} after introducing the Decoupling
approximation we will be able to compute these eigenvalues explicitly,
and continue this discussion.

\paragraph{The Spin Decoupling Approximation\label{sub:TheSDA}}

\begin{figure*}
\begin{centering}
\begin{minipage}[t]{0.5\textwidth}%
\begin{center}
\includegraphics[width=0.8\textwidth]{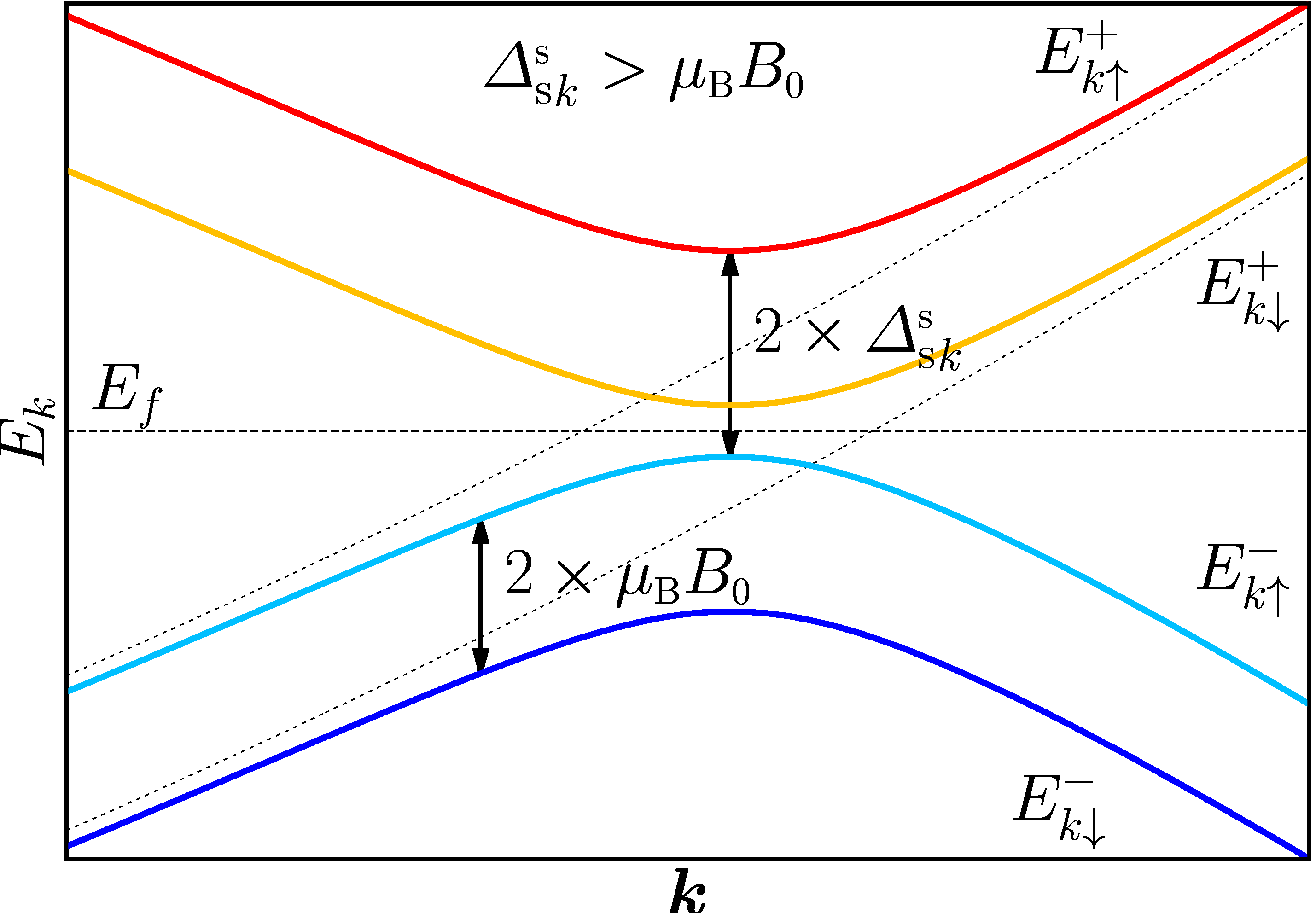}\\
a)
\par\end{center}%
\end{minipage}\nolinebreak%
\begin{minipage}[t]{0.5\textwidth}%
\begin{center}
\includegraphics[width=0.8\textwidth]{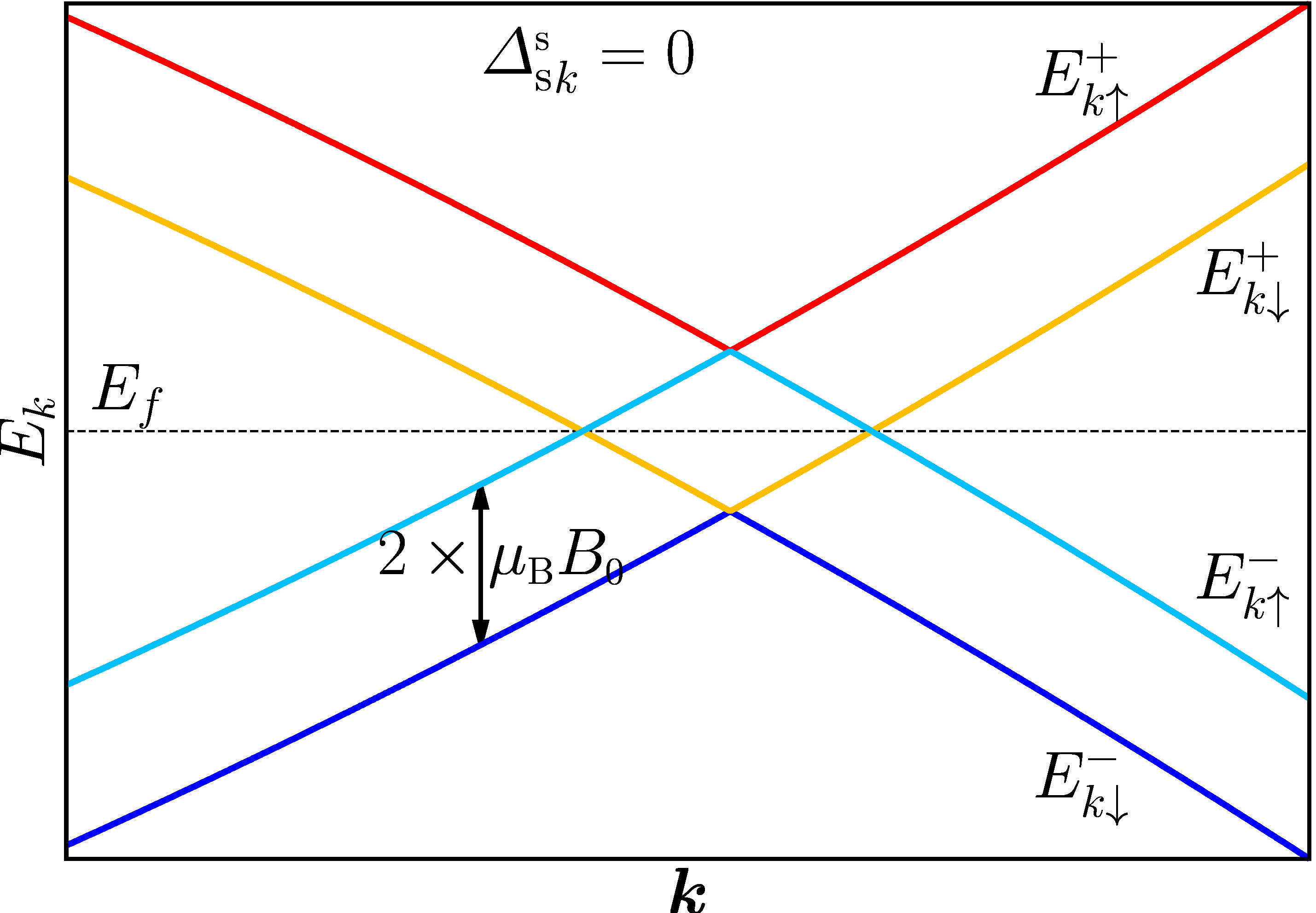}\\
b)
\par\end{center}%
\end{minipage}
\par\end{centering}

\caption{(color online) Sketch of the Bogoliubov eigenvalues $E_{k\sigma}^{\pm}$
for a free electron gas with a homogenous splitting $\varepsilon_{k\sigma}=\frac{1}{2}\boldsymbol{k}^{2}+{\rm sign}(\sigma)\mu_{{\rm {\scriptscriptstyle B}}}B_{{\rm {\scriptscriptstyle 0}}}$.
We choose a constant ${\varDelta_{{\rm s}}^{{\scriptscriptstyle {\rm s}}}}_{k}>\mu_{{\rm {\scriptscriptstyle B}}}B_{{\rm {\scriptscriptstyle 0}}}$
in a) and ${\varDelta_{{\rm s}}^{{\scriptscriptstyle {\rm s}}}}_{k}=0$
in b). We plot the $+$ Bogoliubov branch in red and orange for$\uparrow$and
$\downarrow$ and the - branch in light blue and dark blue for $\uparrow$and
$\downarrow$, respectively. We indicate the $\varepsilon_{k\sigma}$
in a) as thin dashed lines. In a), the $+$ branches are strictly
larger than the Fermi Energy $E_{f}$ and thus constitute the SC KS
particle excitations. On the other hand for ${\varDelta_{{\rm s}}^{{\scriptscriptstyle {\rm s}}}}_{k}<\mu_{{\rm {\scriptscriptstyle B}}}B_{{\rm {\scriptscriptstyle 0}}}$
as in b), the $+$ and $-$ branch partly swap their order. When $E_{k\uparrow}^{-}>E_{f}$
the SC KS particle excitations are from the $-$ branch also.\label{fig:SketchBogolBranches}}
\end{figure*}

It is desirable to reduce the effort to solve the KSBdG Eq.~(\ref{eq:SingletKSBdGEquations})
further. A substantial simplification is the Decoupling approximation \cite{LuedersSCDFTI2005,MarquesSCDFTIIMetals2005}
(or Anderson approximation \cite{Anderson1959DirtySuperc}). There,
one considers only singlet SC and pairing between a quasi particle
state ($i\sigma$) and its time reversed hole state ($-i,-\sigma$).
Furthermore it is assumed that the basis $\{\vec{\varphi}_{i\sigma}\}$
approximates the true non SC quasi particle structure well enough.
In the language of the our KSBdG Eq.~(\ref{eq:KSBdGEquationsNSKSBasis})
this reads

\begin{equation}
\mathcal{E}_{ij}^{\sigma\sigma^{\prime}}\approx\varepsilon_{i\sigma}\delta_{\sigma\sigma^{\prime}}\delta_{ij}\,,\quad(\boldsymbol{\Phi}\cdot\boldsymbol{\Delta}^{{\scriptscriptstyle {\rm s}}})_{ij}^{\sigma\sigma^{\prime}}\approx\Phi_{1}^{\sigma\sigma^{\prime}}{\varDelta_{{\rm s}}^{{\scriptscriptstyle {\rm s}}}}_{i,-i}\delta_{i,-j}\,.
\end{equation}
This type of approximation is inherent in the Eliashberg equations
as well as SCDFT functionals. It is also straightforward
to include a diagonal correction $\mathcal{R}_{ii}^{\sigma\sigma}$.
In the form presented here we will call it Spin Decoupling Approximation
(SDA). For each $k$ and $\mu$, Eq.~(\ref{eq:KSBdGEquationsNSKSBasis})
reduces to the $2\times2$ equation
\begin{eqnarray}
 &  & \left(\begin{array}{cc}
u_{k\sigma}^{k} & v_{-k\sigma}^{k\ast}\\
v_{k-\sigma}^{-k} & u_{-k-\sigma}^{-k\ast}
\end{array}\right)^{\dagger}\!\!\!\cdot\!\left(\!\begin{array}{cc}
\varepsilon_{k\sigma} & \hspace{-0.5cm}{\rm sign}(\sigma){\varDelta_{{\rm s}}^{{\scriptscriptstyle {\rm s}}}}_{k,-k}\\
{\rm sign}(\sigma){\varDelta_{{\rm s}}^{{\scriptscriptstyle {\rm s}}}}_{-k,k}^{\ast} & \hspace{-0.5cm}-\varepsilon_{-k-\sigma}
\end{array}\!\right)\!\!\cdot\nonumber \\
 &  & \qquad\qquad\cdot\left(\begin{array}{cc}
u_{k\sigma}^{k} & v_{-k\sigma}^{k\ast}\\
v_{k-\sigma}^{-k} & u_{-k-\sigma}^{-k\ast}
\end{array}\right)=\left(\begin{array}{cc}
E_{k\sigma}^{+} & 0\\
0 & E_{k\sigma}^{-}
\end{array}\right)\,.\label{eq:SDAKSBdGEquation}
\end{eqnarray}
Here we have introduced a single spin notation $v_{k\sigma}^{-k-\sigma}=v_{k-\sigma}^{-k}$
and $u_{k\sigma}^{k\sigma}=u_{k\sigma}^{k}$. The spin label on the
coefficients of the Bogoliubov transformation always refers to the
normal state KS basis spin label and thus we use the spin notation
$\mu\rightarrow\sigma$. Note however that the spin label cannot be
strictly identified with the spin of a SC KS particle. We will come
back to this point later. From now on we we will use the notation
${\varDelta_{{\rm s}}^{{\scriptscriptstyle {\rm s}}}}_{k}={\varDelta_{{\rm s}}^{{\scriptscriptstyle {\rm s}}}}_{k,-k}={\varDelta_{{\rm s}}^{{\scriptscriptstyle {\rm s}}}}_{-k,k}$.
We may compute the two eigenvalues and eigenvectors analytically.
From the high energy limit $\varepsilon_{k\sigma}+\varepsilon_{-k,-\sigma}\gg\varepsilon_{k\sigma}-\varepsilon_{-k,-\sigma}$
we identify the name $\pm$ for the two branches. The eigenvalues
are
\begin{eqnarray}
\!\!\! E_{k\sigma}^{-} & = & \frac{\varepsilon_{k\sigma}\!-\!\varepsilon_{-k-\sigma}}{2}-\!\!\sqrt{\!\Bigl(\frac{\varepsilon_{k\sigma}\!+\!\varepsilon_{-k-\sigma}}{2}\Bigr)^{2}\!\!\!+\!\vert{\varDelta_{{\rm s}}^{{\scriptscriptstyle {\rm s}}}}_{k}\vert^{2}}\,,\label{eq:SpinDecouplBogoliubovEigenvalueMinus}\\
\!\!\! E_{k\sigma}^{+} & = & \frac{\varepsilon_{k\sigma}\!-\!\varepsilon_{-k-\sigma}}{2}+\!\!\sqrt{\!\Bigl(\frac{\varepsilon_{k\sigma}\!+\!\varepsilon_{-k-\sigma}}{2}\Bigr)^{2}\!\!\!+\!\vert{\varDelta_{{\rm s}}^{{\scriptscriptstyle {\rm s}}}}_{k}\vert^{2}}\,.\label{eq:SpinDecouplBogoliubovEigenvaluePlus}
\end{eqnarray}
In the spin degenerate limit, the $+$ branch has always positive
eigenvalues $E_{k\sigma}^{+}$ and it is clear which of the eigenvectors
belongs to the first column of the Bogoliubov Valatin transformation.
In the spin polarized case the situation is more complicated. Again,
because $E_{k\sigma}^{\pm}=-E_{-k,-\sigma}^{\mp}$ two of the
four Bogoliubov eigenvalues to a given $k$ are positive but without
knowledge of $\varepsilon_{k\sigma}$ and ${\varDelta_{{\rm s}}^{{\scriptscriptstyle {\rm s}}}}_{k}$
one can not tell in advance which ones these are. The general situation
is sketched in Fig.~\ref{fig:SketchBogolBranches} for a constant
${\varDelta_{{\rm s}}^{{\scriptscriptstyle {\rm s}}}}_{k}$ and homogeneously
splitting free electron gas.
 In the next paragraph we give a more detailed discussion of the Bogoliubov
eigenvalues $E_{k\sigma}^{\pm}$.

\paragraph{Eigenvalues in the SDA}

Our first concern is how to interpret the spin quantum number $\sigma$
of $E_{k\sigma}^{\pm}$ in connection with the underlying normal states
$\varepsilon_{k\sigma}$.

First, consider the non-SC limit where
\begin{eqnarray}
{\varDelta_{{\rm s}}^{{\scriptscriptstyle {\rm s}}}}_{k}=0:\,2E_{k\sigma}^{\pm} & = & \varepsilon_{k\sigma}-\varepsilon_{-k-\sigma}\pm\vert\varepsilon_{k\sigma}+\varepsilon_{-k-\sigma}\vert\,.
\end{eqnarray}
This situation is plotted in Fig.~\ref{fig:SketchBogolBranches}~b).
Note that if $\varepsilon_{k\sigma}+\varepsilon_{-k-\sigma}>0$, than $E_{k\sigma}^{-}=-\varepsilon_{-k-\sigma}$
and if $\varepsilon_{k\sigma}+\varepsilon_{-k-\sigma}<0$ we conclude
$E_{k\sigma}^{-}=\varepsilon_{k\sigma}$.

Second, consider the following case that occurs at any $k_{0}$ where
$\varepsilon_{k_{0}\uparrow}+\varepsilon_{-k_{0}\downarrow}=0$. Given
that we have an energy splitting $\varepsilon_{k_{0}\uparrow}-\varepsilon_{-k_{0}\downarrow}>2\vert{\varDelta_{{\rm s}}^{{\scriptscriptstyle {\rm s}}}}_{k}\vert$
we find that both $E_{-k_{0},\downarrow}^{\pm}$ are negative. This
means that according to the definition in Eq.~(\ref{eq:SCDFT-KS digHamil})
to take the positive eigenvalues, both KS particles are from the $\sigma=\uparrow$
branch. It is not possible to construct the Bogoliubov transformations
in this case and in any case the $\hat{\gamma}_{k\uparrow}^{\dagger}$
state cannot be occupied twice. It is, however, possible to give up
the requirement that all SC KS particles are positive and simply always
take the $+$ branch. Then we can say that $\hat{\gamma}_{k\sigma}^{\dagger}$
creates a negative energy excitation which will be occupied in the
ground state. By analogy with BCS, $\hat{\gamma}_{k\sigma}^{\dagger}$
creates an electron like single particle state on the SC vacuum, this leads to the interpretation
that, in the ground state, this $k$ space region is occupied by unpaired electrons. 
A similar discussion can be found (still in the context of BCS theory) in the work of Sarma\cite{SarmaOnTheInfluenceOfAUniformExchangeFieldActingOnSC1963}.
Similar to Eq.~(\ref{eq:SCDFT-KS digHamil}) we can redefine electron
to hole operators at the price of changing the ground state energy.
Because the ground state energy, in turn, cancels from the thermal
averages, the expectation values computed with this theory do not
depend on this interpretation. We want to point out that this
discussion only applies when the splitting is larger than the
pair potential.

\paragraph{Eigenvectors in the SDA}

Furthermore we can analytically compute the normalized eigenvectors
$g_{k\mu}^{\alpha}$ to the eigenvalues $E_{k\mu}^{\alpha}$ ($\alpha=\pm$). We introduce
the notation
\begin{equation}
g_{k\sigma}^{\alpha}=\left(\begin{array}{c}
u_{k\sigma}^{k\alpha}\\
v_{k-\sigma}^{-k\alpha}
\end{array}\right)
\end{equation}
to label the components which are given in terms of the eigenvalues
and components of the matrix by
\begin{eqnarray}
v_{k-\sigma}^{-k\alpha} & = & \sqrt{\frac{\vert E_{k\mu}^{\alpha}-\varepsilon_{k\sigma}\vert}{\vert E_{k\sigma}^{+}+E_{-k-\sigma}^{+}\vert}}\,,\label{eq:SpingSCDFT_vmu_decoupl}\\
u_{k\sigma}^{k\alpha} & = & \frac{\text{sign}(\sigma)}{\text{sign}(\alpha)}\frac{{\varDelta_{{\rm s}}^{{\scriptscriptstyle {\rm s}}}}_{k}}{\vert{\varDelta_{{\rm s}}^{{\scriptscriptstyle {\rm s}}}}_{k}\vert}\sqrt{\frac{\vert\varepsilon_{-k-\sigma}+E_{k\sigma}^{\alpha}\vert}{\vert E_{k\sigma}^{\alpha}+E_{-k-\sigma}^{\alpha}\vert}}\,.\label{eq:SpingSCDFT_umu_decoupl}
\end{eqnarray}
Starting from a converged zero temperature normal state calculation,
within the SDA the only remaining variable is thus the matrix elements
of the pair potential ${\varDelta_{{\rm s}}^{{\scriptscriptstyle {\rm s}}}}_{k}$
because the SC KS wavefunctions as well as the Bogoliubov eigenvalues
are explicitly given in terms of it. 

It is important to point out that within the SDA ${\varDelta_{{\rm s}}^{{\scriptscriptstyle {\rm s}}}}_{k}$
can be chosen to be real\cite{NambuQParticlesGaugeInSuperconductivity1960,Anderson1958RPAInTheoryOfSuperconductivity}.
This can be proved by exploiting the gauge symmetry of Eq.~(\ref{eq:SDAKSBdGEquation})
under rotation about the $\tau_{z}$ axis. If the rotation is applied
with a $k$ dependent angle $\theta_{k}$ of 
\begin{equation}
\theta_{k}=\text{arctan}\Bigl(\frac{\Im{\varDelta_{{\rm s}}^{{\scriptscriptstyle {\rm s}}}}_{k}}{\Re{\varDelta_{{\rm s}}^{{\scriptscriptstyle {\rm s}}}}_{k}}\Bigr)
\end{equation}
we get:
\begin{eqnarray}
 &  & \text{e}^{-\text{i}\tau_{z}\frac{\theta_{k}}{2}}\left(\begin{array}{cc}
\varepsilon_{k\sigma} & {\rm sign}(\sigma){\varDelta_{{\rm s}}^{{\scriptscriptstyle {\rm s}}}}_{k}\\
{\rm sign}(\sigma){\varDelta_{{\rm s}}^{{\scriptscriptstyle {\rm s}}}}_{k}^{\ast} & -\varepsilon_{-k-\sigma}
\end{array}\right)\text{e}^{\text{i}\tau_{z}\frac{\theta_{k}}{2}}\nonumber \\
 & = & \left(\begin{array}{cc}
\varepsilon_{k\sigma} & {\rm sign}(\sigma){\tilde{\varDelta_{{\rm s}}^{{\scriptscriptstyle {\rm s}}}}}_{k}\\
{\rm sign}(\sigma){\tilde{\varDelta_{{\rm s}}^{{\scriptscriptstyle {\rm s}}}}}_{k} & -\varepsilon_{-k-\sigma}
\end{array}\right)
\end{eqnarray}
where ${\tilde{\varDelta_{{\rm s}}^{{\scriptscriptstyle {\rm s}}}}}_{k}={\rm sign}(\Re{\varDelta_{{\rm s}}^{{\scriptscriptstyle {\rm s}}}}_{k})\vert{\varDelta_{{\rm s}}^{{\scriptscriptstyle {\rm s}}}}_{k}\vert\in\mathbb{R}$.
Thus the $(k,-k)$ matrix elements of our general complex decoupled
pair potential are gauge equivalent to purely real ones. We still
keep a general complex notation for ${\varDelta_{{\rm s}}^{{\scriptscriptstyle {\rm s}}}}_{k}$
first, to investigate explicitly if self-energy corrections affect this
conclusion and, second, to make it easier to extent the formalism
to the case where the gauge symmetry does not have enough freedom
to make all matrix elements real.

\subsubsection{Competition between SC and Magnetism in the SDA\label{sub:Competition-between-SC}}

The SDA, as introduced so far, assumes that we compute SC on top of a (magnetic)
quasi particle structure. Thus, for example, it does not allow magnetism to be suppressed 
when a weakly magnetic system becomes SC. In conventional
SCDFT \cite{LuedersSCDFTI2005,MarquesSCDFTIIMetals2005} this type of feedbacks can
be safely neglected because SC changes the dispersion only for states
very close to the Fermi level. The effect on the electronic density
is thus negligible and so is the change in the normal state $xc$
potential. However, since the contributions to $\boldsymbol{m}(\boldsymbol{r})$
are in general more localized at the Fermi level, assuming 
quasi particle energies $\varepsilon_{i\sigma}$ to be unaffected when
SC sets in may not be reasonable for magnetic systems.

We want to point out in here that it is also possible to keep the simple form of the
SDA and include competition of SC and magnetism at the same time, by means of the following 
iterative scheme:
\begin{enumerate}
\item Take the normal KS states $\{\vec{\varphi}_{i\sigma}\}$ and eigenvalues
$\varepsilon_{k\sigma}$ as starting orbitals.
\item Solve the KS-BdG equations in the SDA
\item Recompute the densities $n(\boldsymbol{r})$ and $\boldsymbol{m}(\boldsymbol{r})$
according to the Eqs.~(\ref{eq:SpinSCDFT_Ndens_ij}) and (\ref{eq:MagneticDensityMatrixElements})
\item Re-diagonalize the normal state KS equations with the updated densities
(in particular changes in $\boldsymbol{m}(\boldsymbol{r})$ may be
of relevance)
\item iterate from point 2. until self consistence is reached
\end{enumerate}
This procedure changes the meaning of the SDA during the iteration
because we are self consistently updating the orbitals $\{\vec{\varphi}_{i\sigma}\}$
it refers to.

\subsection{The Sham-Schl\"uter Equation of SpinSCDFT\label{sub:The-Sham-Schluter-Equation}}

So far we have presented the structure of SpinSCDFT with the focus
on the electronic SC KS system. However explicit functionals for the
$xc$-pairing potential ${\varDelta_{{\rm s}}^{{\scriptscriptstyle {\rm s}}}}_{k}$
have not yet been discussed. The derivation of the approximations
for the $xc$-potentials generalizes one proposed by Marques\cite{MarquesPhDThesis2000}
in SCDFT and uses the Sham-Schl\"uter equation of SpinSCDFT. This
equation is based on the observation that the parts of the KS GF and
the interaction GF that correspond to the densities must be equal.
Using the Dyson equation for a SC in a magnetic field starting from
the SC KS system as the formally non interaction one we can relate
the $xc$-potentials to an approximation for the self energy. Here
and in the next section we present a derivation of an $xc$-potential
for SpinSCDFT that generalizes the ones of Marques\cite{MarquesPhDThesis2000}
and Sanna and Gross\cite{SannaMigdalFunctionalSCDFT2014}.

We introduce the GF with the $\tau$ ordering symbol $\bar{{\rm T}}$
and the field operators in the Heisenberg picture
\begin{equation}
\bar{G}(\boldsymbol{r}\tau,\boldsymbol{r}^{\prime}\tau^{\prime})=-\langle\bar{{\rm T}}\hat{\varPsi}(\boldsymbol{r}\tau)\otimes\hat{\varPsi}^{\dagger}(\boldsymbol{r}^{\prime}\tau^{\prime})\rangle\,.\label{eq:DefinitionNambuGreensfunction}
\end{equation}
The imaginary time ordering symbol in Nambu space $\bar{{\rm T}}$
is defined to act on every of the $(4\times4)$ components individually
which can be achieved by transposing in Nambu-spin space 
\begin{eqnarray}
\bar{{\rm T}}\hat{\varPsi}(\boldsymbol{r}\tau)\!\otimes\!\hat{\varPsi}^{\dagger}(\boldsymbol{r}^{\prime}\tau^{\prime}) & = & \uptheta(\tau\!-\!\tau^{\prime})\hat{\varPsi}(\boldsymbol{r}\tau)\!\otimes\!\hat{\varPsi}^{\dagger}(\boldsymbol{r}^{\prime}\tau^{\prime})\nonumber \\
 &  & \hspace{-1.5cm}-\uptheta(\tau^{\prime}\!\!-\!\tau)\bigl(\hat{\varPsi}^{\dagger}(\boldsymbol{r}^{\prime}\tau^{\prime})\!\otimes\!\hat{\varPsi}(\boldsymbol{r}\tau)\bigr)^{{\rm T}_{\text{sn}}}\,.\label{eq:NambuTauOrderingSymbol}
\end{eqnarray}
We define the equal time limit in the $-1,-1$ component different
to the usual one (that we use in the $1,1$ component). The equal
time limit of the time ordering symbol should be defined to recover
the density matrix operator but according to the usual rule where
the creation operator is taken infinitesimally before the annihilator
would lead to the form $\psi\psi^{\dagger}$ in the $-1,-1$ component.
From the equation of motion we derive the Dyson equation starting
from the SC KS system as a formally non interacting system
\begin{eqnarray}
 &  & \hspace{-0.5cm}\bar{G}(\boldsymbol{r},\boldsymbol{r}^{\prime},\omega_{n})=\bar{G}^{{\scriptscriptstyle {\rm KS}}}(\boldsymbol{r},\boldsymbol{r}^{\prime},\omega_{n})+\nonumber \\
 &  & \int\hspace{-0.18cm}{\rm d}\boldsymbol{r}_{1}\!\int\hspace{-0.18cm}{\rm d}\boldsymbol{r}_{1}^{\prime}\bar{G}^{{\scriptscriptstyle {\rm KS}}}\!(\boldsymbol{r}\!,\!\boldsymbol{r}_{1}\!,\!\omega_{n})\!\cdot\!\bar{\varSigma}^{{\scriptscriptstyle {\rm s}}}(\boldsymbol{r}_{1}\!,\!\boldsymbol{r}_{1}^{\prime}\!,\!\omega_{n})\!\cdot\!\bar{G}(\boldsymbol{r}_{1}^{\prime}\!,\!\boldsymbol{r}^{\prime}\!\!,\omega_{n}),\nonumber\\
 &  & \label{eq:DysonEquation}
\end{eqnarray}
with
\begin{equation}
\bar{\varSigma}^{{\scriptscriptstyle {\rm s}}}(\boldsymbol{r},\boldsymbol{r}^{\prime},\omega_{n})=\bar{\varSigma}(\boldsymbol{r},\boldsymbol{r}^{\prime},\omega_{n})\!-\bar{v}_{{\scriptscriptstyle {\rm xc}}}(\boldsymbol{r},\boldsymbol{r}^{\prime})\,.\label{eq:SelfEnergyXCMinusVXC}
\end{equation}
Here $\bar{\varSigma}$ is the irreducible Nambu self-energy, where
the electronic Hartree diagram was subtracted, and $\bar{v}_{{\scriptscriptstyle \text{xc}}}$
is the Nambu $xc$ potential
\begin{eqnarray}
 &  & \hspace{-0.25cm}\bar{v}_{{\scriptscriptstyle {\rm xc}}}(\boldsymbol{r}\!,\!\boldsymbol{r}^{\prime})=\nonumber \\
 &  & \hspace{-0.25cm}\left(\!\begin{array}{cc}
\updelta(\boldsymbol{r}\!-\!\boldsymbol{r}^{\prime})\bigl(\sigma_{0}v_{{\scriptscriptstyle {\rm xc}}}(\boldsymbol{r})\!-\!\mathbf{S}\!\cdot\!\boldsymbol{B}_{{\scriptscriptstyle {\rm xc}}}(\boldsymbol{r})\bigr) & \boldsymbol{\Phi}\cdot\boldsymbol{\varDelta}^{{\scriptscriptstyle {\rm xc}}}(\boldsymbol{r}\!,\!\boldsymbol{r}^{\prime})\\
\hspace{-0.75cm}\hspace{-0.75cm}\hspace{-0.75cm}-\boldsymbol{\Phi}\cdot\boldsymbol{\varDelta}^{{\scriptscriptstyle {\rm xc}}\ast}(\boldsymbol{r}\!,\!\boldsymbol{r}^{\prime}) & \hspace{-0.5cm}\hspace{-0.75cm}\hspace{-0.75cm}-\updelta(\boldsymbol{r}\!-\!\boldsymbol{r}^{\prime})\bigl(\sigma_{0}v_{{\scriptscriptstyle {\rm xc}}}(\boldsymbol{r})\!-\!\mathbf{S}^{\ast}\!\!\cdot\!\boldsymbol{B}_{{\scriptscriptstyle {\rm xc}}}(\boldsymbol{r})\bigr)
\end{array}\!\right)\nonumber \\
\label{eq:GreensFunctions_VxcPotential}
\end{eqnarray}
The SC KS Greens function satisfies
\begin{eqnarray}
 &  & \int\hspace{-0.18cm}{\rm d}\boldsymbol{r}_{1}\Bigl({\rm i}\omega_{n}\updelta(\boldsymbol{r}-\boldsymbol{r}_{1})\tau_{0}\sigma_{0}-\bar{H}_{{\scriptscriptstyle {\rm KS}}}(\boldsymbol{r}\!,\!\boldsymbol{r}_{1})\Bigr)\!\!\cdot\!\bar{G}^{{\scriptscriptstyle {\rm KS}}}(\boldsymbol{r}_{1},\boldsymbol{r}^{\prime}\!,\omega_{n})\nonumber \\
 &  & \qquad=\updelta(\boldsymbol{r}-\boldsymbol{r}^{\prime})\tau_{0}\sigma_{0}\,.\label{eq:EquationOfMotionKSGreensfunction}
\end{eqnarray}
From the equation of motion we can compute the SC KS GF. Because by
construction the SC KS GF yields the same densities as the interacting
system we can cancel the respective parts of the GFs in the Dyson
Eq.~(\ref{eq:DysonEquation}) that correspond to the densities. The
result is the Sham-Schl\"uter equation
\begin{eqnarray}
 &  & \hspace{-1.5cm}\frac{1}{\beta}\sum_{n}\int\hspace{-0.18cm}{\rm d}\boldsymbol{r}_{1}\hspace{-0.18cm}\int\hspace{-0.18cm}{\rm d}\boldsymbol{r}_{1}^{\prime}\Bigl[\bar{G}^{{\scriptscriptstyle {\rm KS}}}(\boldsymbol{r},\boldsymbol{r}_{1},\omega_{n})\cdot\nonumber \\
 &  & \cdot\bar{\varSigma}^{{\scriptscriptstyle {\rm s}}}(\boldsymbol{r}_{1},\boldsymbol{r}_{1}^{\prime},\omega_{n})\cdot\bar{G}(\boldsymbol{r}_{1}^{\prime}\!,\boldsymbol{r}^{\prime}\!\!,\omega_{n})\Bigr]_{\alpha,-\alpha}=0\\
 &  & \hspace{-1.5cm}\frac{1}{\beta}\sum_{n}\int\hspace{-0.18cm}{\rm d}\boldsymbol{r}_{1}\hspace{-0.18cm}\int\hspace{-0.18cm}{\rm d}\boldsymbol{r}_{1}^{\prime}\Bigl[\bar{G}^{{\scriptscriptstyle {\rm KS}}}(\boldsymbol{r},\boldsymbol{r}_{1},\omega_{n})\cdot\nonumber \\
 &  & \cdot\bar{\varSigma}^{{\scriptscriptstyle {\rm s}}}(\boldsymbol{r}_{1},\boldsymbol{r}_{1}^{\prime},\omega_{n})\cdot\bar{G}(\boldsymbol{r}_{1}^{\prime}\!,\boldsymbol{r}\!\!,\omega_{n})\Bigr]_{\alpha,\alpha}=0\,.
\end{eqnarray}
 For convenience the self energy is decomposed in a phononic part
$\bar{\varSigma}_{{\scriptscriptstyle {\rm ph}}}(\omega_{n})$ and
a Coulomb part $\bar{\varSigma}_{{\scriptscriptstyle {\rm C}}}(\omega_{n})$
: 
\begin{equation}
\bar{\varSigma}(\omega_{n})=\bar{\varSigma}_{{\scriptscriptstyle {\rm ph}}}(\omega_{n})+\bar{\varSigma}_{{\scriptscriptstyle {\rm C}}}(\omega_{n})\,.
\end{equation}
$\bar{\varSigma}(\omega_{n})$ has a diagrammatic expansion in terms
of $\bar{G}(\omega_{n})$ \cite{VonsovskySuperconductivityTransitionMetals}
and can be even viewed as part of a Hedin cycle for a SC including
phononic and Coulomb interactions \cite{LinscheidHedinEquationsForSC2014}.
We do not consider vertex corrections, thus the Coulomb self energy
part $\bar{\varSigma}_{{\scriptscriptstyle {\rm C}}}$ is the electronic
GW diagram\\
\begin{minipage}[c]{1\columnwidth}%
\begin{center}
\begin{minipage}[t][1\totalheight][b]{0.8\columnwidth}%
\begin{center}
$\bar{\varSigma}_{{\scriptscriptstyle {\rm C}}}(\omega_{n})\approx$\includegraphics[width=0.3\columnwidth]{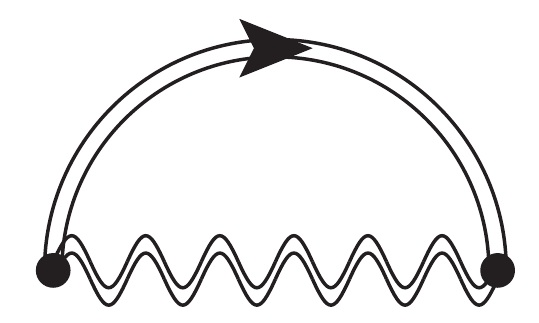}~.
\par\end{center}%
\end{minipage}\nolinebreak%
\begin{minipage}[t][1\totalheight][b]{0.1\columnwidth}%
\begin{flushleft}
\begin{equation}
\label{eq:ElectronicGWDiagram}
\end{equation}

\par\end{flushleft}%
\end{minipage}
\par\end{center}%
\end{minipage}\\
\\
As an interesting extension we could include parts of the vertex corrections
that lead to spin fluctuations. These, in the form of an effective
spin interaction, are discussed by Essenberger \textit{et al.}\cite{EssenbergerSCPairingMediatedBySpinFluctuationsFromFirstPrinciples2015}
and the extension to the present spin dependent formalism is straightforward. 
As compared to the polarization corrections of the same order,
the phononic vertex corrections are negligible \cite{MigdalInteractionBetweenElAndLatticeVibrInANormalMetal1958}. 

Moreover due to the quality of the phonon spectra one obtains with
density functional perturbation theory \cite{BaroniPhononrev2001,GiannozziAbInitioCalculationOfPhDispersions1991}
we do not consider further diagrammatic electronic screening and treat
the phononic interaction in the Hartree-Fock approximation\\
\begin{minipage}[c]{1\columnwidth}%
\begin{center}
\begin{minipage}[t][1\totalheight][b]{0.8\columnwidth}%
\begin{center}
$\bar{\varSigma}_{{\scriptscriptstyle {\rm ph}}}(\omega_{n})\approx$\includegraphics[width=0.3\columnwidth]{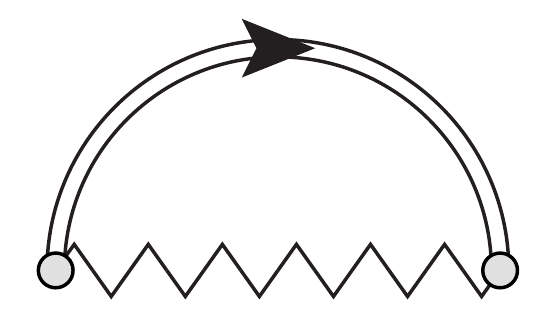}\nolinebreak$+$\nolinebreak\includegraphics[width=0.15\columnwidth]{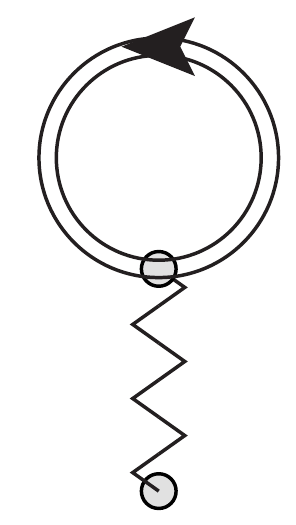}~.
\par\end{center}%
\end{minipage}\nolinebreak%
\begin{minipage}[t][1\totalheight][b]{0.1\columnwidth}%
\begin{flushleft}
\vspace{0.3cm}
\begin{equation}
\label{eq:PhononicHartreeFockDiagram}
\end{equation}

\par\end{flushleft}%
\end{minipage}
\par\end{center}%
\end{minipage} \\
\\
It has been observed that computing the GW quasi particle band structure
in a metal gives usually small corrections to the KS bands (compare
Ref.~\onlinecite{MariniQPAndElectronicStructureGWVsKSBands2002} Fig.~2),
also densities result to be almost identical. Thus, at least in the
spin degenerate case, the GW corrections on a KS band structure of
a metal are usually neglected. For convenience we use a similar assumption
for the spin part. This way we can drop the Nambu diagonal $\bar{v}_{{\scriptscriptstyle {\rm xc}}}$
construction from the Sham-Schl\"uter equation. Representing $\bar{G}^{{\scriptscriptstyle {\rm KS}}}(\boldsymbol{r},\boldsymbol{r}^{\prime},\omega_{n})$
and $\bar{G}(\boldsymbol{r},\boldsymbol{r}^{\prime},\omega_{n})$
in the same basis as the Bogoliubov-Valatin transformations, i.e.~essentially
the normal state KS orbitals $\{\varPsi_{i\sigma\alpha}^{{\scriptscriptstyle {\rm KS}}}(\boldsymbol{r})\}$
with the pure Nambu and spin spinor wavefunctions 
\begin{equation}
\varPsi_{i\sigma\alpha}^{{\scriptscriptstyle {\rm KS}}}(\boldsymbol{r})=\left(\begin{array}{c}
\updelta_{\alpha,1}\vec{\varphi}_{i\sigma}(\boldsymbol{r})\\
\updelta_{\alpha,-1}\vec{\varphi}_{i\sigma}^{\ast}(\boldsymbol{r})
\end{array}\right)\,.
\end{equation}
Sorting the expansion coefficients of $\bar{G}^{{\scriptscriptstyle {\rm KS}}}(\boldsymbol{r},\boldsymbol{r}^{\prime},\omega_{n})=\sum_{\alpha\alpha^{\prime}\sigma\sigma^{\prime}ij}\bar{G}_{i\alpha\sigma j\alpha^{\prime}\sigma^{\prime}}^{{\scriptscriptstyle {\rm KS}}}(\omega_{n})\varPsi_{i\alpha\sigma}^{{\scriptscriptstyle {\rm KS}}}(\boldsymbol{r})\otimes\varPsi_{j\alpha^{\prime}\sigma^{\prime}}^{{\scriptscriptstyle {\rm KS}}}(\boldsymbol{r}^{\prime})$
in similar Nambu and spin form we obtain the $4N\times4N$ matrix
equation
\begin{eqnarray}
 &  & \frac{1}{\beta}\sum_{n}\bar{G}^{{\scriptscriptstyle {\rm KS}}}(\omega_{n})\cdot\left(\begin{array}{cc}
0 & \boldsymbol{\Phi}\cdot\boldsymbol{\varDelta}^{{\scriptscriptstyle {\rm s}}}\\
(\boldsymbol{\Phi}\cdot\boldsymbol{\varDelta}^{{\scriptscriptstyle {\rm s}}})^{\dagger} & 0
\end{array}\right)\cdot\bar{G}(\omega_{n})\nonumber \\
 &  & =\frac{1}{\beta}\sum_{n}\bar{G}^{{\scriptscriptstyle {\rm KS}}}(\omega_{n})\cdot\Biggl(\biggl(\begin{array}{cc}
0 & \bar{\varSigma}_{{\scriptscriptstyle {\rm C}}}^{{\scriptscriptstyle 1,-1}}(\omega_{n})\\
\bar{\varSigma}_{{\scriptscriptstyle {\rm C}}}^{{\scriptscriptstyle -1,1}}(\omega_{n}) & 0
\end{array}\biggr)\nonumber \\
 &  & \qquad\qquad\qquad+\bar{\varSigma}_{{\scriptscriptstyle {\rm ph}}}(\omega_{n})\Biggr)\cdot\bar{G}(\omega_{n})\label{eq:ShamSchlueterEquation}
\end{eqnarray}
that we need to solve for $\boldsymbol{\Phi}\cdot\boldsymbol{\varDelta}^{{\scriptscriptstyle {\rm s}}}$.
From here on we use $\boldsymbol{\Phi}\cdot\boldsymbol{\varDelta}^{{\scriptscriptstyle {\rm s}}}$ and 
$\boldsymbol{\Phi}\cdot\boldsymbol{\varDelta}^{{\scriptscriptstyle {\rm xc}}}$ synonymously, i.e.~the external
pair potential is assumed to be infinitesimal.

In the next section we reduce the problem to the singlet case and
employ the SDA. Because we can solve the KSBdG equations analytically
we obtain a potential functional theory and arrive at a functional
form that is formally similar to the BCS gap equation. We stress that
the methods presented here and in the next section could also be applied
without the restriction to the SDA. However in that case the equations
would have an implicit form and require a numerical solution of the
KSBdG equations. Such a general form would be of importance in considering
triplet superconductivity or to account for pairings beyond the usual
one of time reversed states (as would be needed for example to describe
the FFLO state \cite{FuldeFerrellSuperconductivityInAStrongSpinExchangeField1964,LarkinOvchinnikovInhomogeneousStateOfSuperconductors1965}).
A further discussion can be found in Ref.~\onlinecite{LinscheidDFTofSuperconductivityInThePresenceOfAMagneticField2014}.

\subsection{Derivation $xc-$Potential\label{sub:Derivation-xc-Potential}}

The Sham-Schl\"uter Eq.~(\ref{eq:ShamSchlueterEquation}) involves
the interacting GF which is usually only available after solving the
Dyson equation. In an approximate scheme this step can be avoided.
The straightforward way is to replace the matrix $\bar{G}(\omega_{n})$ with $\bar{G}^{{\scriptscriptstyle \text{KS}}}(\omega_{n})$
on all occurrences. As was realized before \cite{LuedersSCDFTI2005}
this violates Migdal's theorem because there the vertex is compared
with the polarization diagram of the same order. Thus the phonon vertex
corrections are only negligible as compared to the Hartree exchange
diagram with the full GF. To circumvent this problem some of the Authors
introduced a procedure to construct a self-energy that does satisfy
Migdal's theorem \cite{SannaMigdalFunctionalSCDFT2014}. Starting
from an electron gas model with a phononic Hartree exchange diagram,
this leads to excellent agreement with experiment while still retaining
the numerically simple form of the Sham-Schl\"uter equation that
is independent on $\bar{G}(\omega_{n})$ and involves only Matsubara
sums that can be evaluated analytically. The self-energy $\bar{\varSigma}^{{\scriptscriptstyle {\rm KS}}}(\omega_{n})$
with $\bar{G}(\omega_{n})$ replaced by $\bar{G}_{ij}^{{\scriptscriptstyle \text{KS}}}(\omega_{n})$
is the basis of all further improvements. In this work, however, we
will not investigate the parametrization procedure. We will limit
the complexity of the derivation by using assuming $\bar{\varSigma}(\omega_{n})\simeq\bar{\varSigma}^{{\scriptscriptstyle {\rm KS}}}(\omega_{n})$,
where in $\bar{\varSigma}^{{\scriptscriptstyle {\rm KS}}}(\omega_{n})$
the $\bar{G}(\omega_{n})$ is replaced by $\bar{G}_{ij}^{{\scriptscriptstyle \text{KS}}}(\omega_{n})$.
This will give inaccurate critical temperatures but qualitatively
correct results. Thus we are left to solve the equation:
\begin{eqnarray}
 &  & \frac{1}{\beta}\sum_{n}\bar{G}^{{\scriptscriptstyle {\rm KS}}}(\omega_{n})\cdot\left(\begin{array}{cc}
0 & \boldsymbol{\Phi}\cdot\boldsymbol{\varDelta}^{{\scriptscriptstyle {\rm s}}}\\
(\boldsymbol{\Phi}\cdot\boldsymbol{\varDelta}^{{\scriptscriptstyle {\rm s}}})^{\dagger} & 0
\end{array}\right)\cdot\bar{G}^{{\scriptscriptstyle {\rm KS}}}(\omega_{n})\nonumber \\
 &  & =\frac{1}{\beta}\sum_{n}\bar{G}^{{\scriptscriptstyle {\rm KS}}}(\omega_{n})\cdot\Biggl(\biggl(\begin{array}{cc}
0 & \bar{\varSigma}_{{\scriptscriptstyle {\rm C}}}^{{\scriptscriptstyle {\rm KS}}{\scriptscriptstyle 1,-1}}(\omega_{n})\\
\bar{\varSigma}_{{\scriptscriptstyle {\rm C}}}^{{\scriptscriptstyle {\rm KS}}{\scriptscriptstyle -1,1}}(\omega_{n}) & 0
\end{array}\biggr)\nonumber \\
 &  & \qquad\qquad\qquad+\bar{\varSigma}_{{\scriptscriptstyle {\rm ph}}}^{{\scriptscriptstyle {\rm KS}}}(\omega_{n})\Biggr)\cdot\bar{G}^{{\scriptscriptstyle {\rm KS}}}(\omega_{n})\,.\label{eq:ShamSchlueterOnlyGKS}
\end{eqnarray}
In this form the matrix elements of the SC KS GF in the normal state
KS basis are given by
\begin{eqnarray}
\bar{G}_{ij}^{{\scriptscriptstyle {\rm KS}}}(\omega_{n}) & = & \sum_{k}\frac{1}{\mbox{i}\omega_{n}-E_{k}}\left(\begin{array}{cc}
\vec{u}_{k}^{i}\otimes\vec{u}_{k}^{j\ast} & \vec{u}_{k}^{i}\otimes\vec{v}_{k}^{j\ast}\\
\vec{v}_{k}^{i}\otimes\vec{u}_{k}^{j\ast} & \vec{v}_{k}^{i}\otimes\vec{v}_{k}^{j\ast}
\end{array}\right)+\nonumber \\
 &  & \hspace{-1.5cm}+\sum_{k}\frac{1}{\mbox{i}\omega_{n}+E_{k}}\left(\begin{array}{cc}
\vec{v}_{k}^{i\ast}\otimes\vec{v}_{k}^{j} & \vec{v}_{k}^{i\ast}\otimes\vec{u}_{k}^{j}\\
\vec{u}_{k}^{i\ast}\otimes\vec{v}_{k}^{j} & \vec{u}_{k}^{i\ast}\otimes\vec{u}_{k}^{j}
\end{array}\right)\,.\label{eq:KSGreensfunctionNambu}
\end{eqnarray}
We use $\vec{u}_{k}^{i}=(\begin{array}{cc}
u_{k}^{i\uparrow} & u_{k}^{i\downarrow})^{{\rm T}}\end{array}$ with the expansion coefficients $u_{k}^{i\sigma}$ of $\vec{u}_{k}(\mathbf{r})$
in $\vec{\varphi}_{i\sigma}(\mathbf{r})$ given in Eq.~(\ref{eq:ExpansionUVk}).
Similar for $\vec{v}_{k}^{i}$. Further we assume the SDA for the
rest of this paper. Results beyond the SDA are discussed in the PhD thesis
Ref.~\onlinecite{LinscheidDFTofSuperconductivityInThePresenceOfAMagneticField2014}.
In the SDA the SC KS GF simplifies to 
\begin{eqnarray}
 &  & \hspace{-0.25cm}\bar{G}_{ij}^{{\scriptscriptstyle \text{KS}}}(\omega_{n})\!=\nonumber \\
 &  & \hspace{-0.18cm}\sum_{\alpha}\!\!\left(\begin{array}{cccc}
\hspace{-0.18cm}\frac{\vert u_{i\uparrow}^{i\alpha}\vert^{2}\delta_{ij}}{\mbox{i}\omega_{n}-E_{i\uparrow}^{\alpha}} & \hspace{-0.75cm}0 & \hspace{-0.25cm}0 & \hspace{-0.75cm}\frac{u_{i\uparrow}^{i\alpha}(v_{i\downarrow}^{-i\alpha})^{\ast}\delta_{i,-j}}{\mbox{i}\omega_{n}-E_{i\uparrow}^{\alpha}}\\
\hspace{-0.18cm}0 & \hspace{-0.75cm}\frac{\vert u_{i\downarrow}^{i\alpha}\vert^{2}\delta_{ij}}{\mbox{i}\omega_{n}-E_{i\downarrow}^{\alpha}} & \hspace{-0.25cm}\frac{u_{i\downarrow}^{i\alpha}(v_{i\uparrow}^{-i\alpha})^{\ast}\delta_{i,-j}}{\mbox{i}\omega_{n}-E_{i\downarrow}^{\alpha}} & \hspace{-0.75cm}0\\
\hspace{-0.18cm}0 & \hspace{-0.75cm}\frac{(u_{i\uparrow}^{i\alpha})^{\ast}v_{i\downarrow}^{-i\alpha}\delta_{i,-j}}{\mbox{i}\omega_{n}+E_{i\uparrow}^{\alpha}} & \hspace{-0.25cm}\frac{\vert u_{i\uparrow}^{i\alpha}\vert^{2}\delta_{ij}}{\mbox{i}\omega_{n}+E_{i\uparrow}^{\alpha}} & \hspace{-0.75cm}0\\
\hspace{-0.18cm}\frac{(u_{i\downarrow}^{i\alpha})^{\ast}v_{i\uparrow}^{-i\alpha}\delta_{i,-j}}{\mbox{i}\omega_{n}+E_{i\downarrow}^{\alpha}} & \hspace{-0.75cm}0 & \hspace{-0.25cm}0 & \hspace{-0.75cm}\frac{\vert u_{i\downarrow}^{i\alpha}\vert^{2}\delta_{ij}}{\mbox{i}\omega_{n}+E_{i\downarrow}^{\alpha}}
\end{array}\hspace{-0.18cm}\right)\nonumber \\
\label{eq:KSNambuGreenfunctionSpinDecoupl}
\end{eqnarray}
This form and any further formula based on it use the components of
the SC KS wavefunction as given in the Eqs.~(\ref{eq:SpingSCDFT_vmu_decoupl})
and (\ref{eq:SpingSCDFT_umu_decoupl}). In the Dyson equation $\bar{G}^{-1}={\mbox{\ensuremath{\bar{G}^{{\scriptscriptstyle \text{KS}}}}}}^{-1}-\bar{\varSigma}$
we see that we need to compare the self-energy contributions with
the inverse SC KS GF. Inverting $\bar{G}_{ij}^{{\scriptscriptstyle \text{KS}}}(\omega_{n})$
with obtain
\begin{eqnarray}
(\bar{G}^{{\scriptscriptstyle \text{KS}}})_{ij}^{-1}(\omega_{n}) & = & \delta_{ij}\Bigl(\text{i}\omega_{n}\tau_{0}\sigma_{0}-\bigl(\frac{\varepsilon_{i\uparrow}+\varepsilon_{-i\downarrow}}{2}\bigr)\tau_{z}\sigma_{0}\nonumber \\
 &  & \hspace{-0.75cm}\hspace{-0.75cm}-\bigl(\frac{\varepsilon_{i\uparrow}-\varepsilon_{-i\downarrow}}{2}\bigr)\tau_{z}\sigma_{z}\Bigr)+\delta_{i,-j}\bigl((\mbox{i}\tau_{y})(\mbox{i}\sigma_{y})\Re{\varDelta_{{\rm s}}^{{\scriptscriptstyle {\rm s}}}}_{i}\nonumber \\
 &  & +\tau_{x}(\mbox{i}\sigma_{y})\mbox{i}\Im{\varDelta_{{\rm s}}^{{\scriptscriptstyle {\rm s}}}}_{i}\bigr)\,.\label{eq:InverseSCKSGF}
\end{eqnarray}
Here we see that self-energy contributions $\propto\tau_{z}\sigma_{0}$
change the average spin Fermi level $\frac{\varepsilon_{i\uparrow}+\varepsilon_{-i\downarrow}}{2}=0$.
Similarly contributions $\propto\tau_{z}\sigma_{z}$ change the splitting of single particle levels. 
It has to be understood that these are
global properties of the band structure, meaning that the full
$\varepsilon_{i\sigma}$ dispersion has to be integrated to obtain
$N$ electrons per unit cell. If the interaction changes dispersion and occupations
far away from the Fermi level this may still cause a shift of the
original Fermi level. An clear cut example is the following:
In the context of SC one often employs the Eliashberg function $\alpha^{\!2}\! F(\varOmega)$ which is the Fermi-surface average of the electron-phonon interaction \cite{EliashbergInteractionBetweenElAndLatticeVibrInASC1960,AllenTheoryOfSCTc1983,CarbottePropertiesOfBosonExchangeSC1990}, to describe the electron phonon interaction. 
This function is assumed to apply equally to all states, also those away from the
Fermi level. This is a good approximation only if corrections of the Fermi level are excluded a priori (electron-hole symmetry), otherwise under this assumption the correction to the Fermi level $\frac{\varepsilon_{i\uparrow}+\varepsilon_{-i\downarrow}}{2}$ and the splitting $\frac{\varepsilon_{i\uparrow}-\varepsilon_{-i\downarrow}}{2}$ would show a logaritmic divergece. 
As commonly done in Eliashberg theory, where the same effect occurs, one then excludes self-energy contributions $\propto\tau_{z}$. We will assume the same approximation. As the Hartree diagram is proportional to $\tau_{z}$ is thus not considered. While the expected Fermi energy shift is negligible, corrections to the spin splitting $\frac{\varepsilon_{i\uparrow}-\varepsilon_{-i\downarrow}}{2}$ could be of relevance. However due to the extreme additional numerical complexity of considering the true full electronic state dependence of the electron phonon interaction we leave this to a future project. We compute the
self-energy matrix elements in the SDA from the Eq.~(\ref{eq:PhononicHartreeFockDiagram})
\begin{eqnarray}
 &  & \hspace{-0.75cm}{\mbox{\ensuremath{\bar{\varSigma}_{{\scriptscriptstyle {\rm ph}}}^{{\scriptscriptstyle {\rm KS}}}}}}_{i\sigma j\sigma^{\prime}}^{{\scriptscriptstyle 1,1}}(\omega_{n})=\updelta_{\sigma\sigma^{\prime}}\sum_{qk\alpha}g_{ik\sigma}^{q}g_{kj\sigma}^{-q}\times\nonumber \\
 &  & \times\vert u_{k\sigma}^{k\alpha}\vert^{2}M_{{\scriptscriptstyle {\rm ph}}}(\varOmega_{q},E_{k\sigma}^{\alpha},\omega_{n})\label{eq:PhononSelfEnergyDecoul11}\\
 &  & \hspace{-0.75cm}{\mbox{\ensuremath{\bar{\varSigma}_{{\scriptscriptstyle {\rm ph}}}^{{\scriptscriptstyle {\rm KS}}}}}}_{i\sigma j\sigma^{\prime}}^{{\scriptscriptstyle 1,-1}}(\omega_{n})=-\updelta_{\sigma,-\sigma^{\prime}}\sum_{qk\alpha}g_{ik\sigma}^{q}g_{j,-k,-\sigma}^{-q}\times\nonumber \\
 &  & \times u_{k\sigma}^{k\alpha}(v_{k-\sigma}^{-k\alpha})^{\ast}M_{{\scriptscriptstyle {\rm ph}}}(\varOmega_{q},E_{k\sigma}^{\alpha},\omega_{n})\label{eq:PhononSelfEnergyDecoul12}\\
 &  & \hspace{-0.75cm}{\mbox{\ensuremath{\bar{\varSigma}_{{\scriptscriptstyle {\rm ph}}}^{{\scriptscriptstyle {\rm KS}}}}}}_{i\sigma j\sigma^{\prime}}^{{\scriptscriptstyle -1,1}}(\omega_{n})=-\updelta_{\sigma,-\sigma^{\prime}}\sum_{qk\alpha}g_{ki\sigma}^{q}g_{-k,j,-\sigma}^{-q}\times\nonumber \\
 &  & \times(u_{k\sigma}^{k\alpha})^{\ast}v_{k-\sigma}^{-k\alpha}M_{{\scriptscriptstyle {\rm ph}}}(\varOmega_{q},-E_{k\sigma}^{\alpha},\omega_{n})\label{eq:PhononSelfEnergyDecoul21}\\
 &  & \hspace{-0.75cm}{\mbox{\ensuremath{\bar{\varSigma}_{{\scriptscriptstyle {\rm ph}}}^{{\scriptscriptstyle {\rm KS}}}}}}_{i\sigma j\sigma^{\prime}}^{{\scriptscriptstyle -1,-1}}(\omega_{n})=\updelta_{\sigma\sigma^{\prime}}\sum_{ak\alpha}g_{ki\sigma}^{q}g_{jk\sigma}^{-q}\times\nonumber \\
 &  & \times\vert u_{k\sigma}^{k\alpha}\vert^{2}M_{{\scriptscriptstyle {\rm ph}}}(\varOmega_{q},-E_{k\sigma}^{\alpha},\omega_{n})\,.\label{eq:PhononSelfEnergyDecoul22}
\end{eqnarray}
From the hermiticity of $\hat{H}_{{\scriptscriptstyle {\rm KS}}}^{{\scriptscriptstyle {\rm e-ph}}}$
of Eq.~(\ref{eq:KSElPhInteractionHamiltonian}) comes $g_{-q}^{a}(\boldsymbol{r})=\bigl(g_{q}^{a}(\boldsymbol{r})\bigr)^{\ast}$
and thus the electron phonon interaction matrix elements
\begin{equation}
g_{ij\sigma}^{q}=\int\hspace{-0.18cm}{\rm d}\boldsymbol{r}\sum_{a=0,z}\vec{\varphi}_{i\sigma}^{\ast}(\boldsymbol{r})\cdot\sigma_{a}\cdot\vec{\varphi}_{j\sigma}(\boldsymbol{r})g_{q}^{a}(\boldsymbol{r})
\end{equation}
have the property $g_{ij\sigma}^{q}={g_{ji\sigma}^{-q}}^{\ast}$.
Moreover $g_{ij\sigma}^{q}\propto\updelta_{\boldsymbol{k}_{i},\boldsymbol{k}_{j}+\boldsymbol{q}}$
which is expected from the lattice translational symmetry \cite{BaroniPhononrev2001}.
The Matsubara summation $M_{{\scriptscriptstyle {\rm ph}}}(\varOmega,E,\omega_{n})$
is evaluated with the result
\begin{eqnarray}
Y_{{\scriptscriptstyle {\rm ph}}}(\varOmega,E,\omega_{n}) & = & \frac{1}{\beta}\sum_{n^{\prime}}\frac{1}{\text{i}\omega_{n^{\prime}}-E}\frac{1}{\mbox{i}(\omega_{n}-\omega_{n^{\prime}})+\varOmega}\\
 & = & \frac{n_{\beta}(\varOmega)+f_{\beta}(E)}{\varOmega-E+\text{i}\omega_{n}}\label{eq:MatsubaraIntegralY1}\\
M_{{\scriptscriptstyle {\rm ph}}}(\varOmega,E,\omega_{n}) & = & \frac{n_{\beta}(\varOmega)+f_{\beta}(E)}{\varOmega-E+\text{i}\omega_{n}}+\frac{f_{\beta}(E)+n_{\beta}(-\varOmega)}{\varOmega+E-\text{i}\omega_{n}}\nonumber \\
\label{eq:MatsubaraIntegralIs}\\
 & = & Y_{{\scriptscriptstyle {\rm ph}}}(\varOmega,E,\omega_{n})\!-\! Y_{{\scriptscriptstyle {\rm ph}}}^{\ast}(\varOmega,-E,\omega_{n})\,,\label{eq:ManyBodySpectrum_FreqIntegral}
\end{eqnarray}
where $f_{\beta}(E)$ and $n_{\beta}(\varOmega)$ are Fermi and Bose
functions, respectively. The Coulomb self energy parts on the Nambu
off diagonal with the diagram of Eq.~(\ref{eq:ElectronicGWDiagram})
are
\begin{eqnarray}
 &  & \hspace{-0.75cm}{\mbox{\ensuremath{\bar{\varSigma}_{{\scriptscriptstyle {\rm C}}}^{{\scriptscriptstyle {\rm KS}}}}}}_{\, i\sigma j\sigma^{\prime}}^{\!{\scriptscriptstyle 1,-1}}(\omega_{n})=-\updelta_{\sigma,-\sigma^{\prime}}\sum_{k\alpha}{W_{ikj,-k}^{{\scriptscriptstyle \text{stat}}}}_{\sigma,-\sigma}\times\nonumber \\
 &  & \times u_{k\sigma}^{k\alpha}v_{k-\sigma}^{-k\alpha\ast}f_{\beta}(E_{k\sigma}^{\alpha})\,,\label{eq:CoulombSelfEnergyDecoupl12}\\
 &  & \hspace{-0.75cm}{\mbox{\ensuremath{\bar{\varSigma}_{{\scriptscriptstyle {\rm C}}}^{{\scriptscriptstyle {\rm KS}}}}}}_{\, i\sigma j\sigma^{\prime}}^{\!{\scriptscriptstyle -1,1}}(\omega_{n})=-\updelta_{\sigma,-\sigma^{\prime}}\sum_{k\alpha}{W_{ki,-k,j}^{{\scriptscriptstyle \text{stat}}}}_{\sigma,-\sigma}\times\nonumber \\
 &  & \times u_{k\sigma}^{k\alpha\ast}v_{k-\sigma}^{-k\alpha}f_{\beta}(\!-\! E_{k\sigma}^{\alpha}\!)\,.\label{eq:CoulombSelfEnergyDecoupl21}
\end{eqnarray}
with the static screened Coulomb matrix elements 
\begin{eqnarray}
{W_{k_{1}k_{2}k_{3}k_{4}}^{{\scriptscriptstyle \text{stat}}}}_{\sigma\sigma^{\prime}} & = & \int\hspace{-0.18cm}{\rm d}\boldsymbol{r}\int\hspace{-0.18cm}{\rm d}\boldsymbol{r}^{\prime}\vec{\varphi}_{k_{1}\sigma}^{\ast}(\boldsymbol{r})\cdot\vec{\varphi}_{k_{2}\sigma}(\boldsymbol{r})\times\nonumber \\
 &  & \hspace{-0.75cm}\times\frac{\epsilon^{-1}(\boldsymbol{r},\boldsymbol{r}^{\prime},0)}{\vert\boldsymbol{r}-\boldsymbol{r}^{\prime}\vert}\vec{\varphi}_{k_{3}\sigma^{\prime}}^{\ast}(\boldsymbol{r}^{\prime})\cdot\vec{\varphi}_{k_{4}\sigma^{\prime}}(\boldsymbol{r}^{\prime})\,,
\end{eqnarray}
with the inverse dielectric function $\epsilon^{-1}(\boldsymbol{r},\boldsymbol{r}^{\prime},0)$
that is accessible in many electronic structure codes \cite{QE-2009,elkCode}.
$\epsilon^{-1}(\boldsymbol{r},\boldsymbol{r}^{\prime},0)$ is often
calculated within the RPA which yields very good results for metals
in general. As we have pointed out, terms proportional to $\tau_{z}$
i.e.~contributions $({\mbox{\ensuremath{\bar{\varSigma}_{{\scriptscriptstyle {\rm ph}}}^{{\scriptscriptstyle {\rm KS}}}}}}^{{\scriptscriptstyle 1,1}}-{\mbox{\ensuremath{\bar{\varSigma}_{{\scriptscriptstyle {\rm ph}}}^{{\scriptscriptstyle {\rm KS}}}}}}^{{\scriptscriptstyle -1,-1}})$
are dropped from the functional construction. 

Because of the gauge symmetry discussed in Sec.~\ref{sub:TheSDA},
we expect the equations for ${\varDelta_{{\rm s}}^{{\scriptscriptstyle {\rm s}}}}_{k}$
and ${\varDelta_{{\rm s}}^{{\scriptscriptstyle {\rm s}}}}_{k}^{\ast}$
to be similar. Thus we proceed and evaluate only the $1,-1$ component
of the Sham-Schl\"uter equation (\ref{eq:ShamSchlueterEquation})
in SDA and arrive at 
\begin{eqnarray}
{M}_{k,-k}^{k,-k}{\varDelta_{{\rm s}}^{{\scriptscriptstyle {\rm s}}}}_{k}+{M^{\prime}}_{-k,k}^{k,-k}{\varDelta_{{\rm s}}^{{\scriptscriptstyle {\rm s}}}}_{k}^{\ast} & = & \mathfrak{D}_{k,-k}+\mathfrak{C}_{k,-k}\,.\label{eq:ImplicitGapEquation}
\end{eqnarray}
Here $\mathfrak{D}_{k,-k}$ are the purely phononic contributions
due to the Nambu diagonal self energy parts $\tau_{0}({\mbox{\ensuremath{\bar{\varSigma}_{{\scriptscriptstyle {\rm ph}}}^{{\scriptscriptstyle {\rm KS}}}}}}^{{\scriptscriptstyle 1,1}}+{\mbox{\ensuremath{\bar{\varSigma}_{{\scriptscriptstyle {\rm ph}}}^{{\scriptscriptstyle {\rm KS}}}}}}^{{\scriptscriptstyle -1,-1}})$.
$\mathfrak{C}_{k,-k}$ is due to the Nambu-off-diagonal self energy
contributions and contains the phononic interaction along with the
Coulomb potential on the same footing. The coefficients
\begin{eqnarray}
{M}_{k,-k}^{k,-k} & = & \frac{1}{\beta}\sum_{n\sigma}\bar{G}_{k\sigma,k\sigma}^{{\scriptscriptstyle {\rm KS}}{\scriptscriptstyle 1,1}}(\omega_{n})\bar{G}_{-k,-\sigma,-k,-\sigma}^{{\scriptscriptstyle {\rm KS}}{\scriptscriptstyle -1,-1}}(\omega_{n}) \\
{M^{\prime}}_{-k,k}^{k,-k} & = & \frac{1}{\beta}\sum_{n\sigma}\bar{G}_{k\sigma,-k,-\sigma}^{{\scriptscriptstyle {\rm KS}}{\scriptscriptstyle 1,-1}}(\omega_{n})\bar{G}_{k\sigma,-k,-\sigma}^{{\scriptscriptstyle {\rm KS}}{\scriptscriptstyle -1,1}}(\omega_{n})
\end{eqnarray}
are the Matsubara summed SC KS GF parts. Note that ${M^{\prime}}_{-k,k}^{k,-k}{\varDelta_{{\rm s}}^{{\scriptscriptstyle {\rm s}}}}_{k}^{\ast}\propto\vert{\varDelta_{{\rm s}}^{{\scriptscriptstyle {\rm s}}}}_{k}\vert^{2}{\varDelta_{{\rm s}}^{{\scriptscriptstyle {\rm s}}}}_{k}$
so the Sham-Schl\"uter equation in the SDA is unaffected by the phase
of ${\varDelta_{{\rm s}}^{{\scriptscriptstyle {\rm s}}}}_{k}$, as
expected from the gauge symmetry. 

$\mathfrak{D}_{kk^{\prime}}$ and $\mathfrak{C}_{kk^{\prime}}$ have
non vanishing matrix elements apart from $k,-k$. These are not included
in the SDA. Other SC theories such as Eliashberg and spin degenerate SCDFT are build on similar approximations 
and from the quality of the results one obtains, we conclude that such corrections are in general
not important.

Another interesting aspect of the functional construction to observe is that a
self-energy part showing ${\rm tx}$ triplet symmetry appears, that means the
spin inverted Nambu off diagonal components are not equal and of opposite sign
\begin{equation}
{\mbox{\ensuremath{\bar{\varSigma}^{{\scriptscriptstyle {\rm KS}}}}}}_{k\uparrow k^{\prime}\downarrow}^{{\scriptscriptstyle \alpha,-\alpha}}+{\mbox{\ensuremath{\bar{\varSigma}^{{\scriptscriptstyle {\rm KS}}}}}}_{k\downarrow k^{\prime}\uparrow}^{{\scriptscriptstyle \alpha,-\alpha}}\neq0\,.
\end{equation}
These self-energy part leads to non-vanishing functional contributions
in $\mathfrak{C}_{k,-k}$ in the singlet channel. We call these contribution
intermediate triplet contributions. We have investigated the effect of removing
them and found that this has essentially no consequence in the numerical
calculation for a spin independent coupling (see part {\bf II }). In addition
we note that similar to the matrix elements $k^{\prime}\neq-k$, the
diagrams generate triplet contributions that cannot be incorporated
into the SDA. This also means that the terms
\begin{eqnarray}
\sum_{\sigma}\bigl(\frac{1}{\beta}\sum_{n}\bar{G}^{{\scriptscriptstyle {\rm KS}}{\scriptscriptstyle 1,1}}(\omega_{n})\cdot\bar{\varSigma}^{{\scriptscriptstyle {\rm KS}}{\scriptscriptstyle 1,-1}}(\omega_{n})\cdot 
& & \nonumber\\
\cdot\bar{G}^{{\scriptscriptstyle {\rm KS}}{\scriptscriptstyle -1,-1}}(\omega_{n})\bigr)_{\sigma,-\sigma} & \neq & 0 \\
\sum_{\sigma}\bigl(\frac{1}{\beta}\sum_{n}\bar{G}^{{\scriptscriptstyle {\rm KS}}{\scriptscriptstyle 1,-1}}(\omega_{n})\cdot\bar{\varSigma}^{{\scriptscriptstyle {\rm KS}}{\scriptscriptstyle -1,1}}(\omega_{n})
& & \nonumber\\
\cdot\bar{G}^{{\scriptscriptstyle {\rm KS}}{\scriptscriptstyle 1,-1}}(\omega_{n})\bigr)_{\sigma,-\sigma} & \neq & 0
\end{eqnarray}
are not zero as, on the other hand, one would expect for a singlet SC. This fact simply
means that ignoring the triplet components from the external potential
is not consistent, in presence of a magnetic field, because a triplet-singlet
coupling exists at the level of the $xc$-potential. As discussed
earlier (Sec.~\ref{par:Singlet-Superconductivity}), it is not clear
in which cases triplet effects become relevant. However, since experimentally
triplet SC is only observed at very low temperature, in high temperature
regimes disregarding all triplet components should be safe, we will
show in {\bf II} when we investigate the influence of intermediate
triplet contributions, at least, that they are small.

Within the SDA the SC KS wavefunction components $v_{k-\sigma}^{-k\alpha},u_{k\sigma}^{k\alpha}$
are explicit functionals of the potential $\varDelta_{{\rm s}}^{{\scriptscriptstyle {\rm s}}}$.
Thus, left and right hand side of the Sham-Schl\"uter equation (\ref{eq:ImplicitGapEquation})
are equally non-linear functionals of the potential $\varDelta_{{\rm s}}^{{\scriptscriptstyle {\rm s}}}$.
We interpret the Sham-Schl\"uter condition (\ref{eq:ImplicitGapEquation})
as
\begin{equation}
S_{\beta}[\varDelta_{{\rm s}}^{{\scriptscriptstyle {\rm s}}}]\cdot\varDelta_{{\rm s}}^{{\scriptscriptstyle {\rm s}}}=0\quad S_{\beta}=S_{\beta}^{{\scriptscriptstyle \text{M}}}+S_{{\scriptscriptstyle \text{ph}}\beta}^{{\scriptscriptstyle \mathfrak{C}}}+S_{{\scriptscriptstyle \text{C}}\beta}^{{\scriptscriptstyle \mathfrak{C}}}+S_{\beta}^{{\scriptscriptstyle \mathfrak{D}}}\,.\label{eq:ShamSchlueterInOperatorForm}
\end{equation}
Here $S_{\beta}^{{\scriptscriptstyle \text{M}}}\cdot\varDelta_{{\rm s}}^{{\scriptscriptstyle {\rm s}}}$
is equivalent to $-({\varDelta_{{\rm s}}^{{\scriptscriptstyle {\rm s}}}}_{k}{M}_{k,-k}^{k,-k}+{M^{\prime}}_{-k,k}^{k,-k}{\varDelta_{{\rm s}}^{{\scriptscriptstyle {\rm s}}}}_{k}^{\ast})$,
$S_{\beta}^{{\scriptscriptstyle \mathfrak{D}}}\cdot\varDelta_{{\rm s}}^{{\scriptscriptstyle {\rm s}}}=\mathfrak{D}_{k,-k}$
and $(S_{{\scriptscriptstyle \text{ph}}\beta}^{{\scriptscriptstyle \mathfrak{C}}}+S_{{\scriptscriptstyle \text{C}}\beta}^{{\scriptscriptstyle \mathfrak{C}}})\cdot\varDelta_{{\rm s}}^{{\scriptscriptstyle {\rm s}}}=\mathfrak{C}_{k,-k}$.
The non-linear Sham-Schl\"uter operator contributions are given by
\begin{eqnarray}
{S_{\beta}^{{\scriptscriptstyle \text{M}}}}_{kk^{\prime}} & = & -\updelta_{kk^{\prime}}\sum_{\sigma}\Bigl(\frac{(\varepsilon_{k\sigma}+\varepsilon_{-k-\sigma})^{2}}{\vert E_{i\sigma}^{+}-E_{i\sigma}^{-}\vert^{2}}P_{{\rm s}}(E_{k\sigma}^{+},E_{k\sigma}^{-})\nonumber \\
 &  & \hspace{-0.75cm}+2\vert u_{k\sigma}^{k+}\vert^{2}\vert v_{k-\sigma}^{-k+}\vert^{2}\sum_{\alpha}P_{{\rm s}}(E_{k\sigma}^{\alpha},E_{k\sigma}^{\alpha})\Bigr)\,,\label{eq:PotentialTermSDA}
\end{eqnarray}
and
\begin{eqnarray}
{S_{\beta}^{{\scriptscriptstyle \mathfrak{D}}}}_{kk^{\prime}} & = & \frac{1}{2}\updelta_{kk^{\prime}}\sum_{qk_{2}\sigma}\!\sum_{\alpha_{1}\alpha_{2}\alpha_{3}}\frac{\text{sign}(\alpha_{3})}{\vert E_{k\sigma}^{+}-E_{k\sigma}^{-}\vert}\times\nonumber \\
 &  & \hspace{-0.75cm}\times\Bigl(\bigl(\vert v_{k_{2}-\sigma}^{-k_{2}\alpha_{2}}\vert^{2}\vert v_{k-\sigma}^{-k\alpha_{1}}\vert^{2}\vert g_{-k_{2},-k,-\sigma}^{q}\vert^{2}+\nonumber \\
 &  & \hspace{-0.75cm}+\vert u_{k\sigma}^{k\alpha_{1}}\vert^{2}\vert u_{k_{2}\sigma}^{k_{2}\alpha_{2}}\vert^{2}\vert g_{kk_{2}\sigma}^{q}\vert^{2}\bigr)L(\varOmega_{q},E_{k\sigma}^{\alpha_{1}},E_{k_{2}\sigma}^{\alpha_{2}},E_{k\sigma}^{\alpha_{3}})+\nonumber \\
 &  & \hspace{-0.75cm}+\bigl(\vert v_{k_{2}-\sigma}^{-k_{2}\alpha_{2}}\vert^{2}\vert v_{k-\sigma}^{-k\alpha_{1}}\vert^{2}\vert g_{-k,-k_{2},-\sigma}^{q}\vert^{2}+\nonumber \\
 &  & \hspace{-0.75cm}+\vert u_{k\sigma}^{k\alpha_{1}}\vert^{2}\vert u_{k_{2}\sigma}^{k_{2}\alpha_{2}}\vert^{2}\vert g_{k_{2}k\sigma}^{q}\vert^{2}\bigr)L(\varOmega_{q},E_{k\sigma}^{\alpha_{1}},-E_{k_{2}\sigma}^{\alpha_{2}},E_{k\sigma}^{\alpha_{3}})\Bigr)\,.\nonumber \\
\label{eq:NonLinearDTermSymmetrized}
\end{eqnarray}
The term ${S_{\beta}^{{\scriptscriptstyle \mathfrak{D}}}}_{kk^{\prime}}$
due to the Nambu diagonal acts to reduce the critical temperature.
In the Refs.~\onlinecite{LuedersSCDFTI2005,MarquesSCDFTIIMetals2005} this term was
scaled down by a factor of $1/2$ in the functional construction to
compensate for a systematic underestimation as compared to the Eliashberg critical temperature
in the phonon only case.
In Ref.~\onlinecite{SannaMigdalFunctionalSCDFT2014}
a SCDFT functional is constructed, by using a proper interacting GF in the exchange self-energy of Eq.~(\ref{eq:PhononicHartreeFockDiagram}),
therefore removing the necessity to reduce the repulsive ${S_{\beta}^{{\scriptscriptstyle \mathfrak{D}}}}_{kk^{\prime}}$.
Having in mind to generalize this functional to SpinSCDFT, in the
present work we decided not to use the scale factor. In part {\bf II} we
find further indications that this scaling may also effect the robustness
of the SC state against a magnetic splitting. The predicted critical
temperature will be too low as compared to experiment but the correctness
of the qualitative behavior of the theory will be preserved. The Nambu
off-diagonal contributions that derives from the phonon interaction then reads
\begin{eqnarray}
{S_{{\scriptscriptstyle \text{ph}}\beta}^{{\scriptscriptstyle \mathfrak{C}}}}_{kk^{\prime}} & = & -\!\sum_{q\sigma}\!\sum_{\alpha_{1}\alpha_{2}\alpha_{3}}\frac{g_{kk^{\prime}\sigma}^{q}g_{-k,-k^{\prime},-\sigma}^{-q}}{\text{sign}(\alpha_{2})\vert\! E_{k^{\prime}\sigma}^{+}\!-\! E_{k^{\prime}\sigma}^{-}\!\vert}\times\nonumber \\
 &  & \hspace{-0.75cm}\times\Bigl(\!\vert u_{k\sigma}^{k\alpha_{1}}\vert^{2}\vert v_{k-\sigma}^{-k\alpha_{3}}\vert^{2}+u_{k\sigma}^{k\alpha_{1}}v_{k-\sigma}^{-k\alpha_{1}\ast}u_{k\sigma}^{k\alpha_{3}}v_{k-\sigma}^{-k\alpha_{3}\ast}\times\nonumber \\
 &  & \hspace{-0.75cm}\times\frac{{\varDelta_{{\rm s}}^{{\scriptscriptstyle {\rm s}}}}_{k^{\prime}}^{\ast}}{{\varDelta_{{\rm s}}^{{\scriptscriptstyle {\rm s}}}}_{k^{\prime}}}\!\Bigr)L\!(\varOmega_{q},E_{k\sigma}^{\alpha_{1}},E_{k^{\prime}\sigma}^{\alpha_{2}},E_{k\sigma}^{\alpha_{3}})\,,\label{eq:PhononCTerm}
\end{eqnarray}
and the contribution that derives from the static Coulomb interaction reads
\begin{eqnarray}
 &  & \hspace{-0.75cm}{S_{{\scriptscriptstyle \text{C}}\beta}^{{\scriptscriptstyle \mathfrak{C}}}}_{kk^{\prime}}=-\sum_{\sigma}\sum_{\alpha_{1}\alpha_{2}\alpha_{3}}\frac{\text{sign}(\alpha_{2})}{\vert E_{k^{\prime}\sigma}^{+}-E_{k^{\prime}\sigma}^{-}\vert}\times\nonumber \\
 &  & \hspace{-0.5cm}\times\Bigl({W_{kk^{\prime},-k,-k^{\prime}}^{{\scriptscriptstyle \text{stat}}}}_{\sigma,-\sigma}\vert u_{k\sigma}^{k\alpha_{1}}\vert^{2}\vert v_{k-\sigma}^{-k\alpha_{3}}\vert^{2}+\nonumber \\
 &  & \hspace{-0.5cm}+u_{k\sigma}^{k\alpha_{1}}v_{k-\sigma}^{-k\alpha_{1}\ast}u_{k\sigma}^{k\alpha_{3}}v_{k-\sigma}^{-k\alpha_{3}\ast}\frac{{\varDelta_{{\rm s}}^{{\scriptscriptstyle {\rm s}}}}_{k^{\prime}}^{\ast}}{{\varDelta_{{\rm s}}^{{\scriptscriptstyle {\rm s}}}}_{k^{\prime}}}{W_{kk^{\prime},-k,-k^{\prime}}^{{\scriptscriptstyle \text{stat}}}}_{\sigma,-\sigma}^{\ast}\Bigr)\times\nonumber \\
 &  & \hspace{-0.5cm}\times L_{{\scriptscriptstyle {\rm C}}}(\! E_{k\sigma}^{\alpha_{1}}\!,\! E_{k^{\prime}\sigma}^{\alpha_{2}}\!,\! E_{k\sigma}^{\alpha_{3}})\,.\label{eq:CoulombStatCTerm}
\end{eqnarray}
The functions $P_{{\rm s}}$, $L$ and $L_{{\scriptscriptstyle {\rm C}}}$
coming from analytic Matsubara summations, are given in the Appendix
\ref{sec:FormulasMatsubaraSums}, together with a discussion on some
limiting cases.

\subsubsection{Description of the Second Order Phase Transition}

If the SC transition to the normal state is of second order, $\chi(\boldsymbol{r},\boldsymbol{r}^{\prime})$
is assumed to go to zero continuously upon approaching the critical
temperature. From earlier work \cite{SarmaOnTheInfluenceOfAUniformExchangeFieldActingOnSC1963}
in the BCS framework, we expect this to be the case in the low magnetic field
part of the phase diagram. The formalism in the SDA is built on the
potential $\varDelta_{{\rm s}}^{{\scriptscriptstyle {\rm s}}}$ not
the order parameter $\chi$. We thus need to proof that a second order
phase transition implies also a continuous vanishing of the potential
$\varDelta_{{\rm s}}^{{\scriptscriptstyle {\rm s}}}$. We note that
in the SDA it is sufficient to show that the expansion coefficients of
$\chi$ and $\varDelta_{{\rm s}}^{{\scriptscriptstyle {\rm s}}}$
in our normal state basis are of the form
\begin{equation}
{\chi_{{\rm s}}}_{k\sigma,-k,-\sigma}=a_{k}^{\sigma,-\sigma}{\varDelta_{{\rm s}}^{{\scriptscriptstyle {\rm s}}}}_{k}\,\label{eq:LinearChiDeltaConnection},
\end{equation}
where $a_{k}^{\sigma,-\sigma}$ is some function of $\varDelta_{{\rm s}}^{{\scriptscriptstyle {\rm s}}}$
and show that $\lim_{\vert\varDelta_{{\rm s}}^{{\scriptscriptstyle {\rm s}}}\vert\rightarrow0}a_{k}^{\sigma,-\sigma}(\varDelta_{{\rm s}}^{{\scriptscriptstyle {\rm s}}})\neq0$.
Given that this is the case, in the limit $\vert\varDelta_{{\rm s}}^{{\scriptscriptstyle {\rm s}}}\vert\rightarrow0$
only linear order terms in the Sham-Schl\"uter
equation are relevant. Then, at a second order phase transition $T_{{\scriptscriptstyle {\rm c}}}$
can be computed from the condition that the matrix $\lim_{\vert\varDelta_{{\rm s}}^{{\scriptscriptstyle {\rm s}}}\vert\rightarrow0}S_{\beta}[\varDelta_{{\rm s}}^{{\scriptscriptstyle {\rm s}}}]$
is singular. 

\begin{figure}
\begin{centering}
\includegraphics[width=0.7\columnwidth]{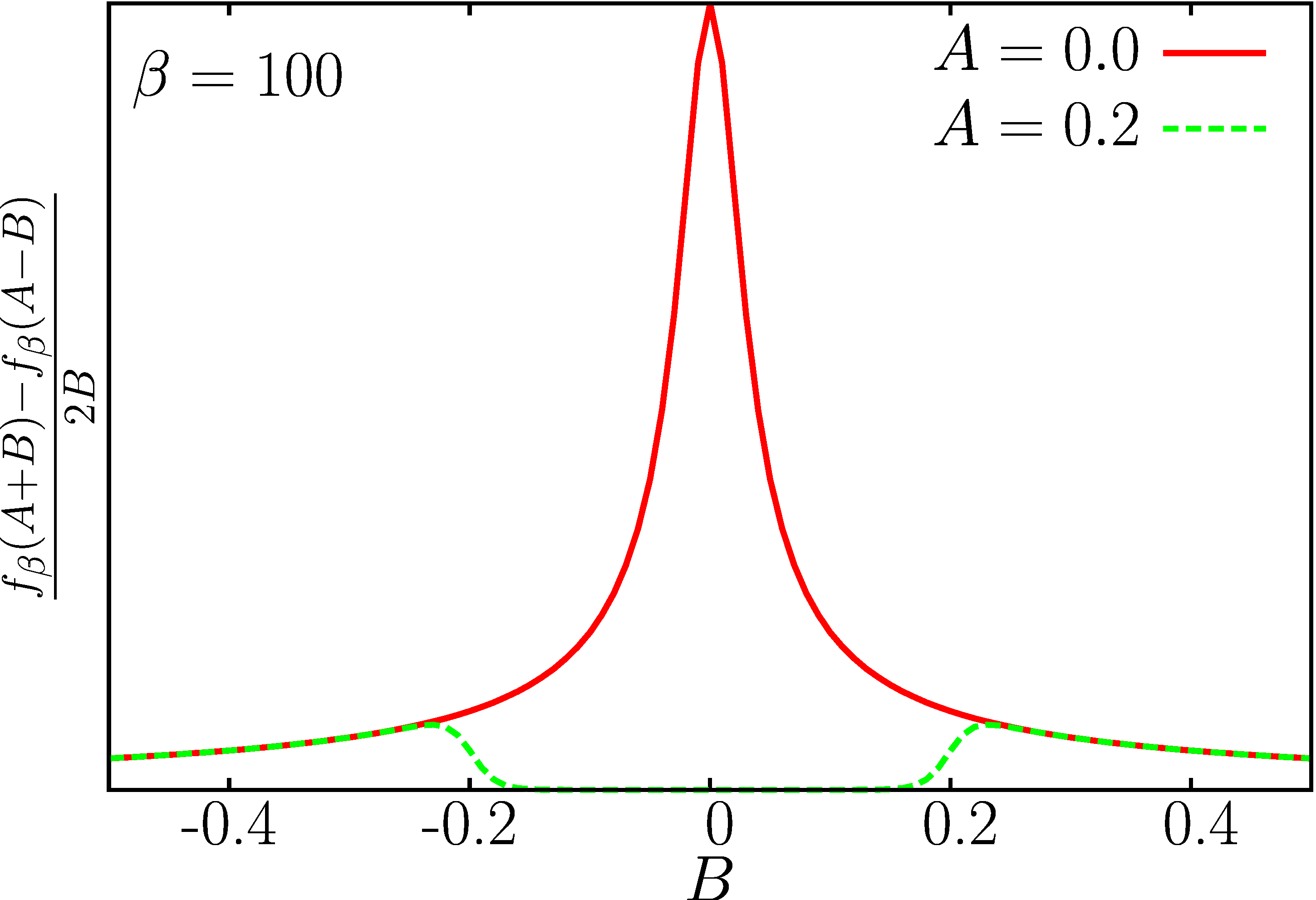}
\par\end{centering}

\caption{Sketch of the function $\beta a_{k}^{\sigma,-\sigma}(\varDelta_{{\rm s}}^{{\scriptscriptstyle {\rm s}}})$
multiplying ${\varDelta_{{\rm s}}^{{\scriptscriptstyle {\rm s}}}}_{k}$
into the coefficients for $\chi_{{\rm s}}(\boldsymbol{r},\boldsymbol{r}^{\prime})$.
While being always non-zero, the coefficient function at $B=0$ behaves
as $-(\beta/2)/\cosh\bigl(\beta(E_{k\sigma}^{+}+E_{k\sigma}^{-})/2\bigr)$
and thus becomes exponentially small with decreasing temperature.\label{fig:SketchFunctionForDeltaIntoChi} }
\end{figure}

Coming back to Eq.~\eqref{eq:LinearChiDeltaConnection} and
using the SDA together with Eq.~\eqref{eq:OrderParameterKSSpaceInTermsOfUV}
we see
\begin{eqnarray}
a_{k}^{\sigma,-\sigma} & = & \frac{f_{\beta}(E_{k\sigma}^{+})-f_{\beta}(E_{k\sigma}^{-})}{\vert E_{k\sigma}^{+}-E_{k\sigma}^{-}\vert}\,.
\end{eqnarray}
Clearly, at $T>0$ $a_{k}^{\sigma,-\sigma}$ can only be zero if $E_{k\sigma}^{+}-E_{k\sigma}^{-}\rightarrow0$.
Taking the respective limit
\begin{eqnarray}
\lim_{E_{k\sigma}^{+}-E_{k\sigma}^{-}\rightarrow0}a_{k}^{\sigma,-\sigma} & = & -\frac{\beta}{2}\frac{1}{\cosh\bigl(\beta(E_{k\sigma}^{+}+E_{k\sigma}^{-})/2\bigr)}\\
 & & < 0
\end{eqnarray}
which is the desired result. We may thus use $\vert\varDelta_{{\rm s}}^{{\scriptscriptstyle {\rm s}}}\vert\rightarrow0$
instead of $\vert\chi_{{\rm s}}\vert\rightarrow0$ at the point of
a second order phase transition. We sketch the function $a_{k}^{\sigma,-\sigma}$
using $A=\beta(E_{k\sigma}^{+}+E_{k\sigma}^{-})/2$ and $B=\beta(E_{k\sigma}^{+}-E_{k\sigma}^{-})/2$
in Fig.~\ref{fig:SketchFunctionForDeltaIntoChi}

Note that while $a_{k}^{\sigma,-\sigma}$ is strictly non-zero if
$\beta(E_{k\sigma}^{+}+E_{k\sigma}^{-})/2\gg1$ then $a_{k}^{\sigma,-\sigma}$
is exponentially small in the range $\vert B\vert\ll\vert A\vert$.
. We thus observe that the order parameter $\chi_{{\rm s}}(\boldsymbol{r},\boldsymbol{r}^{\prime})$
is only weakly dependent on the potential matrix elements ${\varDelta_{{\rm s}}^{{\scriptscriptstyle {\rm s}}}}_{i}$
that correspond to states below the splitting energy $A$. Still,
this does not invalidate the conclusion that at any finite temperature
a continuously vanishing order parameter implies a continuously vanishing
pair potential. We thus expect that (at low splitting) we can use
the linearized Sham-Schl\"uter equation \eqref{eq:ShamSchlueterInOperatorForm}.
In the following, we use a breve on top of linearized entities such as
$\breve{S}_{\beta_{{\scriptscriptstyle {\rm c}}}}=\lim_{\vert\varDelta_{{\rm s}}^{{\scriptscriptstyle {\rm s}}}\vert\rightarrow0}S_{\beta}[\varDelta_{{\rm s}}^{{\scriptscriptstyle {\rm s}}}]$
and Eq.~\eqref{eq:ShamSchlueterInOperatorForm} can be solved from the condition
\begin{equation}
\det\breve{S}_{\beta_{{\scriptscriptstyle {\rm c}}}}=0\,,
\end{equation}
where $\beta_{{\scriptscriptstyle {\rm c}}}=1/T_{{\scriptscriptstyle {\rm c}}}$.
The right eigenvector of $\breve{S}_{\beta_{{\scriptscriptstyle {\rm c}}}}$
is proportional to $\varDelta_{{\rm s}}^{{\scriptscriptstyle {\rm s}}}$.
To compute the small $\varDelta_{{\rm s}}^{{\scriptscriptstyle {\rm s}}}$
limit of $\breve{S}_{\beta_{{\scriptscriptstyle {\rm c}}}}$ we first
investigate the behavior of $\vert u_{k\sigma}^{k\alpha}\vert^{2}$,
$\vert v_{k-\sigma}^{-k\alpha}\vert^{2}$ and $E_{k\sigma}^{\alpha}$
separately where we find 
\begin{eqnarray}
\lim_{\vert\varDelta_{{\rm s}}^{{\scriptscriptstyle {\rm s}}}\vert\rightarrow0}\vert u_{k\sigma}^{k\alpha}\vert^{2} & = & \updelta_{\alpha,{\rm sign}(\varepsilon_{k\sigma}+\varepsilon_{-k-\sigma})}\,,\\
\lim_{\vert\varDelta_{{\rm s}}^{{\scriptscriptstyle {\rm s}}}\vert\rightarrow0}\vert v_{k-\sigma}^{-k\alpha}\vert^{2} & = & \updelta_{\alpha,-{\rm sign}(\varepsilon_{k\sigma}+\varepsilon_{-k-\sigma})}\,,\\
\lim_{\vert\varDelta_{{\rm s}}^{{\scriptscriptstyle {\rm s}}}\vert\rightarrow0}E_{k\sigma}^{\alpha} & = & \updelta_{\alpha,{\rm sign}(\varepsilon_{k\sigma}+\varepsilon_{-k-\sigma})}\varepsilon_{k\sigma}\nonumber \\
 & -& \updelta_{\alpha,-{\rm sign}(\varepsilon_{k\sigma}+\varepsilon_{-k-\sigma})}\varepsilon_{-k,-\sigma}\,.
\end{eqnarray}
Also we see that
\begin{equation}
\lim_{\vert\varDelta_{{\rm s}}^{{\scriptscriptstyle {\rm s}}}\vert\rightarrow0}\vert u_{k\sigma}^{k\alpha}\vert^{2}\vert v_{k-\sigma}^{-k\alpha}\vert^{2}=\lim_{\vert\varDelta_{{\rm s}}^{{\scriptscriptstyle {\rm s}}}\vert\rightarrow0}u_{k\sigma}^{k\alpha}{v_{k-\sigma}^{-k\alpha}}^{\ast}=0\,.
\end{equation}
Thus it is straightforward to arrive at
\begin{equation}
{\mbox{\ensuremath{\breve{S}_{\beta}^{{\scriptscriptstyle \text{M}}}}}}_{kk^{\prime}}=-2\delta_{kk^{\prime}}P_{{\rm s}}(\varepsilon_{k\uparrow},-\varepsilon_{-k\downarrow})\label{eq:PotentialTermLinear}
\end{equation}
and
\begin{eqnarray}
 &  & \hspace{-0.5cm}{\mbox{\ensuremath{\breve{S}_{\beta}^{{\scriptscriptstyle \mathfrak{D}}}}}}_{kk^{\prime}}=\nonumber\\
 &  & \frac{\delta_{kk^{\prime}}}{\varepsilon_{k\uparrow}\!+\!\varepsilon_{-k,\downarrow}}\sum_{ql}\Bigl(\vert g_{kl\uparrow}^{q}\vert^{2}\bigl(L(\varOmega_{q},\varepsilon_{k\uparrow},\varepsilon_{l\uparrow},\varepsilon_{k\uparrow})+\nonumber \\
 &  & \hspace{-0.25cm}+L(\varOmega_{q},\varepsilon_{k\uparrow},-\varepsilon_{l\uparrow},\varepsilon_{k\uparrow})-L(\varOmega_{q},\varepsilon_{k\uparrow},\varepsilon_{l\uparrow},-\varepsilon_{-k,\downarrow})\nonumber \\
 &  & \hspace{-0.25cm}-L(\varOmega_{q},\varepsilon_{k\uparrow},-\varepsilon_{l\uparrow},-\varepsilon_{-k,\downarrow})\bigr)+\nonumber \\
 &  & \hspace{-0.25cm}+\vert g_{-k,-l\downarrow}^{q}\vert^{2}\bigl(L(\varOmega_{q},\varepsilon_{-k\downarrow},\varepsilon_{-l\downarrow},\varepsilon_{-k\downarrow})+\nonumber \\
 &  & \hspace{-0.25cm}+L(\varOmega_{q},\varepsilon_{-k\downarrow},-\varepsilon_{-l\downarrow},\varepsilon_{-k\downarrow})-L(\varOmega_{q},\varepsilon_{-k\downarrow},\varepsilon_{-l\downarrow},-\varepsilon_{k\uparrow})\nonumber \\
 &  & \hspace{-0.25cm}-L(\varOmega_{q},\varepsilon_{-k\downarrow},-\varepsilon_{-l\downarrow},-\varepsilon_{k\uparrow})\bigr)\,.
\end{eqnarray}
Moreover
\begin{eqnarray}
 &  & \hspace{-0.5cm}{\mbox{\ensuremath{\breve{S}_{{\scriptscriptstyle \text{ph}}\beta}^{{\scriptscriptstyle \mathfrak{C}}}}}}_{kk^{\prime}} = \nonumber\\
 &  & -2\!\sum_{q}\!\frac{g_{kk^{\prime}\uparrow}^{q}g_{-k,-k^{\prime},\downarrow}^{-q}}{\vert\varepsilon_{k^{\prime}\uparrow}+\varepsilon_{-k^{\prime},\downarrow}\vert}\bigl(L(\varOmega_{q},\varepsilon_{k\uparrow},\breve{E}_{k^{\prime}\uparrow}^{+},-\varepsilon_{-k\downarrow})+\nonumber\\
 &  & +L(\varOmega_{q},\varepsilon_{-k\downarrow},\breve{E}_{-k^{\prime}\downarrow}^{+},-\varepsilon_{k\uparrow})\bigr)\label{eq:LinearCTermPhonon}
\end{eqnarray}
and
\begin{eqnarray}
 &  & \hspace{-0.75cm}{\mbox{\ensuremath{\breve{S}_{{\scriptscriptstyle \text{C}}\beta}^{{\scriptscriptstyle \mathfrak{C}}}}}}_{kk^{\prime}}=-2{W_{kk^{\prime},-k,-k^{\prime}}^{{\scriptscriptstyle \text{stat}}}}_{\uparrow,\downarrow}\times\nonumber \\
 &  & \times P_{{\rm s}}(\breve{E}_{k^{\prime}\uparrow}^{+},-\breve{E}_{-k^{\prime}\downarrow}^{+})P_{{\rm s}}(\varepsilon_{k\uparrow},-\!\varepsilon_{-k\downarrow})\,.
\end{eqnarray}

\subsubsection{Non-linear Gap-Equation}

Far from $T_{{\scriptscriptstyle {\rm c}}}$ or in those parts
of the phase diagram where the SC transition is of first order we
need to use the non-linear Sham-Schl\"uter equation, because a solution
with small $\vert{\varDelta_{{\rm s}}^{{\scriptscriptstyle {\rm s}}}}_{k}\vert$
may not exist. The most common method to solve an equation of type
Eq.~(\ref{eq:ShamSchlueterInOperatorForm}) is to use an invertible
splitting matrix $\mathcal{S}$ and cast $S_{\beta}[\varDelta_{{\rm s}}^{{\scriptscriptstyle {\rm s}}}]\cdot\varDelta_{{\rm s}}^{{\scriptscriptstyle {\rm s}}}=0$
into a fixed point problem
\begin{equation}
\varDelta_{{\rm s}}^{{\scriptscriptstyle {\rm s}}}=\mathcal{K}_{\mathcal{\mathcal{S}}}[\varDelta_{{\rm s}}^{{\scriptscriptstyle {\rm s}}}]\cdot\varDelta_{{\rm s}}^{{\scriptscriptstyle {\rm s}}}\quad\mathcal{K}_{\mathcal{S}}=\mathcal{S}^{-1}\cdot(S_{\beta}+\mathcal{S})\,.
\end{equation}
This is the gap equation of SpinSCDFT. In the spin degenerate limit
the choice $\mathcal{S}=-\breve{S}_{\beta}^{{\scriptscriptstyle \text{M}}}$
leads to the SCDFT gap equation given in Ref.~\onlinecite{LuedersSCDFTI2005}.
We point out that while we can show that all ${\mbox{\ensuremath{\breve{S}_{\beta}^{{\scriptscriptstyle \text{M}}}}}}_{k,k}<0$
at $\varepsilon_{k\sigma}+\varepsilon_{-k,-\sigma}=0$ ${\mbox{\ensuremath{\breve{S}_{\beta}^{{\scriptscriptstyle \text{M}}}}}}_{k,k}\sim\exp\bigl(-\frac{1}{2}\beta(\varepsilon_{k\sigma}-\varepsilon_{-k,-\sigma})\bigr)$
and is thus a numerically problematic object. In the implementation
that we describe in detail in {\bf II} we find that a good choice is $\mathcal{S}=-S_{\beta}^{{\scriptscriptstyle \text{M}}}(\varepsilon_{k\sigma}=\varepsilon_{-k,-\sigma})$.
Obviously in the spin degenerate limit we recover the formulas given
in Ref.~\onlinecite{LuedersSCDFTI2005}. In part {\bf II} we will also discuss the
properties of the splitting versus temperature diagram for a simple system in detail.

\section{Eliashberg equations\label{sec:The-Eliashberg-equations}}

In the KS-SpinSCDFT formalism, interaction effects are mimicked by
the $xc$-potential that is an (implicit) functional of the densities.
While the functional construction and the additional complications
of the SC KS system pose additional algebraic complexity, the result
is a numerically cheaper computational scheme. This is owed to the
fact that Matzubara summations in the self-energy are not computed
numerically but absorbed into the analytic structure of the $xc$-potential.
Likely, the knowledge of the interacting self-energy is essential
to a future improvement of the presented functional. The self-energy
Eq.~(\ref{eq:SelfEnergyXCMinusVXC}) in turn is constructed via diagrammatic
perturbation theory using the electronic and phononic GF similar to
Sec.~\ref{sub:Derivation-xc-Potential}, and involving the solution of a Dyson equation. 
In the present section, we develop this direct many-body scheme to obtain the electronic GF. 
The final set of equations generalize the ones of Eliashberg \cite{EliashbergInteractionBetweenElAndLatticeVibrInASC1960}
and we refer to them with the same name. Ref.~\onlinecite{VonsovskySuperconductivityTransitionMetals}
discusses similar equations in a different notation with a limitation
to isotropic system with a homogenous splitting parameter.

\subsection{Solving the Dyson Equation}

The starting point of the derivation of the Eliashberg equations is
the Dyson equation of a SC Eq.~(\ref{eq:DysonEquation}). We represent
it in the basis of normal state, zero temperature KS orbitals $\{\vec{\varphi}_{i\sigma}(\boldsymbol{r})\}$
defined in Eq.~(\ref{eq:DefinitionKSOrbitalBasis}). We use the Nambu-Anderson
\cite{NambuQParticlesGaugeInSuperconductivity1960,Anderson1958RPAInTheoryOfSuperconductivity}
notation similar to that used in the functional derivation and in Eq.~\ref{eq:ShamSchlueterEquation}.
The Dyson equation reads
\begin{equation}
\bar{G}_{ij}(\omega_{n})=\bar{G}_{ij}^{{\scriptscriptstyle {\rm KS}}}(\omega_{n})+\!\sum_{kl}\!\!\bar{G}_{ik}^{{\scriptscriptstyle {\rm KS}}}(\omega_{n})\cdot\bar{\varSigma}_{kl}^{{\scriptscriptstyle {\rm s}}}(\omega_{n})\cdot\bar{G}_{lj}(\omega_{n}),\label{eq:TheFullDysonEquation}
\end{equation}
where $\bar{G}_{ij}^{{\scriptscriptstyle {\rm KS}}}$ is the SC KS
GF and $\bar{\varSigma}_{ij}^{{\scriptscriptstyle {\rm s}}}(\omega_{n})=\bar{\varSigma}_{{\scriptscriptstyle {\rm xc}}ij}(\omega_{n})-\bar{v}_{{\scriptscriptstyle {\rm xc}}ij}$
where $\bar{\varSigma}_{{\scriptscriptstyle {\rm xc}}ij}(\omega_{n})$
is the Nambu exchange and correlation self-energy that also includes
the phononic Hartree diagram \cite{LinscheidHedinEquationsForSC2014}.
$\bar{v}_{{\scriptscriptstyle {\rm xc}}ij}$ are the matrix elements
of the $xc$-potential of the SC KS system. Note that the SC KS GF
is not diagonal in the space of $\{\vec{\varphi}_{i\sigma}(\boldsymbol{r})\}$.
Similar to our approach in SpinSCDFT of Section \ref{sec:SpinSCDFT}
we assume that $\{\vec{\varphi}_{i\sigma}(\boldsymbol{r})\}$ is a
good approximation to the quasi particle state%
\footnote{The same scheme for going beyond the decoupling approximation presented
in Sec.~\ref{sub:Competition-between-SC} can be used in this Eliashberg
approach: The KS orbital basis could, in principle, be self consistently
updated with modified densities in the SC state.%
}, i.e.~$\bar{\varSigma}_{kl}^{{\scriptscriptstyle {\rm s}}}(\omega_{n})$
and $\bar{G}_{ij}(\omega_{n})$ are essentially diagonal. We use similar
diagrams (Eq.~(\ref{eq:ElectronicGWDiagram}) and (\ref{eq:PhononicHartreeFockDiagram}))
as for the functional construction of SpinSCDFT in Subsection \ref{sub:Derivation-xc-Potential}
namely the phononic and Coulomb exchange diagram. Again similarly
(compare Sub.~\ref{sub:Derivation-xc-Potential}) we drop the Coulomb
corrections on the Nambu diagonal that add to the $xc$ potential.
Further we assume, as in the SDA of Sec.~\ref{sub:TheSDA}, that
the pairing occurs only between time reversed states \cite{Anderson1959DirtySuperc}.
This means we only consider singlet SC. Starting from Eq.~(\ref{eq:TheFullDysonEquation})
in the form $\bar{G}_{ij}(\omega_{n})=\bigl({\mbox{\ensuremath{\bar{G}_{ij}^{{\scriptscriptstyle {\rm KS}}}}}}^{-1}(\omega_{n})-\bar{\varSigma}_{ij}^{{\scriptscriptstyle {\rm s}}}(\omega_{n})\bigr)^{-1}$,
 under the mentioned approximations, the Dyson equation is a $4\times4$ matrix
equation that can be solved analytically. Note that here we do not substitute
the SC KS GF for the interaction GF in the self-energy (as was done
in the functional construction of SpinSCDFT of Sec.~\ref{sub:Derivation-xc-Potential}).
This is the main difference in the two approaches so far.

\subsubsection{Analytic Inversion of the Dyson Equation}

The easiest way to invert the right hand side of the Dyson equation
\begin{equation}
\bar{G}_{ij}(\omega_{n})=\bigl({\mbox{\ensuremath{\bar{G}_{ij}^{{\scriptscriptstyle {\rm KS}}}}}}^{-1}(\omega_{n})-\bar{\varSigma}_{ij}^{{\scriptscriptstyle {\rm s}}}(\omega_{n})\bigr)^{-1}\,,\label{eq:DysonEqByInversion}
\end{equation}
is to identify contributions of the self-energy that add to a given
variable of the inverse SC KS GF ${\mbox{\ensuremath{\bar{G}_{ij}^{{\scriptscriptstyle {\rm KS}}}}}}^{-1}(\omega_{n})$
of Eq.~(\ref{eq:InverseSCKSGF}). We summarize these self-energy
contributions in Table \ref{tab:Self-energy-contributions}
\begin{table}
\begin{centering}
\begin{tabular}{c|c|c|c}
SE part & ${\mbox{\ensuremath{\bar{G}^{{\scriptscriptstyle {\rm KS}}}}}}^{-1}$
part & Basis vector & Eliashberg\tabularnewline
\hline 
\hline 
$\varSigma_{k}^{\omega}(\omega_{n})$ & ${\rm i}\omega_{n}$ & $\tau_{0}\sigma_{0}$ & $Z_{k}(\omega_{n})$\tabularnewline
\hline 
$A_{k}^{\omega}(\omega_{n})$ & $-$ & $\tau_{0}\sigma_{z}$ & $\tilde{A}_{k}^{\omega}(\omega_{n})$\tabularnewline
\hline 
$\varSigma_{k}^{\varepsilon}(\omega_{n})$ & $(\varepsilon_{k\uparrow}+\varepsilon_{-k\downarrow})/2$ & $\tau_{z}\sigma_{0}$ & $\tilde{\varepsilon}_{k}(\omega_{n})$\tabularnewline
\hline 
$\varSigma_{k}^{J}(\omega_{n})$ & $(\varepsilon_{k\uparrow}-\varepsilon_{-k\downarrow})/2$ & $\tau_{z}\sigma_{z}$ & $\tilde{J}_{k}(\omega_{n})$\tabularnewline
\hline 
$\varSigma_{k}^{{\scriptscriptstyle \Re\Delta}}(\omega_{n})$ & $\Re{\varDelta_{{\rm s}}^{{\scriptscriptstyle {\rm s}}}}_{k}$ & $(\mbox{i}\tau_{y})(\mbox{i}\sigma_{y})$ & $\varDelta_{k}^{{\scriptscriptstyle \text{E}}}(\omega_{n})$\tabularnewline
\cline{1-3} 
$\mbox{i}\varSigma_{k}^{{\scriptscriptstyle \Im\Delta}}(\omega_{n})$ & $\mbox{i}\Im{\varDelta_{{\rm s}}^{{\scriptscriptstyle {\rm s}}}}_{k}$ & $\tau_{x}(\mbox{i}\sigma_{y})$ & $\varDelta_{k}^{{\scriptscriptstyle \text{E}}\star}(\omega_{n})$\tabularnewline
\end{tabular}
\par\end{centering}

\caption{Self-energy contributions, the variable of the inverse SC KS GF which
they add to and the basis vector. E.g.~along the $\tau_{0}\sigma_{0}$
direction in Spin and Nambu space ${\rm i}\omega_{n}+\varSigma_{k}^{\omega}(\omega_{n})$.
In the last column we give the related Eliashberg property. Note that
$\varDelta_{k}^{{\scriptscriptstyle \text{E}}}\sim\varSigma_{k}^{{\scriptscriptstyle \Re\Delta}}+\mbox{i}\varSigma_{k}^{{\scriptscriptstyle \Im\Delta}}$
and $\varDelta_{k}^{{\scriptscriptstyle \text{E}}\star}(\omega_{n})\sim\varSigma_{k}^{{\scriptscriptstyle \Re\Delta}}-\mbox{i}\varSigma_{k}^{{\scriptscriptstyle \Im\Delta}}$.\label{tab:Self-energy-contributions}}
\end{table}
. This means we decompose the Nambu and spin matrix $\bar{\varSigma}_{kl}^{{\scriptscriptstyle {\rm s}}}(\omega_{n})$
along the vectors $\tau_{0}\sigma_{0}$, $\tau_{z}\sigma_{0}$ and
so on. Then, we name the self-energy contributions according to the
property of the SC KS GF they add to in Eq.~(\ref{eq:DysonEqByInversion})
and indicate the property in the superscript. For example the Matsubara
frequency variable of the inverse SC KS GF points along the $\tau_{0}\sigma_{0}$
axis in spin and Nambu space. Correspondingly the self-energy part
along basis vector is referred to as $\varSigma_{k}^{\omega}(\omega_{n})$.
In the following we use $\vert g_{kk^{\prime}\sigma}^{q}\vert^{2}=\vert g_{k^{\prime}k\sigma}^{-q}\vert^{2}$,
$D_{q,-q}^{{\scriptscriptstyle {\rm 0}}}=D_{-q,q}^{{\scriptscriptstyle {\rm 0}}}$
and ${W_{kk^{\prime}k^{\prime}k}^{{\scriptscriptstyle \text{stat}}}}_{\sigma\sigma}={W_{k^{\prime}kkk^{\prime}}^{{\scriptscriptstyle \text{stat}}}}_{\sigma\sigma}$,
Then the equations for the corresponding scalar self-energy components
read
\begin{eqnarray}
\varSigma_{k}^{\omega}(\omega_{n}) & = & \frac{1}{4}\sum_{\sigma\alpha k^{\prime}q}\frac{1}{\beta}\sum_{n^{\prime}}(\tau_{0})_{\alpha\alpha}D_{q,-q}^{{\scriptscriptstyle {\rm 0}}}(\omega_{n^{\prime}}-\omega_{n})\times\nonumber \\
 &  & \times\vert g_{kk^{\prime}\sigma}^{q}\vert^{2}\bar{G}_{k^{\prime}\sigma,k^{\prime}\sigma}^{{\scriptscriptstyle \alpha,\alpha}}(\omega_{n^{\prime}})\label{eq:EliashbergSelfEnergyOmega}\\
A_{k}^{\omega}(\omega_{n}) & = & \frac{1}{4}\sum_{\sigma\alpha k^{\prime}q}\frac{1}{\beta}\sum_{n^{\prime}}\frac{(\tau_{0})_{\alpha\alpha}}{{\rm sign}(\sigma)}D_{q,-q}^{{\scriptscriptstyle {\rm 0}}}(\omega_{n^{\prime}}-\omega_{n})\times\nonumber \\
 &  & \times\vert g_{kk^{\prime}\sigma}^{q}\vert^{2}\bar{G}_{k^{\prime}\sigma,k^{\prime}\sigma}^{{\scriptscriptstyle \alpha,\alpha}}(\omega_{n^{\prime}})\label{eq:EliashbergSelfEnergyA}\\
\varSigma_{k}^{\varepsilon}(\omega_{n}) & = & \frac{1}{4}\sum_{\sigma\alpha k^{\prime}q}\frac{1}{\beta}\sum_{n^{\prime}}(\tau_{z})_{\alpha\alpha}D_{q,-q}^{{\scriptscriptstyle {\rm 0}}}(\omega_{n^{\prime}}-\omega_{n})\times\nonumber \\
 &  & \times\vert g_{kk^{\prime}\sigma}^{q}\vert^{2}\bar{G}_{k^{\prime}\sigma,k^{\prime}\sigma}^{{\scriptscriptstyle \alpha,\alpha}}(\omega_{n^{\prime}})\\
\varSigma_{k}^{J}(\omega_{n}) & = & \frac{1}{4}\sum_{\sigma\alpha k^{\prime}q}\frac{1}{\beta}\sum_{n^{\prime}}\frac{(\tau_{z})_{\alpha\alpha}}{{\rm sign}(\sigma)}D_{q,-q}^{{\scriptscriptstyle {\rm 0}}}(\omega_{n^{\prime}}-\omega_{n})\times\nonumber \\
 &  & \times\vert g_{kk^{\prime}\sigma}^{q}\vert^{2}\bar{G}_{k^{\prime}\sigma,k^{\prime}\sigma}^{{\scriptscriptstyle \alpha,\alpha}}(\omega_{n^{\prime}})\,.\label{eq:EliashbergSelfEnergyJ}
\end{eqnarray}
Note that $A_{k}^{\omega}(\omega_{n})$ stands out in the sense that
the SC KS GF has no contribution along this direction in Nambu and
spin space. On the Nambu-off-diagonal we similarly introduce
\begin{eqnarray}
\varSigma_{k}^{{\scriptscriptstyle \Im\Delta}}(\omega_{n}) & = & \!-\!\sum_{\sigma k^{\prime}}\frac{1}{\beta}\sum_{n^{\prime}}\frac{\text{sign}(\sigma)}{4\mbox{i}}\bigl(\sum_{q}D_{q,-q}^{{\scriptscriptstyle {\rm 0}}}(\omega_{n^{\prime}}-\omega_{n})\times\nonumber \\
 &  & \times g_{kk^{\prime}\sigma}^{q}g_{-k,-k^{\prime},-\sigma}^{-q}+{W_{kk^{\prime},-k,-k^{\prime}}^{{\scriptscriptstyle \text{stat}}}}_{\sigma,-\sigma}\bigr)\times\nonumber \\
 &  & \times\sum_{\alpha}(\tau_{x})_{\alpha\alpha}\bar{G}_{k^{\prime}\sigma,-k^{\prime},-\sigma}^{{\scriptscriptstyle \alpha,-\alpha}}(\omega_{n^{\prime}})\\
\varSigma_{k}^{{\scriptscriptstyle \Re\Delta}}(\omega_{n}) & = & \!-\!\sum_{\sigma k^{\prime}}\frac{1}{\beta}\sum_{n^{\prime}}\frac{\text{sign}(\sigma)}{4}\bigl(\sum_{q}D_{q,-q}^{{\scriptscriptstyle {\rm 0}}}(\omega_{n^{\prime}}-\omega_{n})\times\nonumber \\
 &  & g_{k^{\prime}k\sigma}^{q}g_{-k^{\prime},-k,-\sigma}^{-q}+{W_{k^{\prime}k,-k^{\prime},-k}^{{\scriptscriptstyle \text{stat}}}}_{\sigma,-\sigma}\bigr)\times\nonumber \\
 &  & \times\sum_{\alpha}({\rm i}\tau_{y})_{\alpha\alpha}\bar{G}_{k^{\prime}\sigma,-k^{\prime},-\sigma}^{{\scriptscriptstyle \alpha,-\alpha}}(\omega_{n^{\prime}})\,.
\end{eqnarray}
Here we introduce $B_{k}(\omega_{n})=\varSigma_{k}^{{\scriptscriptstyle \Re\Delta}}(\omega_{n})+{\rm i}\varSigma_{k}^{{\scriptscriptstyle \Im\Delta}}(\omega_{n})$
and $B_{k}^{\star}(\omega_{n})=\varSigma_{k}^{{\scriptscriptstyle \Re\Delta}}(\omega_{n})-{\rm i}\varSigma_{k}^{{\scriptscriptstyle \Im\Delta}}(\omega_{n})$
\begin{eqnarray}
B_{k}(\omega_{n}) & = & -\sum_{\sigma k^{\prime}}\frac{1}{\beta}\sum_{n^{\prime}}\frac{\text{sign}(\sigma)}{2}\bigl(\sum_{q}D_{q,-q}^{{\scriptscriptstyle {\rm 0}}}(\omega_{n^{\prime}}-\omega_{n})\times\nonumber \\
 &  & \times g_{kk^{\prime}\sigma}^{q}g_{-k,-k^{\prime},-\sigma}^{-q}+{W_{kk^{\prime},-k,-k^{\prime}}^{{\scriptscriptstyle \text{stat}}}}_{\sigma,-\sigma}\bigr)\times\nonumber \\
 &  & \times\bar{G}_{k^{\prime}\sigma,-k^{\prime},-\sigma}^{{\scriptscriptstyle 1,-1}}(\omega_{n^{\prime}})\label{eq:EliashbergSelfEnergyB}\\
B_{k}^{\star}(\omega_{n}) & = & \sum_{\sigma k^{\prime}}\frac{1}{\beta}\sum_{n^{\prime}}\frac{\text{sign}(\sigma)}{2}\bigl(\sum_{q}D_{q,-q}^{{\scriptscriptstyle {\rm 0}}}(\omega_{n^{\prime}}-\omega_{n})\times\nonumber \\
 &  & \times g_{k^{\prime}k\sigma}^{q}g_{-k^{\prime},-k,-\sigma}^{-q}+{W_{k^{\prime}k,-k^{\prime},-k}^{{\scriptscriptstyle \text{stat}}}}_{\sigma,-\sigma}\bigr)\times\nonumber \\
 &  & \times\bar{G}_{k^{\prime}\sigma,-k^{\prime},-\sigma}^{{\scriptscriptstyle -1,1}}(\omega_{n^{\prime}})\,.\label{eq:EliashbergSelfEnergyBStar}
\end{eqnarray}
If both $\Sigma_{k}^{{\scriptscriptstyle \Re\Delta}}$ and $\Sigma_{k}^{{\scriptscriptstyle \Im\Delta}}$
are real $B_{k}^{\star}$ is the complex conjugate of $B_{k}$. Further,
for the same reasons discussed in Sec.~\ref{par:Singlet-Superconductivity},
we do not consider the possibility that triplet self-energy contributions
appear. It is important to remark that, just as in the usual derivation
of the spin degenerate Eliashberg equations, the $k$ dependence of
all self-energy parts is generated via the $k$ dependence of the
Couplings $\vert g_{kk^{\prime}\sigma}^{q}\vert^{2}$ and in addition
${W_{k^{\prime}k,-k^{\prime},-k}^{{\scriptscriptstyle \text{stat}}}}_{\sigma,-\sigma}$
on the Nambu off diagonal. 

Introducing the mass renormalization function $Z_{k}(\omega_{n})$
as

\begin{equation}
\hspace{-1cm}Z_{k}(\omega_{n})=1+\text{i}\varSigma_{k}^{\omega}(\omega_{n})\!/\omega_{n}\,,
\end{equation}
we can rewrite some of the above equations by including $Z_{k}^{{\scriptscriptstyle \text{E}}}(\omega_{n})$
into the self-energy parts:

\begin{eqnarray}
\hspace{-1cm}\varDelta_{k}^{{\scriptscriptstyle \text{E}}}(\omega_{n}) & = & B_{k}(\omega_{n})\!/Z_{k}(\omega_{n})\label{eq:DeltaEliashberg1}\\
\hspace{-1cm}\varDelta_{k}^{{\scriptscriptstyle \text{E}}\star}(\omega_{n}) & = & B_{k}^{\star}(\omega_{n})\!/Z_{k}(\omega_{n})\\
\hspace{-1cm}\tilde{\varepsilon}_{k}(\omega_{n}) & = & \bigl((\varepsilon_{k\uparrow}+\varepsilon_{-k\downarrow})\!/2+\varSigma_{k}^{\varepsilon}(\omega_{n})\bigr)\!/Z_{k}(\omega_{n})\\
\hspace{-1cm}\tilde{J}_{k}(\omega_{n}) & = & \bigl((\varepsilon_{k\uparrow}-\varepsilon_{-k\downarrow})\!/2+\varSigma_{k}^{J}(\omega_{n})\bigr)\!/Z_{k}(\omega_{n})\\
\hspace{-1cm}\tilde{A}_{k}^{\omega}(\omega_{n}) & = & A_{k}^{\omega}(\omega_{n})\!/Z_{k}(\omega_{n})\,.\label{eq:AEliashberg5}
\end{eqnarray}
Then by introducing the abbreviation
\begin{eqnarray}
\mathfrak{F}_{k\sigma}(\omega_{n}) & = & \Bigl(\bigl(\tilde{\varepsilon}_{k}(\omega_{n})+\mbox{sign}(\sigma)\tilde{A}_{k}^{\omega}(\omega_{n})\bigr)^{2}\nonumber \\
 &  & +\varDelta_{k}^{{\scriptscriptstyle \text{E}}}(\omega_{n})\varDelta_{k}^{{\scriptscriptstyle \text{E}}\star}(\omega_{n})\Bigr)^{\frac{1}{2}}\,,
\end{eqnarray}
and suppressing the arguments $\omega_{n}$, we arrive at the formulas
for non-vanishing SC GF components
\begin{eqnarray}
\bar{G}_{k\sigma,k\sigma}^{{\scriptscriptstyle 1,1}} & = & \!\frac{1}{2\mathfrak{F}_{k\sigma}Z_{k}}\!\sum_{\alpha}\!\frac{\mathfrak{F}_{k\sigma}+\alpha\bigl(\tilde{\varepsilon}_{k}+\mbox{sign}(\sigma)\tilde{A}_{k}^{\omega}\bigr)}{\text{i}\omega_{n}-\mbox{sign}(\sigma)\tilde{J}_{k}-\alpha\mathfrak{F}_{k\sigma}}\nonumber \\
\\
\bar{G}_{k\sigma,k\sigma}^{{\scriptscriptstyle -1,-1}} & = & \!\frac{1}{2\mathfrak{F}_{k,-\sigma}Z_{k}}\!\sum_{\alpha}\!\frac{\mathfrak{F}_{k,-\sigma}\!+\alpha\bigl(\tilde{\varepsilon}_{k}\!-\!\mbox{sign}(\sigma)\!\tilde{A}_{k}^{\omega}\bigr)}{\text{i}\omega_{n}+\mbox{sign}(\sigma)\tilde{J}_{k}+\alpha\mathfrak{F}_{k,-\sigma}}\nonumber \\
\\
\bar{G}_{k\sigma,-k,-\sigma}^{{\scriptscriptstyle 1,-1}} & = & \!\frac{1}{2\mathfrak{F}_{k\sigma}Z_{k}}\!\sum_{\alpha}\!\frac{\mbox{sign}(\sigma)\alpha\varDelta_{k}^{{\scriptscriptstyle \text{E}}}}{\text{i}\omega_{n}-\mbox{sign}(\sigma)\tilde{J}_{k}-\alpha\mathfrak{F}_{k\sigma}}\nonumber \\
\label{eq:EliashbergGF_F}\\
\bar{G}_{k\sigma,-k,-\sigma}^{{\scriptscriptstyle -1,1}} & = & \!\frac{1}{2\mathfrak{F}_{k,-\sigma}Z_{k}}\!\sum_{\alpha}\!\frac{\mbox{sign}(\sigma)\alpha\varDelta_{k}^{{\scriptscriptstyle \text{E}}\star}}{\text{i}\omega_{n}+\mbox{sign}(\sigma)\tilde{J}_{k}+\alpha\mathfrak{F}_{k,-\sigma}}\,.\nonumber \\
\label{eq:EliashbergGF_Fdagger}
\end{eqnarray}
We have thus expressed the GF in terms of the self-energy components
(Eq.~(\ref{eq:DeltaEliashberg1}) to (\ref{eq:AEliashberg5})) explicitly.
The coupled set of equations Eq.~(\ref{eq:DeltaEliashberg1})
to (\ref{eq:AEliashberg5}) are the Eliashberg equations and have
to be solved according to the scheme:
\begin{enumerate}
\item Start with the coupling matrix elements $g_{k^{\prime}k\sigma}^{q}$
and ${W_{k^{\prime}k,-k^{\prime},-k}^{{\scriptscriptstyle \text{stat}}}}_{\sigma,-\sigma}$
and an initial guess for the self-energy components $\varDelta_{k}^{{\scriptscriptstyle \text{E}}},\varDelta_{k}^{{\scriptscriptstyle \text{E}}\star},\tilde{\varepsilon}_{k},\tilde{J}_{k}$
and $\tilde{A}_{k}^{\omega}$.
\item Evaluate Eq.~(\ref{eq:DeltaEliashberg1}) to (\ref{eq:AEliashberg5}).
They are closed in the sense that inserting the equations of this
section $\varDelta_{k}^{{\scriptscriptstyle \text{E}}},\varDelta_{k}^{{\scriptscriptstyle \text{E}}\star},\tilde{\varepsilon}_{k},\tilde{J}_{k}$
and $\tilde{A}_{k}^{\omega}$ only dependent on each other and the
coupling matrix elements $g_{k^{\prime}k\sigma}^{q}$ and ${W_{k^{\prime}k,-k^{\prime},-k}^{{\scriptscriptstyle \text{stat}}}}_{\sigma,-\sigma}$.
\item  Construct a new self-energy and  iterate from point 2, up to self-consistency. 
\end{enumerate}
$\tilde{A}_{k}^{\omega}$ is a peculiar object because it generates
a spin imbalance in the particle as compared to the hole channel.
To understand the effect of $\tilde{A}_{k}^{\omega}$ consider the
following self-consistent cycle. We start the iteration of these equations
with $\tilde{A}_{k}^{\omega}=0$ and $\varSigma_{k}^{{\scriptscriptstyle \Im\Delta}}=0$.
Then follows $\bar{G}_{k\sigma,-k,-\sigma}^{{\scriptscriptstyle 1,-1}}=\bar{G}_{k,-\sigma,-k\sigma}^{{\scriptscriptstyle -1,1}}$
which results in $B_{k}^{\star}=B_{k}$ and no self-energy part $\varSigma_{k}^{{\scriptscriptstyle \Im\Delta}}$
is generated. Further, because $\bar{G}_{k\sigma,k\sigma}^{{\scriptscriptstyle -1,-1}}=\bar{G}_{k,-\sigma k,-\sigma}^{{\scriptscriptstyle 1,1}}$
we find then that $\tilde{A}_{k}^{\omega}$ is proportional to the
difference of the interaction in the spin channels $\tilde{A}_{k}^{\omega}\propto\vert g_{kk^{\prime}\uparrow}^{q}\vert^{2}-\vert g_{kk^{\prime}\downarrow}^{q}\vert^{2}$.
If now the interaction is independent on the spin channel $\vert g_{kk^{\prime}\uparrow}^{q}\vert^{2}-\vert g_{kk^{\prime}\downarrow}^{q}\vert^{2}=0$
then $\tilde{A}_{k}^{\omega}$ also remains zero and we are at our
starting point. Thus we conclude that for spin independent couplings
$\varSigma_{k}^{{\scriptscriptstyle \Im\Delta}}$ and $\tilde{A}_{k}^{\omega}$
remain zero during iteration. If the interaction \textbf{is} spin
dependent $\vert g_{kk^{\prime}\uparrow}^{q}\vert^{2}-\vert g_{kk^{\prime}\downarrow}^{q}\vert^{2}\neq0$
the self-consistency iteration will generate a spin imbalance in the
GF. This is not surprising because the up and down single particle
spectrum is altered in a different way by the interaction. Then a
non-vanishing $\varSigma_{k}^{{\scriptscriptstyle \Im\Delta}}$ cannot
be excluded.

For future reference we extract the renormalized energy dependence
$\tilde{\varepsilon}_{k}$ of the GF as it appears in the self-energy
Eq.~(\ref{eq:EliashbergSelfEnergyOmega}) to (\ref{eq:EliashbergSelfEnergyJ})
and Eq.~(\ref{eq:EliashbergSelfEnergyB}) and (\ref{eq:EliashbergSelfEnergyBStar}).
With the abbreviation
\begin{eqnarray}
a_{k}(\omega_{n}) & = & (\tilde{A}_{k}^{\omega z})^{2}+\varDelta_{k}^{{\scriptscriptstyle \text{E}}}\varDelta_{k}^{{\scriptscriptstyle \text{E}}\star}+\omega_{n}^{2}-(\tilde{J}_{k})^{2}
\end{eqnarray}
we obtain ($b=0,z$)

\begin{eqnarray}
 &  & \hspace{-0.75cm}\sum_{\alpha}(\tau_{b})_{\alpha\alpha}\bar{G}_{k\sigma,k\sigma}^{{\scriptscriptstyle \alpha,\alpha}}=\nonumber \\
 &  & \hspace{-0.5cm}\!=\!\sum_{\alpha}\!\!\frac{(\tau_{b})_{\alpha\alpha}}{Z_{k}}\frac{\alpha\Bigl(\tilde{\varepsilon}_{k}\!-\!\mbox{sign}(\sigma)\bigl(\tilde{J}_{k}\!+\!\alpha\tilde{A}_{k}^{\omega}\bigr)\Bigr)\!-\!\text{i}\omega_{n}}{a_{k}\!-\!\frac{2\alpha}{\mbox{sign}(\sigma)}\bigl(\text{i}\omega_{n}\tilde{J}_{k}\!+\!\tilde{A}_{k}^{\omega}\tilde{\varepsilon}_{k}\bigr)\!+\!\tilde{\varepsilon}_{k}^{2}}\label{eq:GreensfunctionSum}
\end{eqnarray}
and
\begin{eqnarray}
 &  & \hspace{-0.75cm}\sum_{\sigma}\bar{G}_{k\sigma,-k,-\sigma}^{{\scriptscriptstyle 1,-1}}(\omega_{n})=\nonumber \\
 &  & \hspace{-0.5cm}\!=\!-\frac{1}{Z_{k}}\sum_{\alpha}\frac{\varDelta_{k}^{{\scriptscriptstyle \text{E}}}}{a_{k}-\alpha\bigl(\text{i}\omega_{n}\tilde{J}_{k}+\tilde{A}_{k}^{\omega}\tilde{\varepsilon}_{k}\bigr)+\tilde{\varepsilon}_{k}^{2}}\label{eq:OffDiagonalGFSpinSum}
\end{eqnarray}

\subsection{Analytic Integration of the Energy}

In a numerical solution, the equations (\ref{eq:DeltaEliashberg1})
to (\ref{eq:AEliashberg5}) have to be iterated until self-consistency
is reached. Each self-consistent step requires to perform Matsubara
summations in addition to the $k$ space summations which will be
numerically demanding. 

Note however that the $k$ space summations can be avoided using an
approximation that is very common in the context of Eliashberg theory
which is essentially to replace the couplings with their value at
$\tilde{\varepsilon}_{k}(\omega_{n})=0$. The reason why this is sensible
can be understood from the GF. From the above equation (\ref{eq:GreensfunctionSum})
one can easily see that $\bar{G}_{k\sigma,k\sigma}^{{\scriptscriptstyle 1,1}}(\omega_{n})-\bar{G}_{k\sigma,k\sigma}^{{\scriptscriptstyle -1,-1}}(\omega_{n})$
behaves as $\bigl(\tilde{\varepsilon}_{k}(\omega_{n})\bigr)^{-1}$
for large $\tilde{\varepsilon}_{k}(\omega_{n})$. In turn, $\bar{G}_{k\sigma k\sigma}^{{\scriptscriptstyle 1,1}}(\omega_{n})+\bar{G}_{k\sigma k\sigma}^{{\scriptscriptstyle -1,-1}}(\omega_{n})$
and the Nambu off-diagonal parts $\bar{G}_{k\sigma,-k,-\sigma}^{{\scriptscriptstyle \alpha,-\alpha}}(\omega_{n})$
behave as $\bigl(\tilde{\varepsilon}_{k}(\omega_{n})\bigr)^{-2}$
for large $\tilde{\varepsilon}_{k}(\omega_{n})$. Using this insight
we see from the Eqs.~(\ref{eq:EliashbergSelfEnergyOmega}), (\ref{eq:EliashbergSelfEnergyA}),
(\ref{eq:EliashbergSelfEnergyB}) and (\ref{eq:EliashbergSelfEnergyBStar})
that $Z_{k}(\omega_{n}),\tilde{A}_{k}^{\omega}(\omega_{n}),\varDelta_{k}^{{\scriptscriptstyle \text{E}}}(\omega_{n})$
and $\varDelta_{k}^{{\scriptscriptstyle \text{E}}\star}(\omega_{n})$
are almost independent on the space $k$ belonging to large $\tilde{\varepsilon}_{k}$
because its contributions are suppressed by a factor $\bigl(\tilde{\varepsilon}_{k}(\omega_{n})\bigr)^{-2}$.
Thus these quantities can be computed replacing the couplings with
their value at $\tilde{\varepsilon}_{k}(\omega_{n})=0$.

With the integrand behaving as $\bigl(\tilde{\varepsilon}_{k}(\omega_{n})\bigr)^{-1}$
, the convergence of the Brillouin zone integrals in $\varSigma_{k}^{\varepsilon}(\omega_{n})$
and $\varSigma_{k}^{J}(\omega_{n})$ depend on the $k$-dependence
of the couplings in an essential way, even on $k$ that correspond
to a large $\tilde{\varepsilon}_{k}$. In particular, in absence of
any $k$ dependence of the couplings $\varSigma_{k}^{\varepsilon}(\omega_{n})$
and $\varSigma_{k}^{J}(\omega_{n})$ diverge logarithmically. From
the physical point of view $\varSigma_{k}^{\varepsilon}(\omega_{n})$
shifts the position of the Fermi energy and $\varSigma_{k}^{J}(\omega_{n})$
the magnetic splitting of quasiparticle states due to many-body interactions.
These terms are zero if the system shows particle-hole symmetry and
small in general (see also the discussion in Sec.~\ref{sub:Derivation-xc-Potential}). 
Therefore we will discard these contributions completely and replace the couplings with
their value at $\tilde{\varepsilon}_{k}(\omega_{n})=0$ entirely, reducing the computational costs significantly.

Another very effective simplification of the formalism comes from
assuming the system to be isotropic in $k$. This means that the couplings will
depend on $k$ only via the quasi particle energy $\varepsilon_{k\sigma}$.
Here, we introduce the averaging operation on a generic function $F_{k\sigma}$
on equal center of energy and equal splitting surfaces according to
\begin{eqnarray}
F_{\sigma}(\varepsilon,J) & = & \hat{I}_{k\sigma}(\varepsilon,J)F_{k\sigma}\\
 & = & \frac{1}{\varrho(\varepsilon,J)}\sum_{k}\updelta(\frac{\varepsilon_{{\scriptscriptstyle {\rm sign}(\sigma)}k\uparrow}+\varepsilon_{-{\scriptscriptstyle {\rm sign}(\sigma)}k\downarrow}}{2}-\varepsilon)\times\nonumber \\
 &  & \times\updelta(\frac{\varepsilon_{{\scriptscriptstyle {\rm sign}(\sigma)}k\uparrow}-\varepsilon_{-{\scriptscriptstyle {\rm sign}(\sigma)}k\downarrow}}{2}-J)F_{k\sigma}\,,
\end{eqnarray}
where the number of states on center of energy and splitting surfaces
is given by $\varrho(\varepsilon,J)=\hat{I}_{k\sigma}(\varepsilon,J)\,1$.
The subscript indices ``$k\sigma"$ on $\hat{I}_{k\sigma}(\varepsilon,J)$
indicate the variables that are averaged. Note that we invert the
sign of $\boldsymbol{k}$ for the $\sigma=\downarrow$ part which
makes $\hat{I}_{k\sigma}(\varepsilon,J)F_{k\sigma}=\hat{I}_{-k,-\sigma}(\varepsilon,J)F_{-k,-\sigma}$.
Now we define the analog of the Eliashberg function $\alpha^{\!2}\! F(\omega)$
\cite{CarbottePropertiesOfBosonExchangeSC1990,AllenTheoryOfSCTc1983}.
We are going to keep the state dependence $k$ for a little longer,
and eventually take only those $k$ such that $\tilde{\varepsilon}_{k}(\omega_{n})=0$. 
On the Nambu diagonal it appears the coupling function
\begin{eqnarray}
 &  & \hspace{-0.5cm}\hspace{-0.75cm}\alpha^{\!2}\! F_{\sigma}^{{\scriptscriptstyle {\rm D}}}(\varepsilon,J,\varepsilon^{\prime},J^{\prime},\varOmega)=\nonumber \\
 &  & \hspace{-0.75cm}\varrho(\varepsilon^{\prime}\!,\! J^{\prime})\hat{I}_{k^{\prime}\sigma}(\varepsilon^{\prime}\!,\! J^{\prime})\hat{I}_{k\sigma}(\varepsilon,J)\!\sum_{q}\!\vert g_{kk^{\prime}\sigma}^{q}\vert^{2}\updelta(\varOmega-\varOmega_{q})
\end{eqnarray}
and on the Nambu off diagonal
\begin{eqnarray}
 &  & \hspace{-0.5cm}\hspace{-0.75cm}\alpha^{\!2}\! F(\varepsilon,J,\varepsilon^{\prime},J^{\prime},\varOmega)=\varrho(\varepsilon^{\prime},J^{\prime})\hat{I}_{k^{\prime}\sigma}(\varepsilon^{\prime},J^{\prime})\times\nonumber \\
 &  & \hspace{-0.75cm}\times\hat{I}_{k\sigma}(\varepsilon,J)\sum_{q}g_{kk^{\prime}\sigma}^{q}g_{-k,-k^{\prime},-\sigma}^{-q}\updelta(\varOmega-\varOmega_{q})\label{eq:PhononIsotropicCoupling}\\
 &  & \hspace{-0.5cm}\hspace{-0.75cm}C^{{\rm {\scriptscriptstyle stat}}}(\varepsilon,J,\varepsilon^{\prime},J^{\prime})=\nonumber \\
 &  & \hspace{-0.75cm}\varrho(\varepsilon^{\prime},J^{\prime})\hat{I}_{k^{\prime}\sigma}(\varepsilon^{\prime},J^{\prime})\hat{I}_{k\sigma}(\varepsilon,J){W_{k^{\prime}k,-k^{\prime},-k}^{{\scriptscriptstyle \text{stat}}}}_{\sigma,-\sigma}\,.\label{eq:CoulombIsostropicCoupling}
\end{eqnarray}
Note that in the above equations (\ref{eq:PhononIsotropicCoupling})
and (\ref{eq:CoulombIsostropicCoupling}), the left hand side does
not depend on $\sigma$ because the averaging leads to the same result
for $\sigma=\uparrow$ or $\sigma=\downarrow$. The summation over
$k^{\prime}$ and $q$ in the self-energy Eqs.~(\ref{eq:EliashbergSelfEnergyOmega})
to (\ref{eq:EliashbergSelfEnergyBStar}) are then transformed to integrals
over $\varepsilon^{\prime},J^{\prime}$ and $\varOmega$ respectively.
However, if the couplings loose their center of energy dependence $\varepsilon$,
the following functions only depend on the Matsubara frequency $\omega_{n}$
(that we now indicate as the index $n$) and the splitting: $Z_{n}(J),\tilde{A}_{n}^{\omega}(J),\varDelta_{n}^{{\scriptscriptstyle \text{E}}}(J)$
and $\varDelta_{n}^{{\scriptscriptstyle \text{E}}\star}(J)$. With
$\tilde{\varepsilon}_{k}(\omega_{n})\equiv\varepsilon/Z_{n}$ and
$\tilde{J}_{k}(\omega_{n})=J/Z_{n}$ we can compute analytically the integral over the 
center of energy $\varepsilon$ of Eq.~(\ref{eq:GreensfunctionSum}).
Because the integrand decays faster than $\varepsilon^{-1}$ for large $\varepsilon$, 
we may compute the integral 
\begin{eqnarray}
 &  & \mathfrak{M}_{n\sigma}(J)=\nonumber \\
 &  & \int\hspace{-0.18cm}{\rm d}\varepsilon\sum_{\alpha}\frac{\alpha\Bigl(\varepsilon-\mbox{sign}(\sigma)\bigl(J+\tilde{A}_{n}^{\omega}Z_{n}\bigr)\Bigr)-\text{i}\omega_{n}Z_{n}}{a_{n}(J){Z_{n}}^{2}-\frac{2\alpha}{{\rm sign}(\sigma)}\bigl(\text{i}\omega_{n}Z_{n}J+\tilde{A}_{n}^{\omega}Z_{n}\varepsilon\bigr)+\varepsilon^{2}}\nonumber \\
\end{eqnarray}
as the sum of residues in the upper complex half plane. Since it is not
clear which of the four poles will be in the upper half we compute
all residues. Adding those, we obtain the energy integral in Eq.~(\ref{eq:EliashbergSelfEnergyOmega})
and Eq.~(\ref{eq:EliashbergSelfEnergyA}) with 
\begin{equation}
\mathfrak{S}_{n,\sigma}(J)=\sqrt{-\bigl({Z_{n}}^{2}\varDelta_{n}^{{\scriptscriptstyle \text{E}}}\varDelta_{n}^{{\scriptscriptstyle \text{E}}\star}-\bigl(\text{i}\omega_{n}Z_{n}-\mbox{sign}(\sigma)J\bigr)^{2}\bigr)}
\end{equation}
as
\begin{eqnarray}
 &  & \!\!\!{\it \mathfrak{M}}_{n\sigma}(J)=\nonumber \\
 &  & \!\!\pi\mbox{i}\Bigl(\frac{\text{i}\omega_{n}Z_{n}-\mbox{sign}(\sigma)J}{\mathfrak{S}_{n,\sigma}}-1\Bigr)\uptheta\Bigl(\Im\bigl(-\frac{\tilde{A}_{n}^{\omega}Z_{n}}{{\rm sign}(\sigma)}-\mathfrak{S}_{n,\sigma}\bigr)\Bigr)\nonumber \\
 &  & \!\!-\pi\mbox{i}\Bigl(\frac{\text{i}\omega_{n}Z_{n}-\mbox{sign}(\sigma)J}{\mathfrak{S}_{n,\sigma}}+1\Bigr)\uptheta\Bigl(\Im\bigl(-\frac{\tilde{A}_{n}^{\omega}Z_{n}}{{\rm sign}(\sigma)}+\mathfrak{S}_{n,\sigma}\bigr)\Bigr)\nonumber \\
 &  & \!\!+\pi\mbox{i}\Bigl(\frac{\text{i}\omega_{n}Z_{n}+\mbox{sign}(\sigma)J}{\mathfrak{S}_{n,-\sigma}}+1\Bigr)\uptheta\Bigl(\Im\bigl(\frac{\tilde{A}_{n}^{\omega}Z_{n}}{{\rm sign}(\sigma)}-\mathfrak{S}_{n,-\sigma}\bigr)\Bigr)\nonumber \\
 &  & \!\!-\pi\mbox{i}\Bigl(\frac{\text{i}\omega_{n}Z_{n}+\mbox{sign}(\sigma)J}{\mathfrak{S}_{n,-\sigma}}-1\Bigr)\uptheta\Bigl(\Im\bigl(\frac{\tilde{A}_{n}^{\omega}Z_{n}}{{\rm sign}(\sigma)}+\mathfrak{S}_{n,-\sigma}\bigr)\Bigr)\,.\nonumber \\
\end{eqnarray}
Further, for Eqs.~(\ref{eq:EliashbergSelfEnergyB}) and (\ref{eq:EliashbergSelfEnergyBStar}),
we integrate Eq.~(\ref{eq:OffDiagonalGFSpinSum}) in center of energy
$\varepsilon$. We define
\begin{eqnarray}
\mathfrak{N}_{n}(J) & = & \hspace{-0.18cm}\sum_{\alpha}\hspace{-0.18cm}\int\hspace{-0.18cm}\frac{{\rm d}\varepsilon}{a_{n}Z_{n}^{2}-\alpha\bigl(\text{i}\omega_{n}Z_{n}J+\varepsilon\tilde{A}_{n}^{\omega}\bigr)+\varepsilon^{2}}\,,
\end{eqnarray}
that is evaluated to
\begin{eqnarray}
\mathfrak{N}_{n}(J) & = & \pi\mbox{i}\Bigl({\mathfrak{S}_{n,\uparrow}}^{-1}\uptheta\bigl(\Im(-\tilde{A}_{n}^{\omega}Z_{n}-\mathfrak{S}_{n,\uparrow})\bigr)\nonumber \\
 &  & -{\mathfrak{S}_{n,\uparrow}}^{-1}\uptheta\bigl(\Im(-\tilde{A}_{n}^{\omega}Z_{n}+\mathfrak{S}_{n,\uparrow})\bigr)\nonumber \\
 &  & +{\mathfrak{S}_{n,\downarrow}}^{-1}\uptheta\bigl(\Im(\tilde{A}_{n}^{\omega}Z_{n}-\mathfrak{S}_{n,\downarrow})\bigr)\nonumber \\
 &  & -{\mathfrak{S}_{n,\downarrow}}^{-1}\uptheta\bigl(\Im(\tilde{A}_{n}^{\omega}Z_{n}+\mathfrak{S}_{n,\downarrow})\bigr)\Bigr)\,.
\end{eqnarray}
We obtain the Eliashberg equations similar to their usual, spin-degenerate
form \cite{SannaPittalis2012,ASannaPHD}, that only refer to the GF implicitly 
\begin{eqnarray}
Z_{n}(J) & = & 1+\frac{\text{i}}{4\omega_{n}}\hspace{-0.18cm}\int\hspace{-0.18cm}\mbox{d}J^{\prime}\frac{1}{\beta}\sum_{n^{\prime}\sigma}\mathfrak{K}_{n,n^{\prime}}^{\sigma}(J,J^{\prime})\mathfrak{M}_{n^{\prime}\sigma}(J^{\prime})\nonumber \\
\label{eq:Eliashberg_Eq_Z_1}\\
\tilde{A}_{n}^{\omega}(J) & = & \frac{1}{4Z_{n}(J)}\hspace{-0.18cm}\int\hspace{-0.18cm}\mbox{d}J^{\prime}\frac{1}{\beta}\sum_{n^{\prime}\sigma}\frac{\mathfrak{K}_{n,n^{\prime}}^{\sigma}(J,J^{\prime})}{\text{sign}(\sigma)}\mathfrak{M}_{n^{\prime}\sigma}(J^{\prime})\nonumber \\
\\
\varDelta_{n}^{{\scriptscriptstyle \text{E}}}(J) & = & -\frac{1}{2Z_{n}(J)}\hspace{-0.18cm}\int\hspace{-0.18cm}\mbox{d}J^{\prime}\frac{1}{\beta}\sum_{n^{\prime}}\mathfrak{L}_{n,n^{\prime}}(J,J^{\prime})\times\nonumber \\
 &  & \times Z_{n^{\prime}}(J^{\prime})\varDelta_{n^{\prime}}^{{\scriptscriptstyle \text{E}}}(J^{\prime})\mathfrak{N}_{n^{\prime}}(J^{\prime})\label{eq:Eliashberg_Eq_Delta_3}\\
\varDelta_{n}^{{\scriptscriptstyle \text{E}}\star}(J) & = & -\frac{1}{2Z_{n}(J)}\hspace{-0.18cm}\int\hspace{-0.18cm}\mbox{d}J^{\prime}\frac{1}{\beta}\sum_{n^{\prime}}\mathfrak{L}_{n,n^{\prime}}(J,J^{\prime})\times\nonumber \\
 &  & \times Z_{n^{\prime}}(J^{\prime})\varDelta_{n^{\prime}}^{{\scriptscriptstyle \text{E}}\star}(J^{\prime})\mathfrak{N}_{n^{\prime}}(J^{\prime})\label{eq:Eliashberg_Eq_DeltaStar_4}
\end{eqnarray}
where
\begin{eqnarray}
\mathfrak{K}_{n,n^{\prime}}^{\sigma}(J,J^{\prime}) & = & \int\hspace{-0.18cm}\mbox{d}\varOmega\frac{2\varOmega\,\alpha^{\!2}\! F_{\sigma}^{{\scriptscriptstyle {\rm D}}}(0,J,0,J^{\prime},\varOmega)}{(\omega_{n}-\omega_{n^{\prime}})^{2}+\varOmega^{2}}\\
\mathfrak{L}_{n,n^{\prime}}(J,J^{\prime}) & = & \int\hspace{-0.18cm}\mbox{d}\varOmega\frac{2\varOmega\,\alpha^{\!2}\! F(0,J,0,J^{\prime},\varOmega)}{(\omega_{n}-\omega_{n^{\prime}})^{2}+\varOmega^{2}}+\nonumber \\
 &  & +C^{{\rm {\scriptscriptstyle stat}}}(0,J,0,J^{\prime})\,.
\end{eqnarray}
We point out that the Coulomb interaction is not well suited for the
$k$-constant coupling approximation. The reason is that the function
$\mathfrak{N}_{n}(J)$ behaves as $1/n$ for large $n$ while $Z_{n}(J)$
goes to $1$ and thus the Matsubara integral shows a logarithmic divergence
due to $C^{{\rm {\scriptscriptstyle stat}}}(0,J,0,J^{\prime})$ if
$\varDelta_{n}^{{\scriptscriptstyle \text{E}}}(J)$ does not cut off
the integral. Often the effect of the Coulomb potential is mimicked
by replacing $C^{{\rm {\scriptscriptstyle stat}}}$ with $\mu^{\star}\uptheta(\omega_{c}-\vert\omega_{n}\vert)$
where $\mu^{\star}=\frac{C^{{\rm {\scriptscriptstyle stat}}}}{1+C^{{\rm {\scriptscriptstyle stat}}}\ln(\mathcal{E}/\omega_{c})}$
with $\mathcal{E}$, a parameter of the electronic band structure
and $\omega_{c}$ a phonon frequency cutoff\cite{MorelCalculationSCStateParametersWithRetardedElPhInteraction1962,ScalapinoStrongCouplSC1966}. Usually the so called
Morel-Anderson pseudo potential $\mu^{\star}$ is fitted 
so that the calculated $T_{c}$ matches the experimental one. $\mu^{\star}$
usually ranges between $0.1$ and $0.16$ for conventional SC \cite{CarbottePropertiesOfBosonExchangeSC1990}.
The above equations imply that the coupling is isotropic in the sense
that all states with equal center of energy and equal splitting share
the same coupling matrix elements. Sometimes as in the well known
case of ${\rm MgB}_{2}$ there are significant differences in the
couplings and it is important to group states into bands for the isotropic
approximation to hold. We refer to this case as the multiband approximation
which simply means that all isotropic variables obtain another index
for the band they correspond to.

Comparing the equations for the SC KS GF of Eq.~(\ref{eq:KSNambuGreenfunctionSpinDecoupl})
(noting $u_{k\sigma}^{k\alpha}(v_{k-\sigma}^{-k\alpha})^{\ast}=\alpha{\rm sign}(\sigma){\varDelta_{{\rm s}}^{{\scriptscriptstyle {\rm s}}}}_{k}/F_{k}$
where $F_{k}=\sqrt{\frac{\varepsilon_{k\uparrow}+\varepsilon_{-k\downarrow}}{2}+\vert{\varDelta_{{\rm s}}^{{\scriptscriptstyle {\rm s}}}}_{k}\vert^{2}}$)
with the interacting GF Eq.~(\ref{eq:EliashbergGF_F}) we note that
${\varDelta_{{\rm s}}^{{\scriptscriptstyle {\rm s}}}}_{k}$ takes
the role of $\varDelta_{k}^{{\scriptscriptstyle \text{E}}}(\omega_{n})$
so the similar name is not accidental. However, as we have seen $\varDelta_{k}^{{\scriptscriptstyle \text{E}}}(\omega_{n})$
takes its significant shape in Matsubara space while ${\varDelta_{{\rm s}}^{{\scriptscriptstyle {\rm s}}}}_{k}$
does not have such a $\omega_{n}$ dependence and mimics the SC pairing
in its $k$ dependence in a way that densities of the interacting
system are reproduced.

\section{Summary and Conclusion}

In this work we have developed fully ab-initio methods to compute
the SC phase of a material in a magnetic field Zeeman-coupled to
the spin magnetization. In a unified notation we present a purely
GF based (the Eliashberg approach) and a Density Functional based
scheme.

In our DFT we have employed a SC KS system to reproduce the interacting
densities $n(\boldsymbol{r}),\boldsymbol{m}(\boldsymbol{r}),\boldsymbol{\chi}(\boldsymbol{r},\boldsymbol{r}^{\prime})$
and $\varGamma(\boldsymbol{R}_{1}\ldots\boldsymbol{R}_{N})$. The
SC KS system can be solved analytically using the SDA where we only
consider the singlet pairing of time reversed basis states. We have
derived $xc$-potentials in this case that include the electron-nuclear
interaction on the level of KS phonons and treats the Coulomb interaction
in the same footing without the need for any adjustable parameter.

As a second step we have applied similar approximations to the Dyson
equation starting from the SC KS system as a formally non-interacting
system. This procedure leads to the Eliashberg equations of a SC in
a magnetic field similar to those discussed in Ref.~\onlinecite{VonsovskySuperconductivityTransitionMetals}.

While SpinSCDFT allows to include the full Coulomb potential and promises
  numerically efficient calculations tor real materials, the direct
GF approach is, instead, valuable to get direct physical insights to
develop approximations and further improve the SpinSCDFT scheme. 

The theoretical framework presented in this work
allows to compute the phenomenon of coexistence and competition of
SC with magnetism from first principles. Especially in connection
with the discovery of Fe superconductors this was intensively studied
in recent years.

In the subsequent part {\bf II}, we will discuss a detailed numerical implementation
of the equations presented in this work, i.e. the linear and non-linear functionals and
the Eliashberg equations without Coulomb interactions. Further we will
introduce a $\rm{G}_{0}\rm{W}_{0}$ scheme to obtain the excitation spectrum starting
from a SpinSCDFT calculation. 

\appendix

\section{Formulas For The Matsubara Sums\label{sec:FormulasMatsubaraSums}}

In the potential terms it appears the Matsubara summation

\begin{equation}
P_{s}(E,E^{\prime})=\frac{1}{\beta}\sum_{n}\frac{1}{(\mbox{i}\omega_{n}-E)(\mbox{i}\omega_{n}-E^{\prime})}\label{eq:AppendixPotentialMatzubaraSum}
\end{equation}
This is analytically evaluated with the result
\begin{eqnarray}
P_{s}(E,E^{\prime}) & = & \frac{f_{\beta}(E)-f_{\beta}(E^{\prime})}{E-E^{\prime}}\\
\lim_{E^{\prime}\rightarrow E}P_{s}(E,E^{\prime}) & = & \partial_{E}f_{\beta}(E)=-\beta f_{\beta}(E)f_{\beta}(-E)
\end{eqnarray}
where the symmetries $P_{s}(E,E^{\prime})=P_{s}(-E,-E^{\prime})$
and $P_{s}(E,E^{\prime})=P_{s}(E^{\prime},E)$ hold. 
The Matsubara frequency summation 
\begin{eqnarray}
I(\varOmega,E_{1},E_{2},E_{3}) & \!= & \!\frac{1}{\beta^{2}}\sum_{nn^{\prime}}\frac{1}{\text{i}\omega_{n}-E_{1}}\frac{1}{\text{i}(\omega_{n}-\omega_{n^{\prime}})-\varOmega}\times\nonumber \\
 &  & \times\frac{1}{\text{i}\omega_{n^{\prime}}-E_{2}}\frac{1}{\text{i}\omega_{n}-E_{3}}\label{eq:FirstTerm_Ifreq_int}\\
L(\varOmega,E_{1},E_{2},E_{3}) & \!= & \! I(-\!\varOmega,E_{1},E_{2},E_{3})-\! I(\varOmega,E_{1},E_{2},E_{3})\nonumber \\
\label{eq:LMatzubaraSum}
\end{eqnarray}
is also in principle straightforward. However the resulting formulas are rather large
and computer algebra becomes essential for the evaluation of residues and limiting behaviours,
 necessary for a numerical implementation. Note that a partial summation leads to
\begin{equation}
L(\varOmega,E_{1},E_{2},E_{3})=\frac{1}{\beta}\sum_{n}\frac{M_{{\scriptscriptstyle {\rm ph}}}(\varOmega,E_{2},\omega_{n})}{(\text{i}\omega_{n}-E_{1})(\text{i}\omega_{n}-E_{3})}.\label{eq:ConnectionSelfEnergySumFullSum}
\end{equation}
From the definition we observe the following symmetry relations
\begin{eqnarray}
L(\varOmega,E_{1},E_{2},E_{3}) & = & L(\varOmega,E_{3},E_{2},E_{1})\label{eq:LSymm1To3}\\
L(-\varOmega,E_{1},E_{2},E_{3}) & = & -L(\varOmega,E_{3},E_{2},E_{1})\\
\bigl(L(\varOmega,E_{1},E_{2},E_{3})\bigr)^{\ast} & = & -L(\varOmega,-E_{1},-E_{2},-E_{3})\label{eq:LSymInvAll}
\end{eqnarray}
Evaluation of the Coulomb requires the following summation
\begin{eqnarray}
L_{{\rm {\scriptscriptstyle C}}}(E_{1},E_{2},E_{3}) & = & \frac{1}{\beta^{2}}\sum_{nn^{\prime}}\frac{1}{\text{i}\omega_{n^{\prime}}-E_{2}}\times\nonumber \\
 &  & \times\frac{1}{\text{i}\omega_{n}-E_{1}}\frac{1}{\text{i}\omega_{n}-E_{3}}\\
 & = & f_{\beta}(E_{2})P_{{\rm s}}(E_{1},E_{3})
\end{eqnarray}
Using \textsf{Mathematica}, we evaluate the sums Eqs.~(\ref{eq:FirstTerm_Ifreq_int}) and \eqref{eq:LMatzubaraSum}
to\begin{widetext}
\begin{eqnarray}
L(\varOmega,E_{1},E_{2},E_{3}) & = & \Biggl(\frac{f_{\beta}(E_{2})n_{\beta}(\varOmega)}{(E_{2}-E_{1}+\varOmega)(E_{2}-E_{3}+\varOmega)}+\frac{f_{\beta}(E_{2})\bigl(1+n_{\beta}(\varOmega)\bigr)}{(E_{1}-E_{2}+\varOmega)(E_{3}-E_{2}+\varOmega)}+\frac{f_{\beta}(E_{1})\bigl(1-f_{\beta}(E_{2})+n_{\beta}(\varOmega)\bigr)}{(E_{1}-E_{3})(E_{1}-E_{2}-\varOmega)}\nonumber \\
 &  & +\frac{f_{\beta}(E_{3})\bigl(1-f_{\beta}(E_{2})+n_{\beta}(\varOmega)\bigr)}{(E_{1}-E_{3})(E_{2}-E_{3}+\varOmega)}+\frac{f_{\beta}(E_{1})\bigl(f_{\beta}(E_{2})+n_{\beta}(\varOmega)\bigr)}{(E_{1}-E_{3})(E_{1}-E_{2}+\varOmega)}+\frac{f_{\beta}(E_{3})\bigl(f_{\beta}(E_{2})+n_{\beta}(\varOmega)\bigr)}{(E_{3}-E_{1})(E_{3}-E_{2}+\varOmega)}\Biggr)\,.\label{eq:FirstTerm_LfreqInt}
\end{eqnarray}
Clearly some points, e.g.~$E_{1}=E_{3}$ are numerically problematic,
so whenever $E_{1}\approx E_{3}$ we may have to evaluate the limiting
formula instead. In general, the various limits where the denominators
are zero, all exist and can be computed explicitly, again using \textsf{Mathematica}.
The results are 
\begin{eqnarray}
\lim_{E_{1}\rightarrow E_{3}}L(\varOmega,E_{1},E_{2},E_{3}) & = & f_{\beta}(E_{2})\biggl(\frac{n_{\beta}(\varOmega)}{(E_{2}-E_{3}+\varOmega)^{2}}+\frac{1+n_{\beta}(\varOmega)}{(E_{2}-E_{3}-\varOmega)^{2}}\biggr)-f_{\beta}(E_{3})\biggl(\frac{f_{\beta}(-E_{2})+n_{\beta}(\varOmega)}{(E_{2}-E_{3}+\varOmega)^{2}}+\nonumber \\
 &  & \hspace{-2.5cm}\frac{f_{\beta}(E_{2})+n_{\beta}(\varOmega)}{(E_{2}-E_{3}-\varOmega)^{2}}-\frac{\beta f_{\beta}(-E_{3})}{(E_{2}-E_{3})^{2}-(\varOmega)^{2}}\Bigl(\bigl(f_{\beta}(E_{2})-f_{\beta}(-E_{2})\bigr)\varOmega+\bigl(2n_{\beta}(\varOmega)+1\bigr)(E_{2}-E_{3})\Bigr)\biggr)\,,
\end{eqnarray}
\begin{eqnarray}
\lim_{\varOmega\rightarrow E_{3}-E_{2}}\lim_{E_{1}\rightarrow E_{3}}L(\varOmega,E_{1},E_{2},E_{3}) & = & \beta\frac{\bigl(1+f_{\beta}(E_{2})+n_{\beta}(E_{3}-E_{2})\bigr)f_{\beta}(-E_{3})f_{\beta}(E_{3})}{2(E_{2}-E_{3})}+\nonumber \\
 &  & \hspace{-3cm}+\frac{f_{\beta}(E_{2})+f_{\beta}(E_{3})\bigl(1-2f_{\beta}(E_{2})\bigr)}{4(E_{2}-E_{3})^{2}}+\beta^{2}f_{\beta}(-E_{2})\bigl(2+n_{\beta}(E_{3}-E_{2})\bigr)f_{\beta}(E_{3})\bigl(\frac{1}{2}-f_{\beta}(E_{3})\bigr)\,,
\end{eqnarray}
\begin{eqnarray}
\lim_{\varOmega\rightarrow E_{1}-E_{2}}L(\varOmega,E_{1},E_{2},E_{3}) & = & \frac{f_{\beta}(E_{1})\bigl(f_{\beta}(E_{2})+n_{\beta}(E_{1}-E_{2})\bigr)}{2(E_{1}-E_{2})(E_{1}-E_{3})}+\frac{f_{\beta}(E_{2})\bigl(1+n_{\beta}(E_{1}-E_{2})\bigr)}{2(E_{1}-E_{2})(E_{1}-2E_{2}+E_{3})}+\nonumber \\
 &  & +\frac{f_{\beta}(E_{3})\bigl(n_{\beta}(E_{1}-E_{2})+f_{\beta}(-E_{2})\bigr)-f_{\beta}(E_{2})n_{\beta}(E_{1}-E_{2})}{(E_{1}-E_{3})^{2}}\nonumber \\
 &  & +\frac{f_{\beta}(E_{3})\bigl(f_{\beta}(E_{2})+n_{\beta}(E_{1}-E_{2})\bigr)}{2(E_{3}-E_{1})(E_{1}-2E_{2}+E_{3})}+\beta\frac{f_{\beta}(-E_{1})f_{\beta}(E_{2})n_{\beta}(E_{1}-E_{2})}{E_{3}-E_{1}}\,,
\end{eqnarray}
\begin{eqnarray}
\lim_{E_{1}\rightarrow2E_{2}-E_{3}}\lim_{\varOmega\rightarrow E_{1}-E_{2}}L(\varOmega,E_{1},E_{2},E_{3}) & = & \frac{f_{\beta}(E_{2})\bigl(1+n_{\beta}(E_{2}-E_{3})\bigr)}{2(E_{2}-E_{3})}\biggl(\beta f_{\beta}(-E_{3})-\frac{1}{2(E_{2}-E_{3})}\biggr)\,.
\end{eqnarray}
We point out here that the Limit $\varOmega_{\boldsymbol{q}\lambda}\rightarrow0$
does not exist. It is however unimportant as the $g_{ij}^{\lambda\boldsymbol{q}}$
go to zero in the limit $\varOmega\rightarrow0$ faster than $L$ diverges.\end{widetext}

\bibliographystyle{apsrev4-1}
\bibliography{references}

\end{document}